\documentclass[
  journal=pasa,
  manuscript=research-paper, 
  year=2020,
  volume=37,
]{cup-journal}

\usepackage{microtype,siunitx,booktabs}
\usepackage[nolist]{acronym}
\usepackage{amssymb}
\usepackage{amsmath}
\usepackage{stmaryrd} 

\usepackage{cleveref} 
\usepackage[normalem]{ulem} 

\usepackage{subcaption}

\usepackage{hyphenat}
\lccode`\/`\/
\newcommand{\acrocite}[2]{\acl{#1} \citep[\acs{#1},][]{#2}\acused{#1}}

\newcommand{\arcsec}{$^{\prime\prime}$}
\newcommand{\arcmin}{$^{\prime}$}

\title{Measuring photometric redshifts for high-redshift radio source surveys}

\author{K. J. Luken}
\affiliation{School of Science, Western Sydney University, Second Ave, Kingswood, NSW, 2747, Australia}
\alsoaffiliation{Data61, CSIRO, PO Box 76, Epping, NSW 1710, Australia}
\email[K. J. Luken]{kieran@luken.au}

\author{R. P. Norris}
\affiliation{School of Science, Western Sydney University, Second Ave, Kingswood, NSW, 2747, Australia}
\alsoaffiliation{CSIRO Space \& Astronomy, Australia Telescope National Facility, PO Box 76, Epping, NSW 1710, Australia}

\author{X. R. Wang}
\affiliation{Centre for Research in Mathematics and Data Science, Western Sydney University, 
Australia}
\alsoaffiliation{Data61, CSIRO, PO Box 76, Epping, NSW 1710, Australia}

\author{L. A. F. Park}
\affiliation{Centre for Research in Mathematics and Data Science, Western Sydney University, 
Australia}

\author{Y. Guo}
\affiliation{Data61, CSIRO, PO Box 76, Epping, NSW 1710, Australia}

\author{M. D. Filipovi\'c}
\affiliation{School of Science, Western Sydney University, Second Ave, Kingswood, NSW, 2747, Australia}

\doi{10.1017/pasa.2020.32}

\received {09/04/2023}
\revised  {}
\accepted {}
\published{}

\keywords{galaxies: distances and redshifts < Galaxies; galaxies: high-redshift < Galaxies; radio continuum: galaxies < Sources as a function of wavelength; methods: statistical < Astronomical instrumentation, methods and techniques; methods: analytical < Astronomical instrumentation, methods and techniques} 

\begin{acronym}[ELAIS-S1]
\acro{2dfGRS}{2dF Galaxy Redshift Survey}
\acro{AGN}{Active Galactic Nucleus}
\acrodefplural{AGN}{Active Galactic Nuclei}
\acro{ANOVA}{Analysis of Variance}
\acro{ASKAP}{Australian Square Kilometre Array Pathfinder}
\acro{ATCA}{Australia Telescope Compact Array}
\acro{ATLAS}{Australia Telescope Large Area Survey}
\acro{BIC}{Bayesian Information Criteria}
\acro{COSMOS}{COSMic evOlution Survey}
\acro{DECaLS}{Dark Energy Camera Legacy Survey}
\acro{DES}{Dark Energy Survey}
\acro{DT}{Decision Tree}
\acro{eCDFS}{extended Chandra Deep Field South}
\acro{ELAIS-S1}{European Large Area ISO Survey--South 1}
\acro{EM}{Expectation-Maximization}
\acro{EMU}{Evolutionary Map of the Universe}
\acro{EMU-PS}{Evolutionary Map of the Universe -- Pilot Survey 1}
\acro{FIRST}{Faint Images of the Radio Sky at Twenty-Centimeters}
\acro{GLEAM-X}{GaLactic and Extragalactic All-sky Murchison Widefield Array survey eXtended}
\acro{GLEAM}{GaLactic and Extragalactic All-sky Murchison Widefield Array survey}
\acro{GMM}{Gaussian Mixture Model}
\acro{GP}{Gaussian Process}
\acrodefplural{GP}{Gaussian Processes}
\acro{kNN}{$k$-Nearest Neighbours}
\acro{LMNN}{Large Margin Nearest Neighbour}
\acro{LOFAR}{LOw Frequency ARray}
\acro{LOTSS}{LOFAR Two-metre Sky Survey}
\acro{LSST}{Legacy Survey of Space and Time}
\acro{ML}{Machine Learning}
\acro{MLKR}{Metric Learning for Kernel Regression}
\acro{MOS}{Multi-Object Spectroscopy}
\acro{MSE}{Mean Square Error}
\acro{MWA}{Murchison Widefield Array}
\acro{NMAD}{Normalised Median Absolute Deviation}
\acro{NMI}{Normalised Mutual Information}
\acro{NN}{Neural Network}
\acro{NVSS}{NRAO VLA Sky Survey}
\acro{OzDES}{Australian Dark Energy Survey}
\acro{PSF}{Point Spread Function}
\acro{QSO}{Quasi-Stellar Object}
\acro{RGZ}{Radio Galaxy Zoo}
\acro{RF}{Random Forest}
\acro{SDSS}{Sloan Digital Sky Survey}
\acro{SED}{Spectral Energy Distribution}
\acro{SFG}{Star Forming Galaxy}
\acro{SKA}{Square Kilometre Array}
\acro{SKADS}{Square Kilometre Array Design Survey}
\acro{SST}{Spitzer Space Telescope}
\acro{SWIRE}{Spitzer Wide-Area Infrared Extragalactic Survey}
\acro{VLA}{Very Large Array}
\acro{VLASS}{Very Large Array Sky Survey}
\acro{WAVES}{Wide Area Vista Extragalactic Survey}
\end{acronym}

\begin{document}

\begin{abstract}
    With the advent of deep, all-sky radio surveys, the need for ancillary data to make the most of the new, high-quality radio data from surveys like the \ac{EMU}, \acl{GLEAM-X}, \acl{VLASS} and \acl{LOTSS} is growing rapidly. Radio surveys  produce significant numbers of \acp{AGN}, and have a significantly higher average redshift when compared with optical and infrared all-sky surveys. Thus, traditional methods of estimating redshift are challenged, with spectroscopic surveys not reaching the redshift depth of radio surveys, and \acp{AGN} making it difficult for template fitting methods to accurately model the source. \ac{ML} methods have been used, but efforts have typically been directed towards optically selected samples, or samples at significantly lower redshift than expected from upcoming radio surveys. This work compiles and homogenises a radio-selected dataset from both the northern hemisphere (making use of \acl{SDSS} optical photometry), and southern hemisphere (making use of \acl{DES} optical photometry). We then test commonly used \ac{ML} algorithms such as \ac{kNN}, \acl{RF}, ANNz and GPz on this monolithic radio-selected sample. We show that \ac{kNN} has the lowest percentage of catastrophic outliers, providing the best match for the majority of science cases in the \ac{EMU} survey. We note that the wider redshift range of the combined dataset used allows for estimation of sources up to $z = 3$ before random scatter begins to dominate. When binning the data into redshift bins and treating the problem as a classification problem, we are able to correctly identify $\approx$76\% of the highest redshift sources --- sources at redshift $z > 2.51$ --- as being in either the highest bin ($z > 2.51$), or second highest ($z = 2.25$). 
\end{abstract}

\section{Introduction}
\label{sec:intro}
\acresetall

Radio astronomy is at a cross-roads. With large survey telescopes like the \acrocite{ASKAP}{hotanAustralianSquareKilometre2021}, \acrocite{MWA}{tingayMurchisonWidefieldArray2013}, \acrocite{LOFAR}{vanhaarlemLOFARLOwFrequencyARray2013}, and upgrades to the \acrocite{VLA}{thompsonVeryLargeArray1980} producing catalogues of up to tens of millions of new radio sources, traditional methods of producing science are struggling to keep up. New methods need to be developed to pick up the shortfall.

One of the most essential pieces of knowledge about an astronomical object is its redshift. From this measurement the object's age and distance can be gleaned, and its redshift used in combination with photometric measurements to estimate a myriad of other features. \\

Traditionally, redshift has been measured spectroscopically. However, even with modern \ac{MOS} instrumentation, the tens of millions of radio galaxies expected to be discovered in the coming years will by far  outstrip the world's spectroscopic capacity. For example, the 17$^{\mathrm{th}}$ data release of the \acrocite{SDSS}{sdss_17}\footnote{\url{https://www.sdss.org/dr17/scope/}} is currently the largest source of spectroscopic redshifts, with $\approx4.8$ million redshifts measured -- significantly less than the tens of millions of sources the \acrocite{EMU}{norrisEMUEvolutionaryMap2011}, \acrocite{GLEAM-X}{gleam-x}, \acrocite{LOTSS}{shimwellLOFARTwometreSky2017}, and \acrocite{VLASS}{murphyVlaSkySurvey2015} is expected to deliver, even if all redshifts measured were focused exclusively on radio galaxies. Future spectroscopic surveys like the \acrocite{WAVES}{driverWideAreaVISTA2016} are expected to increase the number of spectroscopically known redshifts by another $\sim$2.5 million sources, but this will still not be enough. 

Alternatively, photometric template fitting \citep{baumPhotoelectricDeterminationsRedshifts1957,lohPhotometricRedshiftsGalaxies1986} has been highly effective at estimating the redshift of sources for many years, and is able to achieve accuracies approaching those of spectroscopically measured redshifts \citep{ilbertCosmosPhotometricRedshifts2009}. However, the breadth and depth of measured photometric bands required for this level of accuracy is unavailable for the majority of sources detected by radio surveys like the \ac{EMU}, \ac{GLEAM-X}, \ac{LOTSS}, and \ac{VLASS}  surveys. Additionally,  radio galaxies in particular suffer in the photometric template fitting regimes, partly due to a lack of specialised templates, and partly due to the difficulty of separating out the star formation emission from the black hole emission \citep{ salvatoManyFlavoursPhotometric2018, norris2019}. 

Finally, like most problems, \ac{ML} techniques have been applied to the problem of estimating redshift. From the simple algorithms like the \acrocite{kNN}{coverNearestNeighborPattern1967} in \citet{ballRobustMachineLearning2007}, \citet{ballRobustMachineLearning2008}, \citet{oyaizuGalaxyPhotometricRedshift2008}, \citet{zhangEstimatingPhotometricRedshifts2013}, \citet{kuglerDeterminingSpectroscopicRedshifts2015}, \citet{cavuotiMETAPHORMachinelearningbasedMethod2017}, \citet{lukenPreliminaryResultsUsing2019}, and  \citet{lukenMissingDataImputation2021} and \acrocite{RF}{hoRandomDecisionForests1995,breimanRandomForests2001} in \citet{cavuotiPhotometricRedshiftsQuasi2012}, \citet{cavuotiPhotometricRedshiftEstimation2015}, \citet{hoyleMeasuringPhotometricRedshifts2016}, \citet{sadehANNz2PhotometricRedshift2016}, \citet{cavuotiMETAPHORMachinelearningbasedMethod2017}, and \citet{pasquet-itamDeepLearningApproach2018}, to more complex algorithms like \acp{NN} in \citet{firthEstimatingPhotometricRedshifts2003}, \citet{tagliaferriNeuralNetworksPhotometric2003}, \citet{collisterANNzEstimatingPhotometric2004}, \citet{brodwinPhotometricRedshiftsIRAC2006}, \citet{oyaizuGalaxyPhotometricRedshift2008}, \citet{hoyleMeasuringPhotometricRedshifts2016}, \citet{sadehANNz2PhotometricRedshift2016}, \citet{curranQSOPhotometricRedshifts2020}, \citet{curranQSOPhotometricRedshifts2021}, \citet{curran_2022_1}, and \citet{curran_2022_2} and \acp{GP} in \citet{duncanPhotometricRedshiftsNext2018}, \citet{duncanPhotometricRedshiftsNext2018a}, and \citet{duncanLOFARTwometerSky2021}, using the GPz software. Some studies --- for example \citet{pasquet-itamDeepLearningApproach2018} and \citet{disantoPhotometricRedshiftEstimation2018} --- make use of the images themselves, rather than photometry measured from the images. Typically though, \ac{ML} algorithms  aren't tested in a manner suitable for large-scale radio surveys --- \ac{ML} algorithms are generally evaluated using data from fields like the \ac{COSMOS}, where there are many (up to 31) different photometric bands measured for each source --- far beyond what is available to all-sky surveys, or on data from the \ac{SDSS}, where either the Galaxy sample is used, containing millions of galaxies with optical photometry  and a spectroscopically measured redshift (but restricted to $z \lesssim 0.8$), or the \ac{QSO} sample is used, containing quasars out to a significantly higher redshift, at the cost of lower source count. 

As noted by \citet{salvatoManyFlavoursPhotometric2018}, \ac{ML}-based methods frequently perform better then traditional template fitting methods when the density of observed filters is lacking, or when the sample being estimated contain rarer sub-types like radio or x-ray \ac{AGN}. The drawback, however, is that \ac{ML} methods still require a representative sample of these galaxies to be able to model the features well enough to acceptably predict their redshift. One of the biggest issues with any \ac{ML} algorithm is finding a representative sample to train the model with. For redshift estimation, this generally requires having spectroscopic surveys containing sources to a similar depth as the sources being predicted (or reliably photometrically estimated redshift -- see \citet{speagle_2019_redshift_photo_spec} for an in-depth investigation). 

\begin{figure}
    \centering
    \includegraphics[trim=0 0 0 0, width=\textwidth]{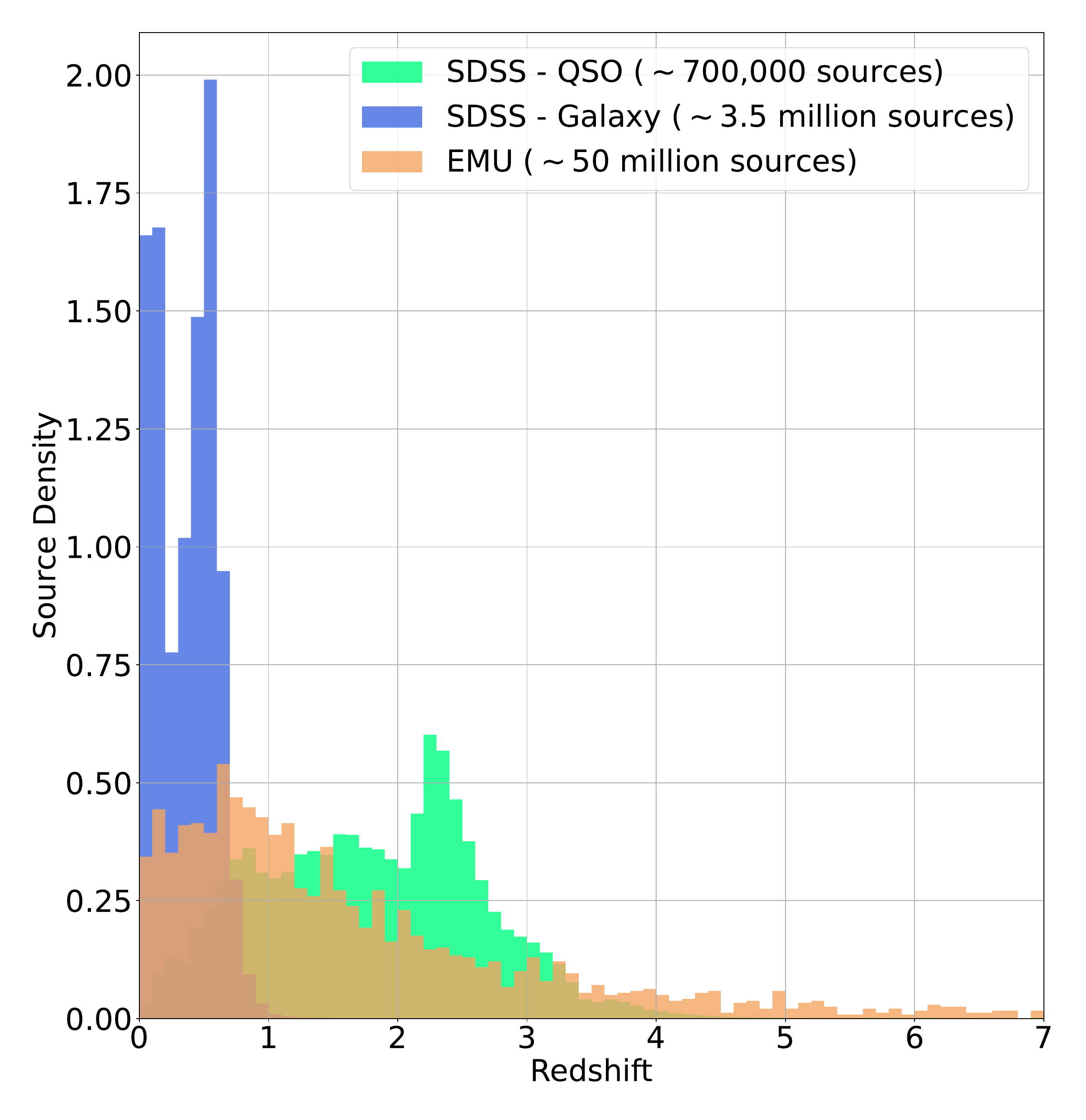}
    \caption{Histogram showing the density of sources at different redshifts in the SDSS Galaxy Sample (blue), SDSS QSO Sample (green) and the \acrocite{SKADS}{levrierMappingSKASimulated2009} simulation trimmed to expected \ac{EMU} depth \citep{norrisEMUEvolutionaryMap2011}.}
    \label{fig:skads_sdss_hist}
\end{figure}

An example of the expected redshift distribution of the \ac{EMU} survey, compared with the \ac{SDSS} Galaxy and \ac{QSO} samples is presented in Figure~\ref{fig:skads_sdss_hist}, demonstrating the differences in redshift distributions -- one reason why radio samples are typically more difficult to estimate than optically selected samples. Training samples are often not entirely representative of the data being predicted. 

Further, \citet{duncan_2022} compares their results with \citet{Duncan_LoTSS_2019}, showing that for most populations of galaxies, taking the additional step of training a \ac{GMM} to split optically selected datasets into more representative samples improves estimates across all measured error metrics. However, \citet{duncan_2022} notes that redshift estimates for optically luminous QSO have lower error estimates when training exclusively on representative data as in \citet{Duncan_LoTSS_2019}, compared with using the \ac{GMM} prior to estimation. Two reasons are postulated for this -- one being the addition of the $i$ and $y$ bands used by \citet{Duncan_LoTSS_2019}, with the additional reason being the specific training on the representative sample, rather than a generalised approach. 

Finally, when \ac{ML} models have been trained on radio selected samples, they have typically been focused on achieving the best possible accuracy, with model parameters optimised based on the average accuracy. While this approach is entirely appropriate for other use cases, the preferred parameter to optimise in this work is the \emph{Outlier Rate} --- the percentage of sources where the estimated redshift is determined to have catastrophically failed (further details in Section~\ref{sec:methods_errors}). 

This subtle change optimises the results for key science goals of surveys such as EMU in which the number of catastrophic outliers is more important than the accuracy of each redshift estimate. For example, constraining non-Gaussianity \citep{raccanelli_cosmology}, or measuring the evolution of the cosmic star formation rate over cosmic time \citep{2006ApJ...651..142H} do not require accurate estimates of redshift, but suffer greatly if redshifts are significantly incorrect.

In light of these struggles using optically selected samples to estimate the redshift of radio-selected samples, we create a new radio-selected training sample, taken from the northern hemisphere (selected from the \acrocite{FIRST}{beckerFIRSTSurveyFaint1995} and \acrocite{NVSS}{condonNRAOVLASky1998}) using \ac{SDSS} spectroscopy and photometry, and combining it with southern hemisphere data (selected from the \acrocite{ATLAS}{norrisDeepATLASRadio2006,franzenATLASThirdRelease2015} and Stripe82 \citep{Stripe_Hodge_2011,Stripe_Prescott_2018}, where the \ac{ATLAS} data contains \acrocite{DES}{darkenergysurveycollaborationDarkEnergySurvey2016} photometry, and the Stripe82 field contains both \ac{SDSS} and \ac{DES} photometry. All fields contain AllWISE infrared photometry. 

With this large radio-selected dataset, we compare four commonly used \ac{ML} algorithms and softwares -- \ac{kNN}, \ac{RF}, GPz, and ANNz. Where possible, we compare these methods using both a regression, and classification mode, as discussed in \citet{lukenMissingDataImputation2021,luken_2022}. In order to better cater to the \ac{EMU} science goals\footnote{\url{http://askap.pbworks.com/w/page/88123540/KeyProjects}}, instead of comparing the overall accuracies of each method, we compare the outlier rates -- the percentage of sources that have catastrophically failed. 

In this work, we pose the research question: Given the upcoming radio surveys (specifically the \ac{EMU} survey), which \ac{ML} algorithm provides the best performance for the estimation of radio galaxy's redshift, where best performance is measured by the outlier rate.

\subsection{Overall Contributions of this Study}
Overall, our contributions for this study include:
\begin{itemize}
    \item An in-depth investigation of the \ac{DES} and \ac{SDSS} optical photometry, and its compatibility, specifically examining the modifications needed to use both surveys for the estimation of redshift using Machine Learning. 
    \item The construction of a representative and homogenous (where possible) training set, available to be used for the estimation of redshift for radio-selected samples.
    \item The comparison of multiple widely used \ac{ML} algorithms, providing a like-for-like comparison on the same dataset. 
    \item The comparison of classification- and regression-based methods where possible. 
\end{itemize}

\subsection{Knowledge Gap}
\begin{itemize}
    \item Current Template Fitting methods require better photometric coverage than is typical for all-sky radio surveys, and are are based on a set of templates that are not well-matched to those of galaxies that host radio sources.
    \item Current ML techniques are typically trained and tested on wide, shallow surveys, limited to z $<$ 0.7, or specific, optically selected samples. Where they are not trained on restricted samples, they are typically optimised for best accuracy, rather than minimising the number of catastrophic failures. 
    \item We are looking at a combination of datasets in order to better match the expected density of sources, as well as comparing against current methods used in literature in order to best prepare for the next generation of radio surveys. 
\end{itemize}

\section{Data}
\label{sec:data}

In this section we outline the photometry used and sources of data (Section~\ref{sec:data_description}), the steps taken to homogenise the northern sky \ac{SDSS} and southern sky \ac{DES} optical photometry (Section~\ref{sec:data_homogenisation}), and the process of binning the data in redshift space in order to test classification modes of the different algorithms (Section~\ref{sec:data_reg_and_class}).

\subsection{Data Description}
\label{sec:data_description}

As noted in Section~\ref{sec:intro}, most \ac{ML}-based techniques are typically focused on optically selected datasets, primarily based around the \ac{SDSS} datasets, providing high source counts of stars, galaxies, and \acp{QSO} with photometry (generally) in $u$, $g$, $r$, $i$, and $z$ bands, with a spectroscopically measured redshift --- generally using the \ac{SDSS} Galaxy or \ac{QSO} datasets, shown in Figure~\ref{fig:skads_sdss_hist}. In this work, the data are selected specifically to better represent the data expected from the upcoming \ac{EMU} Survey \citep{norrisEMUEvolutionaryMap2011} and \acrocite{EMU-PS}{EMUPS}. Towards this end, we only accept \ac{SDSS} objects with a counterpart in the \ac{NVSS} or \ac{FIRST} radio surveys.

\begin{figure}
    \centering
    \includegraphics[trim=0 0 0 0, width=\textwidth]{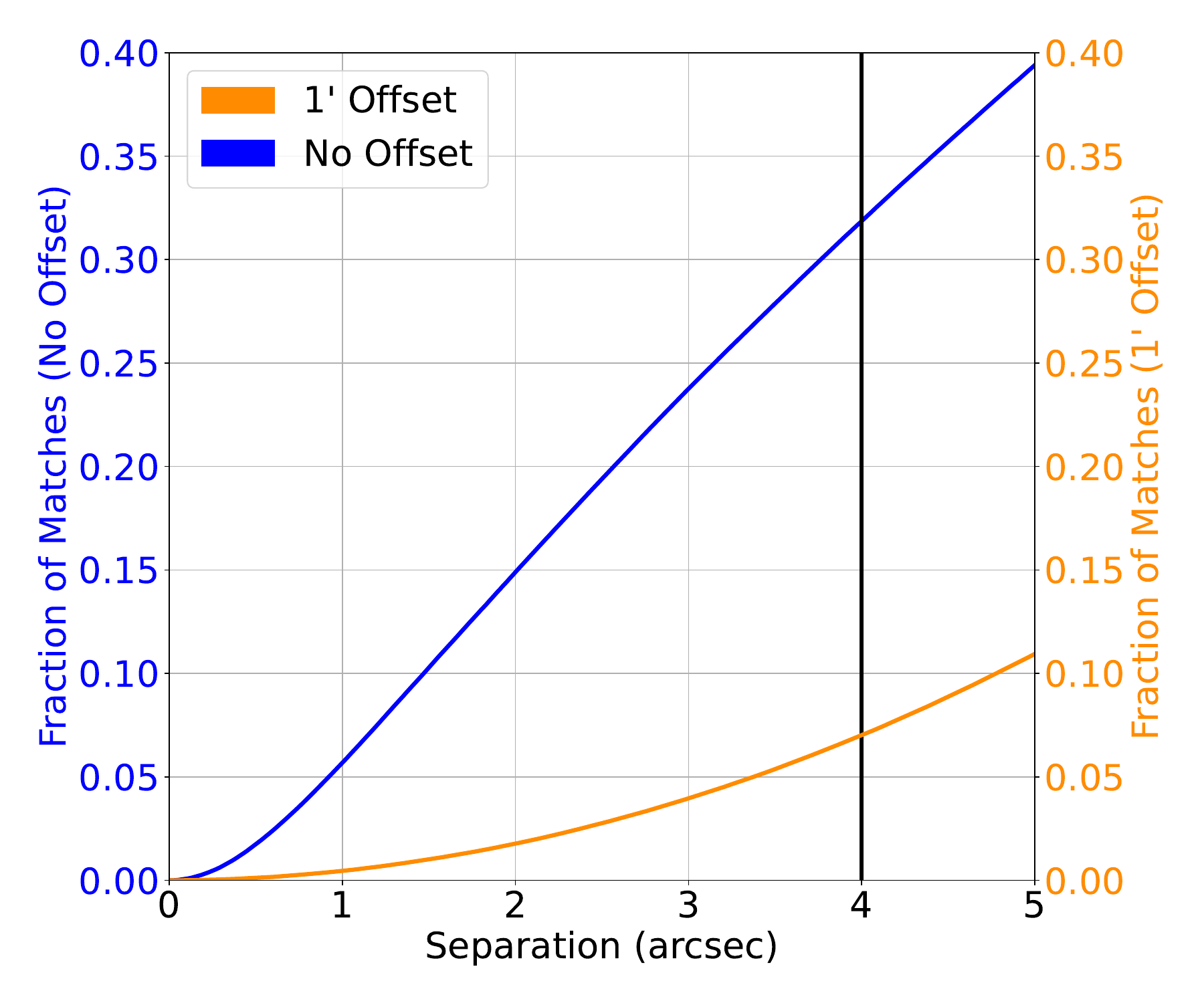}
    \caption{Cross-match completed between the NVSS Radio sample, and the AllWISE Infrared sample. The blue-line represents the straight nearest-neighbour cross-match between the two datasets, and the orange line represents the nearest-neighbour cross-match where 1\arcmin\, has been added to the declination of every radio source. The vertical black line denotes the chosen cutoff. }
    \label{fig:nvss_allwise_crossmatch}
\end{figure}

\begin{figure}
    \centering
    \includegraphics[trim=0 0 0 0, width=\textwidth]{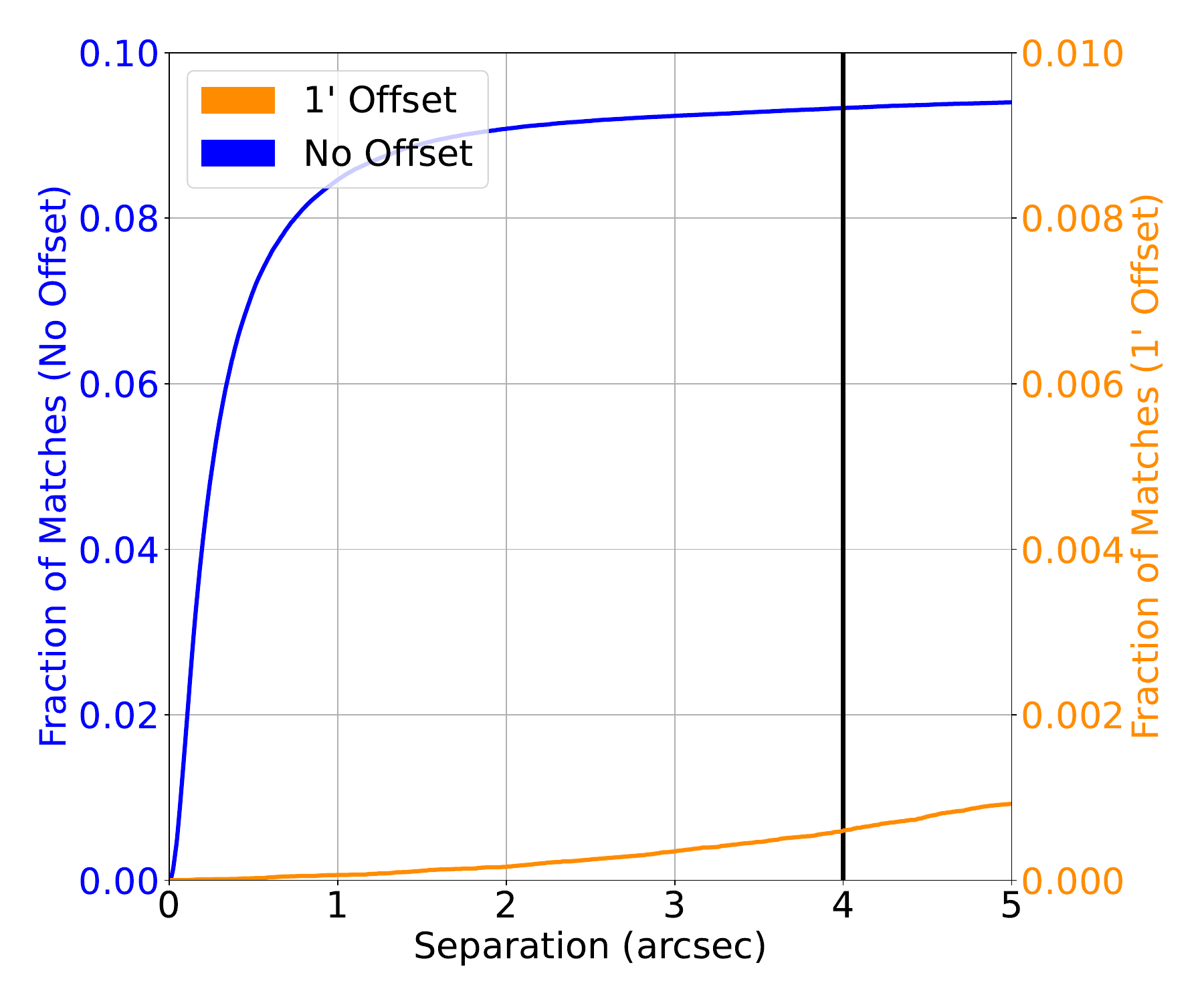}
    \caption{Similar to Figure~\ref{fig:nvss_allwise_crossmatch}, cross-matching the Northern Sky sample based on the \ac{NVSS} and \ac{RGZ} radio catalogues and AllWISE Infrared, with SDSS optical photometry and spectroscopic redshift. Note the different scale of the right-side y-axis.}
    \label{fig:sdss_ir_optical_crossmatch}
\end{figure}

\begin{figure}
    \centering
    \includegraphics[trim=0 0 0 0, width=\textwidth]{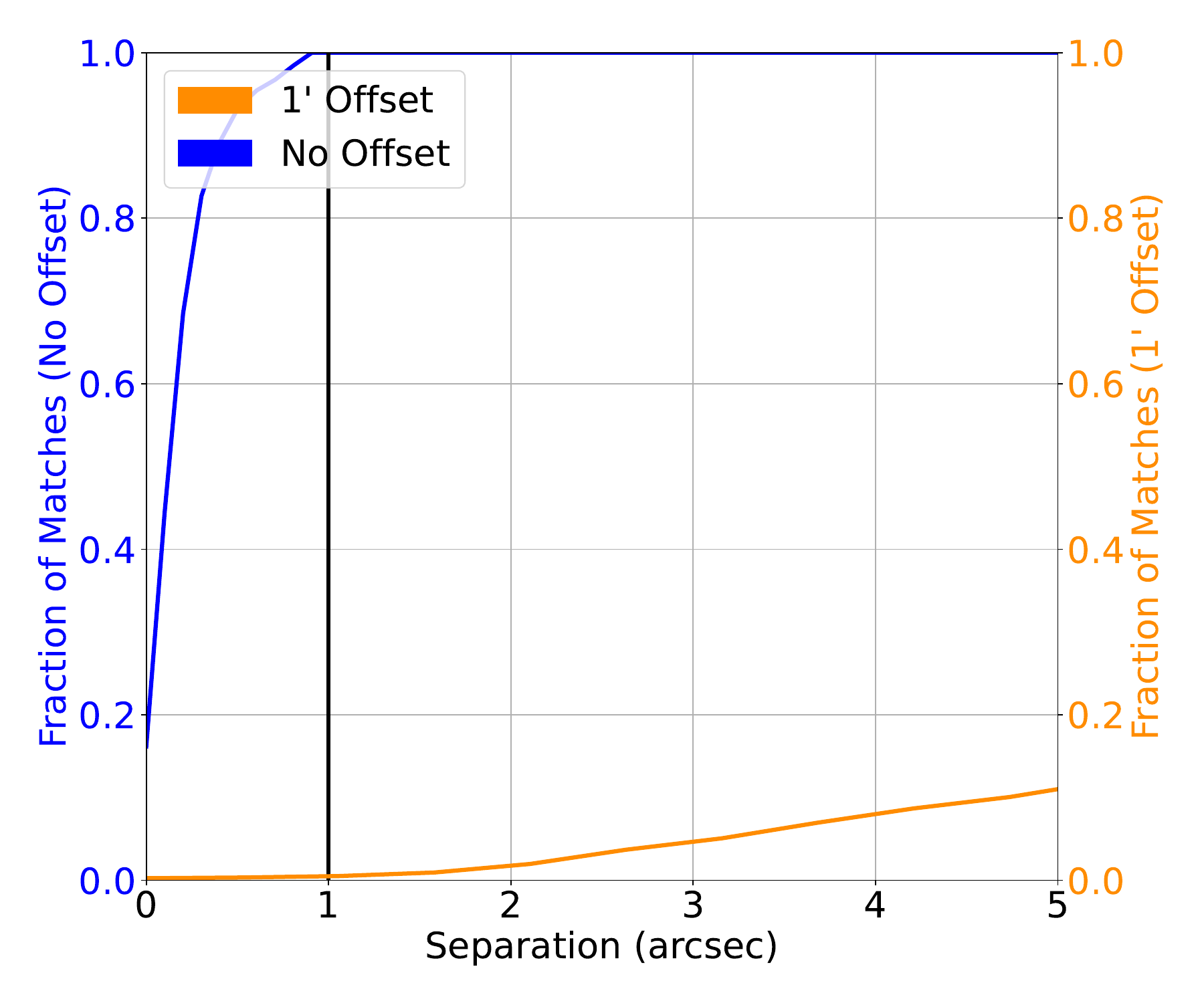}
    \caption{Similar to Figure~\ref{fig:nvss_allwise_crossmatch}, cross-matching the southern sky \ac{ATLAS} sample with the AllWISE catalogue, matching the \ac{SWIRE} positions with the AllWISE positions. }
    \label{fig:atlas_allwise}
\end{figure}

\begin{figure}
    \centering
    \includegraphics[trim=0 0 0 0, width=\textwidth]{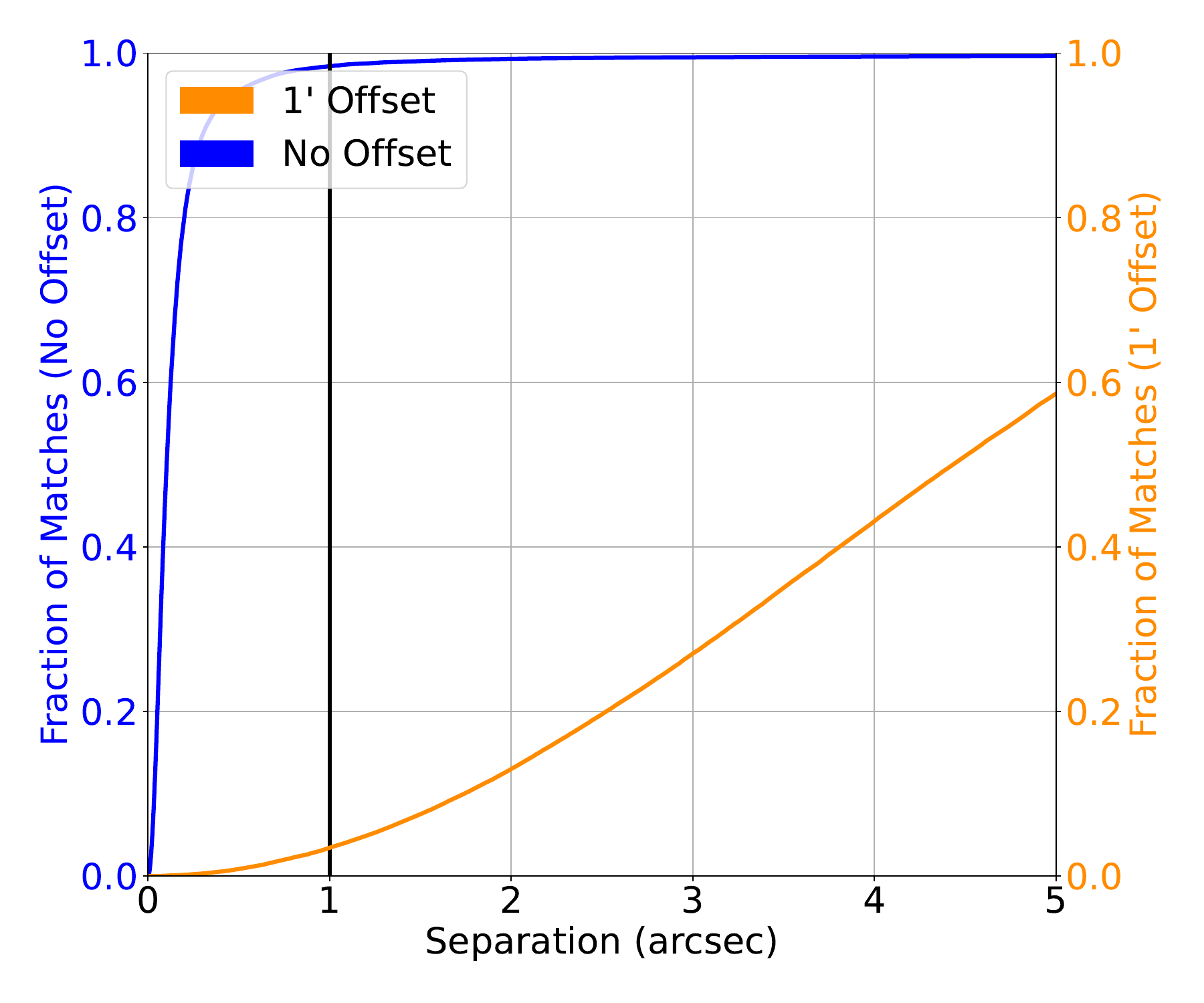}
    \caption{Similar to Figure~\ref{fig:nvss_allwise_crossmatch}, cross-matching the \ac{DES} optical photometry with the \ac{SDSS} optical photometry and spectroscopic redshift catalogues. }
    \label{fig:stripe_optical}
\end{figure}

\begin{figure}
    \centering
    \includegraphics[trim=0 0 0 0, width=\textwidth]{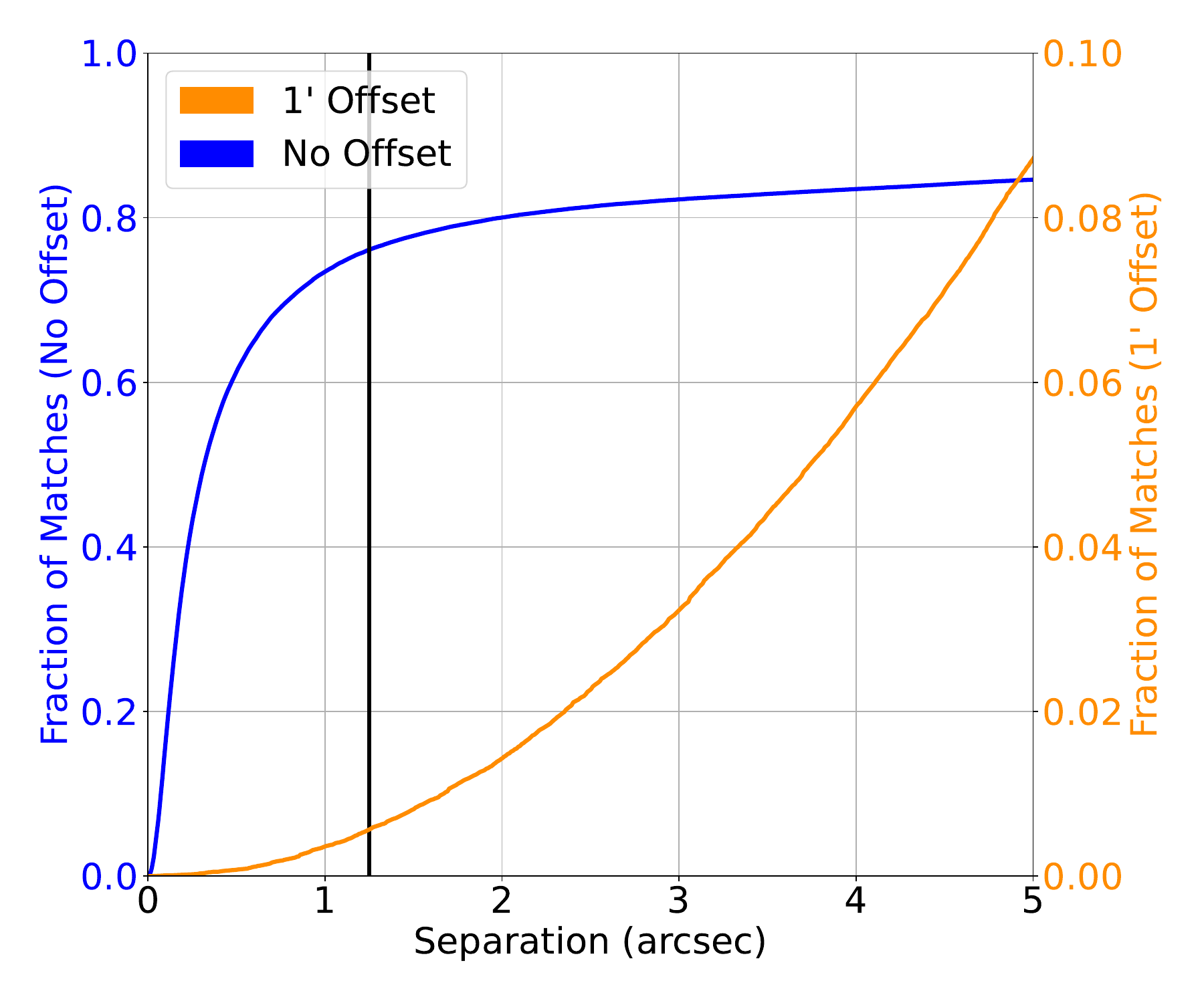}
    \caption{Similar to Figure~\ref{fig:nvss_allwise_crossmatch}, cross-matching the \ac{DES} positions with the AllWISE Infrared catalogue. Note the different scale of the right-side y-axis.}
    \label{fig:stripe_infrared}
\end{figure}

\begin{figure}
    \centering
    \includegraphics[trim=0 0 0 0, width=\textwidth]{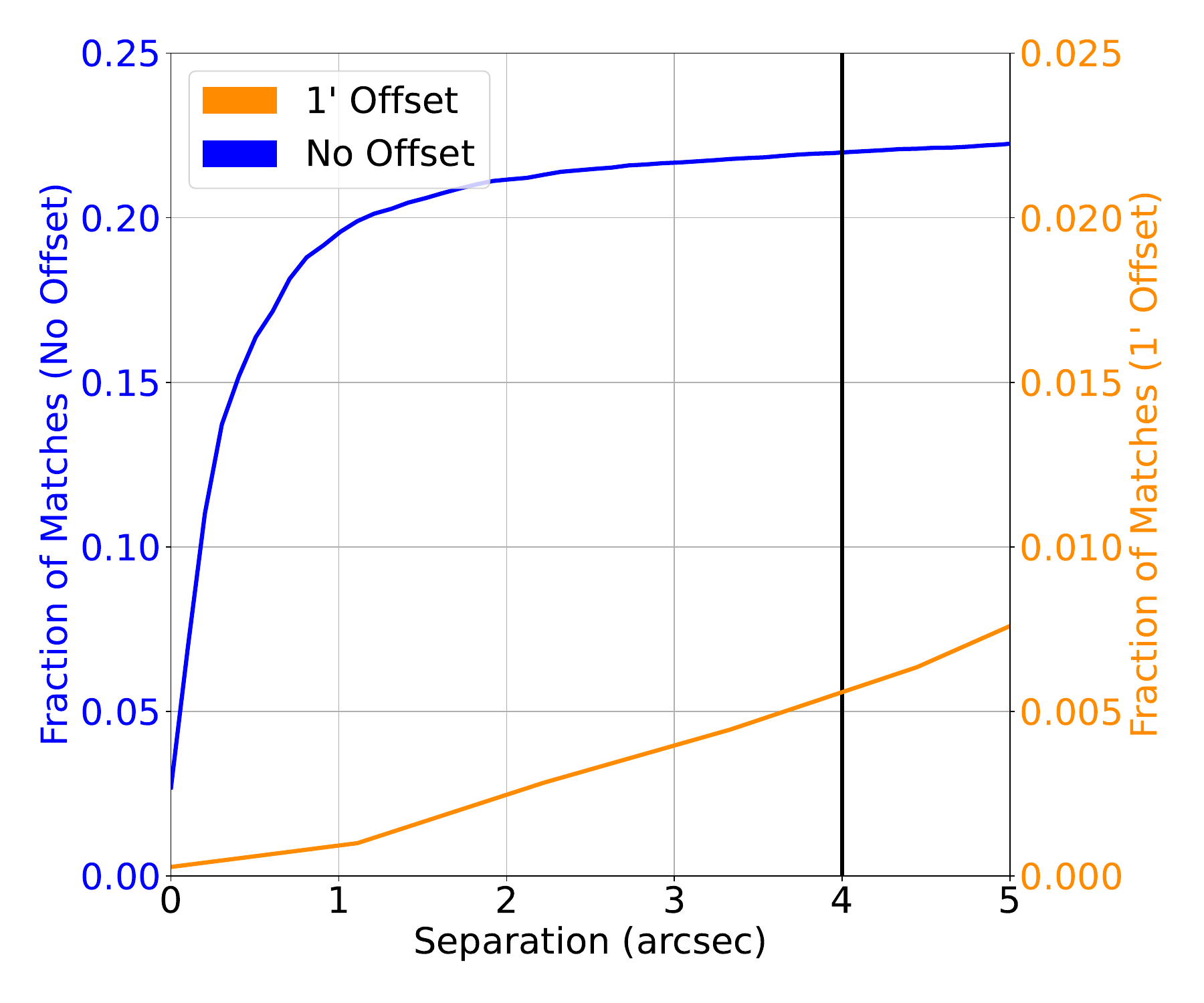}
    \caption{Similar to Figure~\ref{fig:nvss_allwise_crossmatch}, cross-matching the AllWISE positions with the \citet{Stripe_Hodge_2011} Radio catalogue. Note the different scale of the right-side y-axis.}
    \label{fig:stripe_hodge}
\end{figure}

\begin{figure}
    \centering
    \includegraphics[trim=0 0 0 0, width=\textwidth]{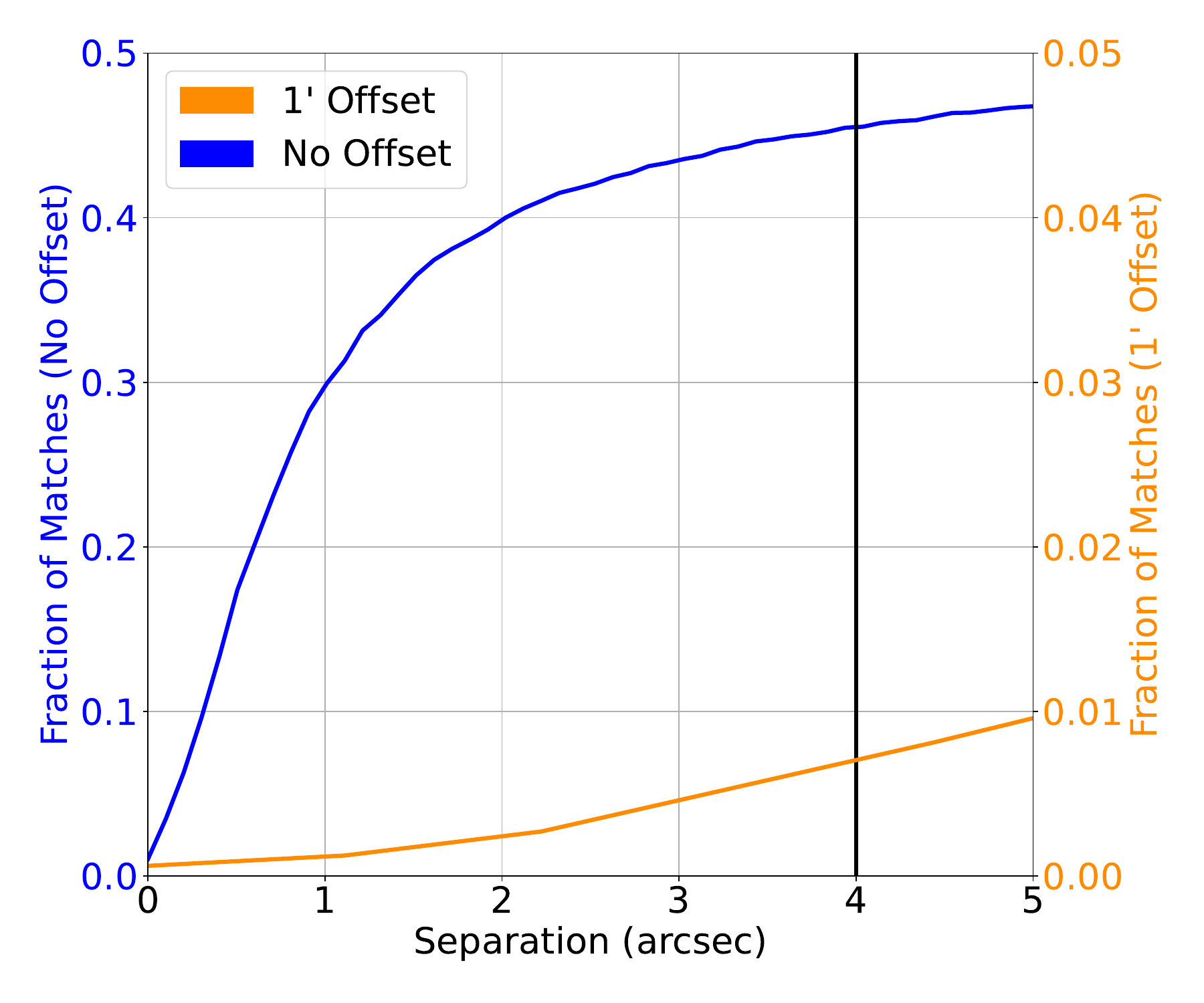}
    \caption{Similar to Figure~\ref{fig:nvss_allwise_crossmatch}, cross-matching the AllWISE positions with the \citet{Stripe_Prescott_2018} Radio catalogue. Note the different scale of the right-side y-axis.}
    \label{fig:stripe_prescott}
\end{figure}

This work compiles three datasets, each containing multiple features for comparison:


\begin{enumerate}
    \item Northern Sky --- \ac{RGZ} and \ac{NVSS}
    \begin{enumerate}
        \item Our Northern Sky dataset contains two radio samples --- the \ac{NVSS} sample, and the \ac{RGZ} \ac{FIRST}-based sample (where the \ac{RGZ} sample has been cross-matched with the AllWISE sample, explained in \citet{rgz} and Wong et al. (in prep.)).
        \item The \ac{NVSS} sample was cross-matched with AllWISE at 4\arcsec, providing 564,799 radio sources --- approximately 32\% of the \ac{NVSS} sample --- with an infrared counterpart. The \ac{NVSS}/AllWISE cross-match has an estimated 7\% --- 123,484 sources -- misclassification rate, where the misclassification rate is quantified by shifting the declination of all sources in the \ac{NVSS} catalogue by 1\arcmin and re-cross-matching based on the new declination, following the process described in \citet{norrisEvolutionaryMapUniverse2021}. Figure~\ref{fig:nvss_allwise_crossmatch} shows the classification/misclassification rates as a function of angular separation, and is used to determine the optimum cross-match radius. 
        \item The \ac{NVSS} and \ac{RGZ} samples were then combined, removing duplicates based on the AllWISE unique identifier, providing 613,551 radio sources with AllWISE detections. 
        \item The northern sky radio/infrared catalogue was then cross-matched against the \ac{SDSS} catalogue (providing both optical photometry, and spectroscopic redshifts) based on the infrared source locations --- a radio-infrared cross-match tends to be more reliable, when compared with the radio-optical cross-match \citep{swanMultifrequencyMatchingClassification2018} --- at 4\arcsec, providing a classification/misclassification rate of 9.33\%/0.06\% (55,716/348 sources) (Figure~\ref{fig:sdss_ir_optical_crossmatch}).
        \item Finally, all sources from the Stripe82 Equatorial region  were removed (sources with an RA between $9^{\circ}$ and $36^{\circ}$ or $330^{\circ} $ and $350^{\circ}$, and DEC between $\text{-}1.5^{\circ}$ and $1.5^{\circ}$). 
        \item The final Northern Sky Sample contains 55,452 radio-selected sources with a spectroscopically measured redshift, SDSS $g$, $r$, $i$ and $z$ magnitudes measured using Model/PSF/Fibre systems, and AllWISE W1, W2, W3 and W4 infrared magnitudes, shown in Table~\ref{table:dataset_sizes}
        \end{enumerate}
    \item Southern Sky --- \ac{ATLAS}
    \begin{enumerate}
        \item Beginning with the \ac{ATLAS} dataset --- described in \citet{luken_2022} --- we cross-match the \ac{SWIRE} infrared positions with AllWISE in order to gain the same infrared bands as the Northern Sky dataset. Cross-matching at 1\arcsec\,produces a 100\% / 0.34\% classification/misclassification rate (1,156 / 4 sources) (Figure~\ref{fig:atlas_allwise}), and a final source count of 1,156 sources, all with $g$, $r$, $i$, and $z$ optical magnitudes in Auto, 2\arcsec, 3\arcsec, 4\arcsec, 5\arcsec, 6\arcsec, and 7\arcsec\,apertures, as well as the W1, W2, W3, and W4 infrared magnitudes. 
    \end{enumerate}
    \item Equatorial --- Stripe82
    \begin{enumerate}
        \item Along the equatorial plane, the Stripe82 field has been extensively studied by both northern- and southern-hemisphere telescopes, providing a field that contains both \ac{SDSS}, and \ac{DES} photometry. Cross-matching the \ac{DES} catalogue with the \ac{SDSS} catalogue (where both catalogues were restricted to the Stripe82 field) at 1\arcsec produces a 98.4\% / 3.36\% (170,622/5,831 sources) classification/misclassification rate (Figure~\ref{fig:stripe_optical}). 
        \item The optical catalogues were then cross-matched against the AllWISE catalogues at 1.25\arcsec producing a 84.09\% / 0.60\% (129,837/932 sources) classification/misclassification rate (Figure~\ref{fig:stripe_infrared}). 
        \item Finally, cross-matching the AllWISE Infrared catalogue against the \citet{Stripe_Hodge_2011} (at 4\arcsec; see Figure~\ref{fig:stripe_hodge}) and \citet{Stripe_Prescott_2018} (at 4\arcsec; see Figure~\ref{fig:stripe_prescott}) gives us a 21.96\% / 0.42\% (3,946 / 75 sources) and 45.46\% / 0.54\% (2,180 / 26 sources) classification/misclassification rates respectively.
        \item After combination of the Stripe82 Radio datasets with duplicates removed (based on AllWISE ID), we have a final dataset of 3,030 radio-selected sources with a spectroscopic redshift, W1, W2, W3, and W4 infrared magnitude, $g$, $r$, $i$, and $z$ optical magnitudes in PSF, Fibre, and Model systems (for \ac{SDSS} photometry), and Auto, 2\arcsec, 3\arcsec, 4\arcsec, 5\arcsec, 6\arcsec, and 7\arcsec apertures (for \ac{DES} photometry).
    \end{enumerate}
\end{enumerate}

\begin{table}
\centering
\caption{The source count in each sample compiled in Section~\ref{sec:data}}
\label{table:dataset_sizes}
\begin{tabular}{cccc}
\toprule
   Dataset & Source Count  \\
\midrule
    Northern Sky & 55,452  \\
    Southern Sky & 1,156  \\
    Equatorial (Stripe82) & 3,030  \\
    \midrule
    Total & 59,638  \\
    \bottomrule
\end{tabular}
\end{table}

\begin{figure}
    \centering
    \includegraphics[trim=0 0 0 0, width=\textwidth]{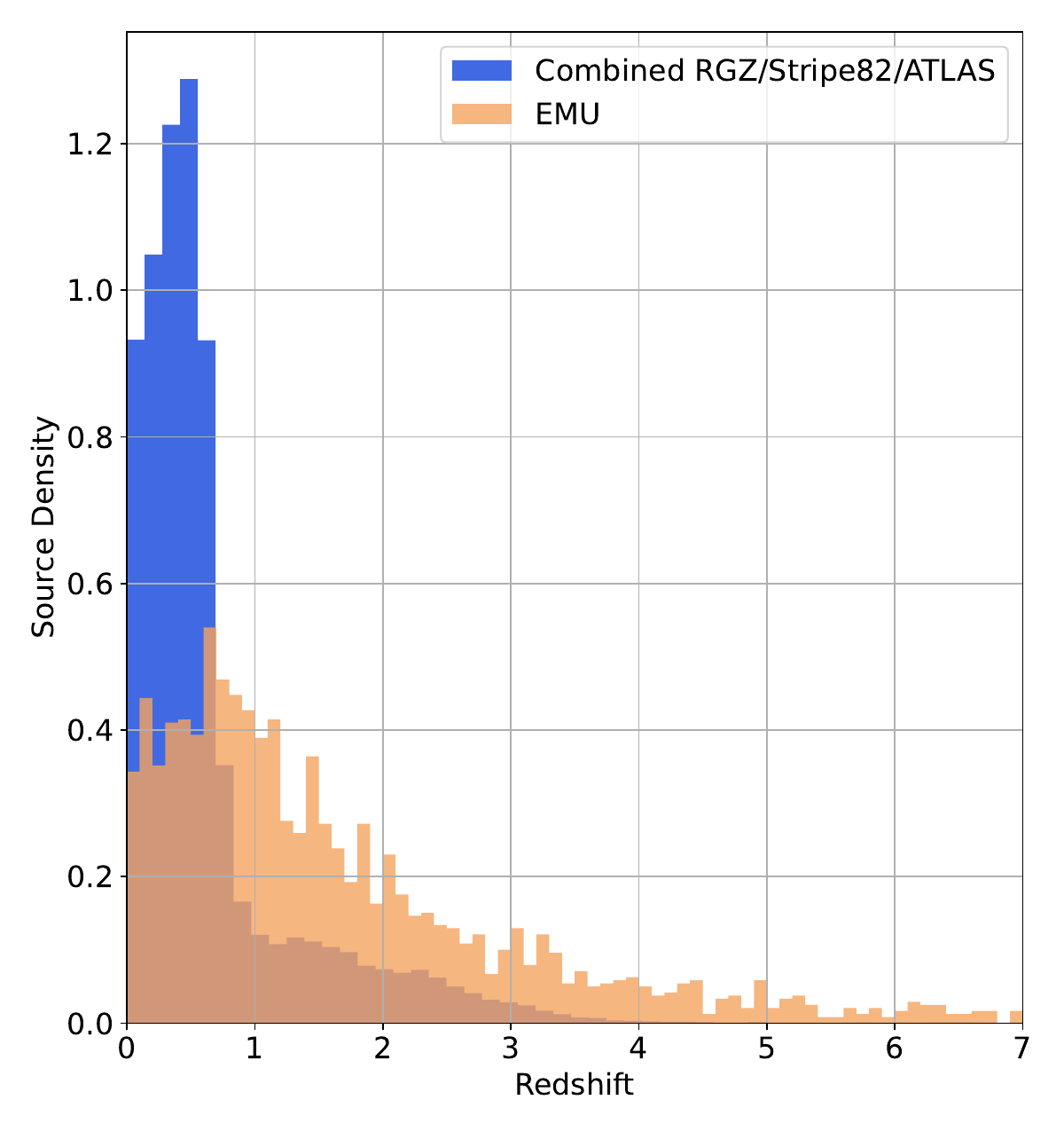}
    \caption{Histogram showing the density of sources at different redshifts in the combined \ac{RGZ} -- North, Stripe82 -- Equatorial, and \ac{ATLAS} -- South -- (blue), and the \acrocite{SKADS}{levrierMappingSKASimulated2009} simulation trimmed to expected \ac{EMU} depth \citep{norrisEMUEvolutionaryMap2011}.}
    \label{fig:combined_redshift_distribution}
\end{figure}

To summarise, all datasets are radio-selected, and contain:

\begin{itemize}
    \item a spectroscopically measured redshift, taken from either the \ac{OzDES}, or \ac{SDSS};
    \item $g$, $r$, $i$, and $z$ optical magnitudes, taken from either the \ac{DES} or \ac{SDSS};
    \item W1, W2, W3, and W4 (3.4, 4.6, 12, 24$\mu\mathrm{m}$ respectively) infrared magnitudes, taken from AllWISE;
\end{itemize}
with a final redshift distribution shown in Figure~\ref{fig:combined_redshift_distribution}. While we are still not matching the expected distribution from the \ac{EMU} survey, we are ensuring all sources have a radio counterpart (and hence, will be dominated by the difficult-to-estimate \acp{AGN}), with the final distribution containing more, higher redshift radio sources than previous works like \citet{luken_2022}.

\begin{figure}
    \centering
    \includegraphics[trim=0 0 0 0, width=\textwidth]{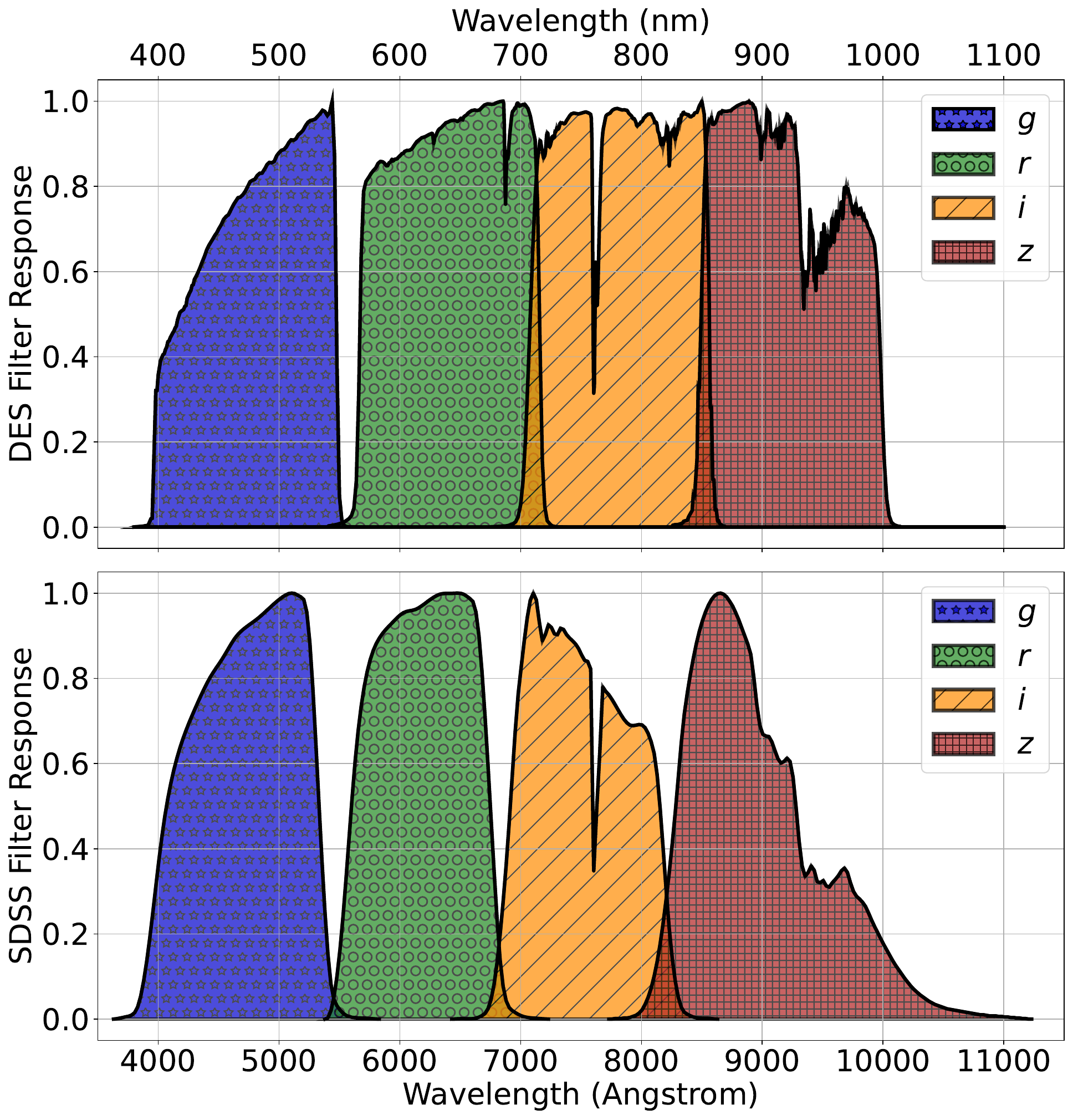}
    \caption{A comparison of the $g$, $r$, $i$, and $z$ filter responses, used by the \ac{DES} (top), and the \ac{SDSS} (bottom). }
    \label{fig:sdss_des_filters}
\end{figure}

The primary difference between the datasets is the source of the optical photometry. Even though both the DECam on the Blanco Telescope at the Cerro Tololo Inter-American Observatory in Chile and the Sloan Foundation 2.5m Telescope at the Apache Point Observatory in New Mexico both use $g$, $r$, $i$, and $z$ filters, the filter responses are slightly different (demonstrated in Figure~\ref{fig:sdss_des_filters}, the \ac{DES} Collaboration notes that there may be up to 10\% difference between the \ac{SDSS} and \ac{DES} equivalent filters\footnote{\url{https://data.darkenergysurvey.org/aux/releasenotes/DESDMrelease.html}}), with different processing methods producing multiple, significantly different measurements for the same sources. For \ac{ML} models, a difference of up to 10\% is significant, and had significant effects on redshift estimations in early tests without correction (sample results with one \ac{ML} algorithm shown in \ref{sec:appendix_data}).

For the \ac{SDSS}, the three measures of magnitude used in this work (\ac{PSF}, Fibre and Model) are all extensively defined by the \ac{SDSS}\footnote{\url{https://www.sdss.org/dr12/algorithms/magnitudes}}. Simply put, the \ac{PSF} magnitude measures the flux within the \ac{PSF} of the telescope for that pointing, the Fibre is a static sized aperture based on a single fibre within the \ac{SDSS} spectrograph (generally 3\arcsec), and the model magnitude tries to fit the source using a variety of models. 

The \ac{DES} pipelines produce statically defined apertures from 2\arcsec to 12\arcsec, as well as an auto magnitude that is fit by a model. 

For our purposes in finding \ac{DES} photometry compatible with \ac{SDSS} photometry, we only examined the \ac{DES} auto, and 2--7\arcsec measurements, as the larger aperture \ac{DES} measurements begin to greatly differ from any measured \ac{SDSS} measurement. We find that the \ac{DES} auto magnitude is most similar to the \ac{SDSS} model magnitude, and hence, exclusively use this pairing. 

\subsection{Optical Photometry Homogenisation}
\label{sec:data_homogenisation}

\begin{figure}
    \centering
    \includegraphics[trim=0 0 0 0, width=\textwidth]{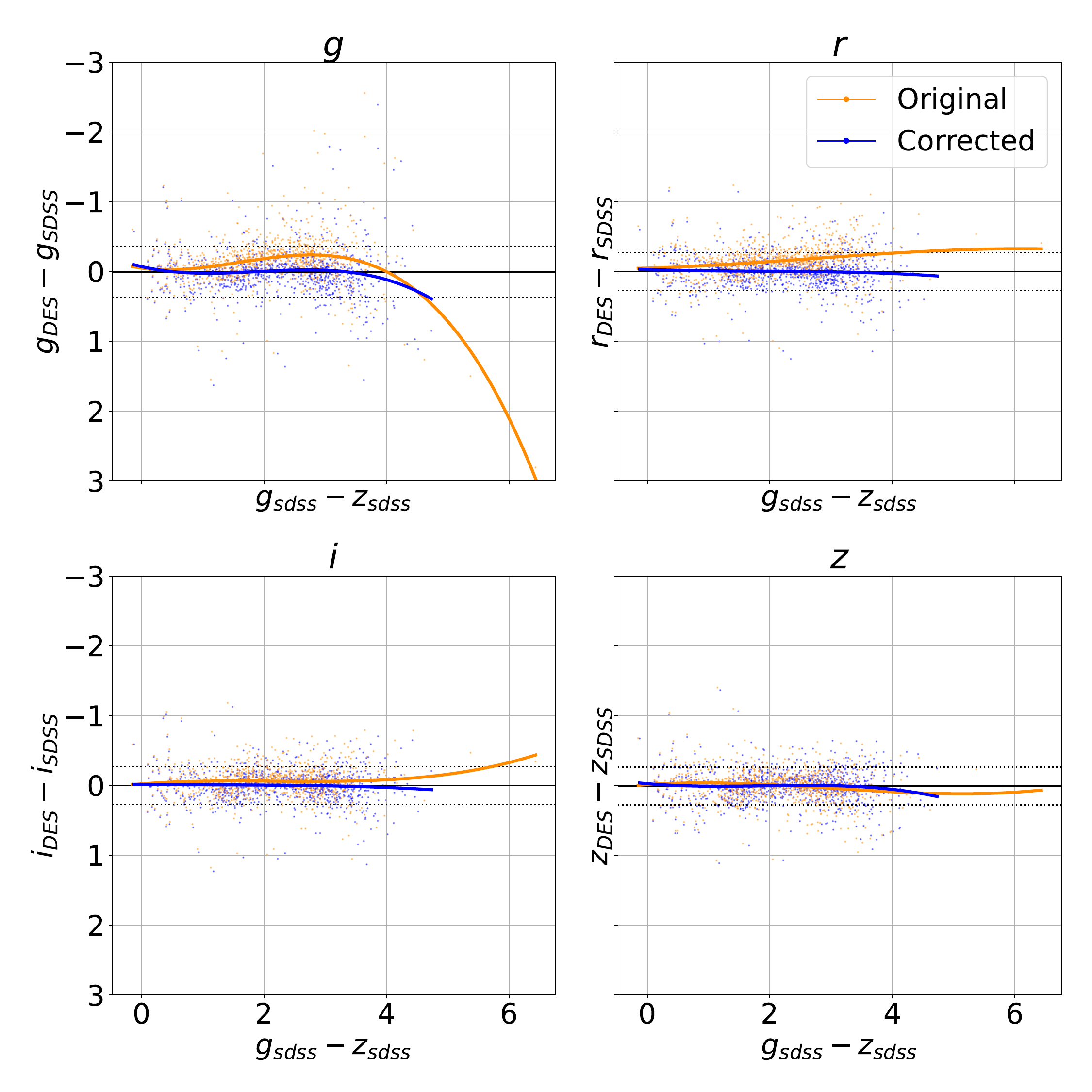}
    \caption{Plot showing the effects of homogenisation on the optical photometry. Each panel shows the original difference between the \ac{DES} and \ac{SDSS} photometry for a given band (with the band noted in the title of the subplot), as a function of $g - z$ colour. The orange scatterplots are the original data, with the orange line showing a 3rd order polynomial fit for to the pre-corrected data. The blue scatterplots are the corrected data, with the blue line showing the post-correction fit, highlighting the improvement the corrections bring.
    }
    \label{fig:des_sdss_homogenisaton}
\end{figure}

The combined dataset discussed above (Section~\ref{sec:data_description}) contains optical photometry measured using the \ac{SDSS} (in the Northern, and Equatorial fields) and the \ac{DES} (Southern and Equatorial fields). As shown in Figure~\ref{fig:sdss_des_filters}, while the \ac{SDSS} and \ac{DES} $g$, $r$, $i$, and $z$ filters are similar, they are not identical, and hence should not be directly compared without modification before use by typical \ac{ML} algorithms. As the Stripe82 Equatorial field contains observations with both optical surveys, we can fit a third order polynomial from the $g - z$ colour, to the difference in the \ac{SDSS} and \ac{DES} measured magnitude for each band for each object, and use the fitted model to homogenise the \ac{DES} photometry to the \ac{SDSS} photometry for the Southern hemisphere data. Figure~\ref{fig:des_sdss_homogenisaton} shows four panels --- one for each of the $g$, $r$, $i$, and $z$ magnitudes --- with the orange points showing the original difference between optical samples against the $g - z$ colour, blue points showing the corrected difference, orange line showing the third order polynomial fitted to the original data, and the blue line showing a third order polynomial fitted to the corrected data. While this homogenisation doesn't adjust for the scatter in the differences, it does shift the average difference, dropping from 0.158, 0.149, 0.061, and 0.006 to 0.004, 0.001, 0.001, and 0.007 for the $g$, $r$, $i$, and $z$ magnitudes respectively. We explore the difference the corrections make to predicting the redshift of sources with \ac{SDSS} and \ac{DES}, using the \ac{kNN} algorithm trained on the opposite optical survey, using corrected, and uncorrected \ac{DES} photometry in \ref{sec:appendix_data}.

\subsection{Regression and Classification}
\label{sec:data_reg_and_class}

The distribution of spectroscopically measured redshifts is highly non-uniform, providing additional difficulties to what is typically a regression problem (a real value --- redshift --- being estimated based on the attributes --- features --- of the astronomical object). As demonstrated in Figure~\ref{fig:skads_sdss_hist}, it also does not follow the expected distribution of the \ac{EMU} survey, partly because the optical source counts of the local universe vastly outnumber those of the high-redshift universe, and partly because high-redshift galaxies are too faint for most optical spectroscopy surveys. The non-uniform distribution means high-redshift sources will be under-represented in training samples, and therefore are less likely to be modelled correctly by \ac{ML} models.

In an attempt to provide a uniform redshift distribution for the \ac{ML} methods to provide better high-$z$ estimations, we quantise the data into 30 redshift bins with equal numbers of sources in each (where the bin edges, and the expected value of the bin --- typically the median redshift of the bin --- are shown in Table~\ref{table:bin_edges}). While binning the data means that it is no longer suitable for regression, it allows us to use the classification modes of the \ac{ML} methods and test whether treating the redshift estimation problem as a classification problem rather than attempt to estimate the redshift of sources as a continuous value aids in the estimation of sources in the high-redshift regime.

\begin{table}[t]
\centering
\caption{Example redshift bin boundaries used in the classification tests, calculated with the first random seed used. We show the bin index, the upper and lower bounds, and the predicted value for the bin. }
\label{table:bin_edges}

\begin{tabular}{cccc}
\toprule
   Bin  & Lower & Predicted & Upper  \\
   Number & Bound & Value & Bound \\
\midrule

1  & 0.0002 & 0.0120 & 0.0394  \\
2  & 0.0394 & 0.0556 & 0.0719  \\
3  & 0.0719 & 0.0903 & 0.1087  \\
4  & 0.1088 & 0.1253 & 0.1420  \\
5  & 0.1420 & 0.1588 & 0.1755  \\
6  & 0.1755 & 0.1910 & 0.2064  \\
7  & 0.2064 & 0.2212 & 0.2359  \\
8  & 0.2360 & 0.2520 & 0.2680  \\
9  & 0.2681 & 0.2830 & 0.2979  \\
10 & 0.2979 & 0.3121 & 0.3262  \\
11 & 0.3262 & 0.3389 & 0.3514  \\
12 & 0.3515 & 0.3645 & 0.3775  \\
13 & 0.3775 & 0.3911 & 0.4047  \\
14 & 0.4047 & 0.4178 & 0.4308  \\
15 & 0.4309 & 0.4440 & 0.4570  \\
16 & 0.4571 & 0.4707 & 0.4844  \\
17 & 0.4844 & 0.4966 & 0.5088  \\
18 & 0.5089 & 0.5223 & 0.5356  \\
19 & 0.5356 & 0.5489 & 0.5621  \\
20 & 0.5621 & 0.5765 & 0.5909  \\
21 & 0.5909 & 0.6075 & 0.6241  \\
22 & 0.6241 & 0.6437 & 0.6633  \\
23 & 0.6633 & 0.6924 & 0.7215  \\
24 & 0.7215 & 0.7739 & 0.8263  \\
25 & 0.8263 & 0.9339 & 1.0416  \\
26 & 1.0416 & 1.1861 & 1.3305  \\
27 & 1.3305 & 1.4804 & 1.6304  \\
28 & 1.6304 & 1.8210 & 2.0114  \\
29 & 2.0116 & 2.2543 & 2.4970  \\
30 & 2.4970 & 2.9163 & 2.9182  \\

    \bottomrule
\end{tabular}
\end{table}

\section{Machine Learning Methods}
\label{sec:methods}

In this section we outline the error metrics we use to compare the results across different \ac{ML} algorithms (Section~\ref{sec:methods_errors}) and the efforts to explain any random variance across our tests (Section~\ref{sec:method_statistical}), before discussing the different algorithms used -- the \ac{kNN} algorithm (using the Mahalanobis distance metric; Section~\ref{sec:method_knn}), the \ac{RF} algorithm (Section~\ref{sec:method_rf}), the ANNz2 algorithm (Section~\ref{sec:method_annz}), and the GPz algorithm (Section~\ref{sec:method_gpz}). Finally, we discuss the training methods used in this work (Section~\ref{sec:ml_training}). In this work we provide an initial explanation of each algorithm. However, we direct the reader to their original papers for a full discussion. 

\subsection{Error Metrics}
\label{sec:methods_errors}

As stated in Section~\ref{sec:intro}, this work differs from the typical training methods that attempt to minimise the average accuracy of the model (defined in Equation~\ref{eqn:outlier_sigma}, or Equation~\ref{eqn:nmad}). Instead, it  is primarily focused on minimising the number of estimates that are incorrect by a catastrophic level --- a metric defined as the Outlier Rate:
    \begin{equation}
        \label{eqn:outlier}
        \eta_{0.15} = \frac{1}{N}{\sum_{z\in Z} \llbracket |\Delta z| > 0.15(1 + z_{spec}) \rrbracket},
    \end{equation} 
where $\eta_{0.15}$ is the catastrophic outlier rate, $Z$ is the set of sources, $|Z| = N$, $\llbracket x \rrbracket$ is the indicator function (1 if $x$ is true, otherwise it is 0), $z_{spec}$ is the measured spectroscopic redshift, and $\Delta z$ is the residual: 
    \begin{equation}
        \label{eqn:residual}
        \Delta z = z_{spec} - z_{photo},
    \end{equation}

Alternative, we provide the 2-$\sigma$ outlier rate as a more statistically sound comparison:
    \begin{equation}
        \label{eqn:outlier_2sigma}
        \eta_{0.15} = \frac{1}{N}{\sum_{z\in Z} \llbracket |\Delta z| > 2 \sigma \rrbracket},
    \end{equation}
where $\eta_{2\sigma}$ is the 2-$\sigma$ outlier rate, and $\sigma$ is the residual standard deviation:        
        \begin{equation}
            \label{eqn:outlier_sigma}
            \sigma = \sqrt{\frac{1}{N} \sum_{i=1}^N(y_i - \hat{y}_i)^2},
        \end{equation}
        where $\sigma$ is the residual standard deviation, $y_i$ is an individual spectroscopic redshift, and $\hat{y_i}$ is the corresponding estimate for source $i$. The residual standard deviation gives an indication of the average accuracy of the estimates.
    
The \ac{NMAD} gives a similar metric to the Residual Standard Deviation, but is more robust to outliers as it relies on the median, rather than the mean of the residuals: 
        \begin{equation}
            \label{eqn:nmad}
            \sigma_{\mathrm{NMAD}} = 1.4826 \times (\mathrm{median}(|X_i - \mathrm{median}(X)|),
        \end{equation}
        $\sigma_{\mathrm{NMAD}}$ is the \ac{NMAD}, $X$ is a set of residuals (where the individual values are calculated in Equation~\ref{eqn:residual} as $\Delta z$), from which $x_i$ is an individual observation.

The \ac{MSE}, is only used in Regression-based tests, provides the average squared error of the estimates
    \begin{equation}
        \label{eqn:mse}
        \operatorname{MSE} =\frac {1}{N} \sum _{i=1}^{N}(y_{i}-\hat {y_{i}})^{2},
    \end{equation}
where $\operatorname{MSE}$ is the \acl{MSE}, $y_i$ is an individual spectroscopic redshift, and  $\hat {y_{i}}$ is the corresponding estimated redshift for source $i$.

The Accuracy is only used in Classification-based tests, and provides the percentage of sources predicted in the correct ``class'', where the class is a particular redshift bin. This metric is provided for completeness only, as the accuracy is only accepting of perfect classifications, whereas the aim of this work is provide redshift estimates that are approximately correct --- i.e.\ we are inherently accepting of classifications in nearby redshift bins, which would be considered incorrect classifications by the Accuracy metric. 
    \begin{equation}
        \label{eqn:accuracy}
        \mathrm{Accuracy}(y, \hat{y}) = \frac{1}{N} \sum_{i=0}^{N-1} \llbracket \hat{y}_i = y_i \rrbracket,
    \end{equation}
where $y$ is a vector of spectroscopic redshifts, and $\hat{y}$ is the corresponding vector of estimated redshifts.

\subsection{Statistical Significance}
\label{sec:method_statistical}

In order to measure the potential random variation within our results, all tests were conducted 100 times, with different random seeds  --- creating 100 different training/test sets to train and test each algorithm on. All values presented are the average of the results gained, with the associated standard error:
\begin{equation}
    \label{eqn:std_error}
    \sigma_{\bar{x}} = \frac{\sigma_{x}}{\sqrt{n}},
\end{equation}
where $\sigma_{\bar{x}}$ is the standard error of $\bar{x}$ which is calculated from the the standard deviation of the 100 repetitions of the experiment using different random seeds (denoted as $\sigma_x$), $\bar{x}$ is the mean classification/regression error, and $n$ is the number of repetitions --- 100 in this case. 

We note that the classification bin distribution is calculated for each random initialisation --- this means that while each of the 100 random training sets will have roughly the same redshift distribution, there will be slight differences in the bin distributions calculated for classification.

\subsection{\acl{kNN}}
\label{sec:method_knn}
The \ac{kNN} algorithm is one of the oldest \citep{coverNearestNeighborPattern1967}, as well as one of the simplest machine learning algorithms. Using some kind of distance metric --- typically Euclidean distance --- a similarity matrix is computed between every source in the training set, comparing the observed photometry between sources. The photometry of sources in the test set --- sources with ``unknown'' redshift ---  can then be compared to the photometry in the training set, and find the `$k$' (hereafter $k_n$) sources with most similar photometry. The mean or mode (depending on whether regression, or classification is performed respectively) of the most similar sources redshift from the training set is taken as the redshift of the unknown source. Following \citet{luken_2022} who have shown that Euclidean distance is far from optimal for redshift estimation, here we use the Mahalanobis distance metric \citep[Equation~\ref{eqn:mahalanobis}; ][]{mahalanobisGeneralizedDistanceStatistics1936}:

\begin{equation} 
	\label{eqn:mahalanobis} 
    d(\vec{p},\vec{q}) = \sqrt{(\vec{p} - \vec{q})^\mathrm{T}S^{-1}(\vec{p} - \vec{q})}, 
\end{equation}
where $d(\vec{p},\vec{q})$ is the Mahalanobis distance between two feature vectors $\vec{p}$ and $\vec{q}$, and $S$ is the covariance matrix. 

The value of $k_n$ is optimised using $k$-fold cross-validation, a process where the training set is split into $k$ (hereafter $k_f$ and is assigned a value of 5 for this work) subsets, allowing the parameter being optimised to be trained and tested on the entire training set.

\subsection{\acl{RF}}
\label{sec:method_rf}

The \ac{RF} algorithm is an ensemble \ac{ML} algorithm, meaning that it combines the results of many other algorithms (in this case \acp{DT}) to produce a final estimation. \acp{DT} to split the data in a tree-like fashion until the algorithm arrives at a single answer (when the tree is fully grown). These decisions are calculated by optimising over the impurity at the proposed split using Equation~\ref{eqn:rf_impurity}:

\begin{equation}
    \label{eqn:rf_impurity}
    G(Q_m, \theta) = \frac{n_{left}}{n_m} H(Q_{left}(\theta)) + \frac{n_{right}}{n_m} H(Q_{right}(\theta)) ,
\end{equation}
where $Q_m$ is the data at node $m$, $\theta$ is a subset of data, $n_m$ is the number of objects at node $m$, $n_{left}$ and $n_{right}$ are the numbers of objects on the left and right sides of the split, $Q_{left}$ and $Q_{right}$ are the objects on the left and right sides of the split, and the $H$ function is an impurity function that differs between classification and regression. For Regression, the Mean Square Error is used (defined in Equation~\ref{eqn:mse}), whereas Classification often uses the Gini Impurity (defined in Equation~\ref{eqn:rf_gini}).

\begin{equation}
    \label{eqn:rf_gini}
    H(X_m) = \sum_{k \in J} p_{mk} (1 - p_{mk}),
\end{equation}
where $p_{mk}$ is the proportion of split $m$ that are class $k$ from the set  of classes $J$, defined formally in Equation~\ref{eqn:rf_gini_pmk}:

\begin{equation}
    \label{eqn:rf_gini_pmk}
    p_{mk} = \frac{1}{n_m} \sum_{y \in Q_m} \llbracket y = k\rrbracket ,
\end{equation}
where $\llbracket x \rrbracket$ is the indicator function identifying the correct classifications.

\subsection{ANNz2}
\label{sec:method_annz}

The ANNz2\footnote{\url{https://github.com/IftachSadeh/ANNZ}} software \citep{sadehANNz2PhotometricRedshift2016} is another ensemble method, combining the results of many (in this work we use 100) randomly assigned machine learning models as a weighted average from the pool of \acp{NN} and boosted decision trees, using settings noted in \citet{bilickiPhotometricRedshiftsKiloDegree2018,bilickiBrightGalaxySample2021}. However, whereas \citet{bilickiBrightGalaxySample2021} uses the ANNz functionality to weight the training set by the test set feature distributions, here we do not use this option for two reasons. First, this work is designed for larger surveys to be completed, for which we do not know  the distributions, so we are unable to effectively weight the training samples towards future samples. Secondly, when attempted, the final outputs were not significantly different, whether the training sets were weighted or not.

\subsection{GMM+GPz}
\label{sec:method_gpz}

The GPz algorithm is based upon \acl{GP} Regression, a \ac{ML} algorithm that takes a slightly different track than traditional methods. Whereas most algorithms model an output variable from a set of input features using a single, deterministic function, \acp{GP} use a series of Gaussians to model a probability density function to map the input features to output variable. The \ac{GP} algorithm is extended further in the GPz algorithm to handle missing and noisy input data, through the use of sparse \acp{GP} and additional basis functions modelling the missing data \citep{gpz_2,gpz_1}.

Following \citet{duncan_2022}, we first segment the data into separate clusters using a \acl{GMM} before training a GPz model (without providing the redshift to the \ac{GMM} algorithm) on each cluster, the idea being that if the training data better reflects the test data, a better redshift estimate can be made. We emphasise that no redshift information has been provided to the \ac{GMM} algorithm, and the clusters determined by the algorithm is solely based on the $g$ -- $z$, and W1-W4 optical and infrared photometry  --- the same photometry used for the estimation of redshift. 

The \ac{GMM} uses the \ac{EM} algorithm to optimise the centers of each cluster it defines by using an iterative approach, adjusting the parameters of the models being learned in order to maximise the likelihood of the data belonging to the clusters assigned. The \ac{EM} algorithm does not optimise the number of clusters, which must be balanced between multiple competing interests:
\begin{itemize}
    \item The size of the data --- the greater the number of clusters, the more chance the \ac{GMM} will end up with insufficient source counts in a cluster to adequately train a redshift estimator
    \item The number of distinct source types within the data --- If the number of clusters is too small, there will be too few groupings that adequately split the data up into it's latent structure, whereas if it is too high, the \ac{GMM} will begin splitting coherent clusters
\end{itemize}

This means that the number of components used by the \ac{GMM} to model the data is a hyper-parameter to be fine-tuned. Ideally, the number of components chosen should be physically motivated --- the number of classes of galaxy we would expect to be within the dataset would be an ideal number, so the \ac{ML} model is only training on sources of the same type to remove another source of possible error. However, this is not necessarily a good option, as, due to the unsupervised nature of the \ac{GMM}, we are not providing class labels to the \ac{GMM}, and hence cannot be sure that the \ac{GMM} is splitting the data into the clusters we expect. On the other hand, being unsupervised means the \ac{GMM} is finding its own physically motivated clusters which don't require the additional --- often human derived --- labels. The lack of labels can be a positive, as human-decided labels may be based less on the actual source properties, and more on a particular science case (see \citet{rudnick_tags} for further discussion).

In this work, we optimise the number of components hyper-parameter, where the number of components $\mathrm{n_{comp}}$ is drawn from $ \mathrm{n_{comp}} \in  \{1, 2, 3, 5, 10, 15, 20, 25, 30\}$. We emphasise that the number of components chosen is not related to the number of redshift bins used for classification, and has an entirely separate purpose. The primary metric being optimised is the \acrocite{BIC}{schwarz_bic}:
\begin{equation}
    \label{eqn:bic}
    \operatorname{BIC} = -2\log(\hat{L}) + \log(N)d,
\end{equation}
where $\log(\hat{L})$ is the log likelihood of seeing a single point drawn from a Gaussian Mixture Model, defined in Equation~\ref{eqn:log_liklihood_gauss}, and $d$ is the number of parameters. 

\begin{equation}
\label{eqn:log_liklihood_gauss}
\log{\hat{L}} = {\sum_i \log{\sum_j \lambda_j \mathcal{N}(y_i|\mu_j, \sigma_j)}}
\end{equation}
where $y_i$ is an individual observation, $\mathcal{N}$ is the Normal density with parameters $\sigma_j^2$ and $\mu_j$ (the sample variance and mean of a single Gaussian component), and $\lambda_j$ is the mixture parameter, drawn from the mixture model.

The \ac{BIC} operates as a weighted likelihood function, penalising higher numbers of parameters. The lower the \ac{BIC}, the better.

\begin{figure}
    \centering
    \includegraphics[trim=0 0 0 0, width=\textwidth]{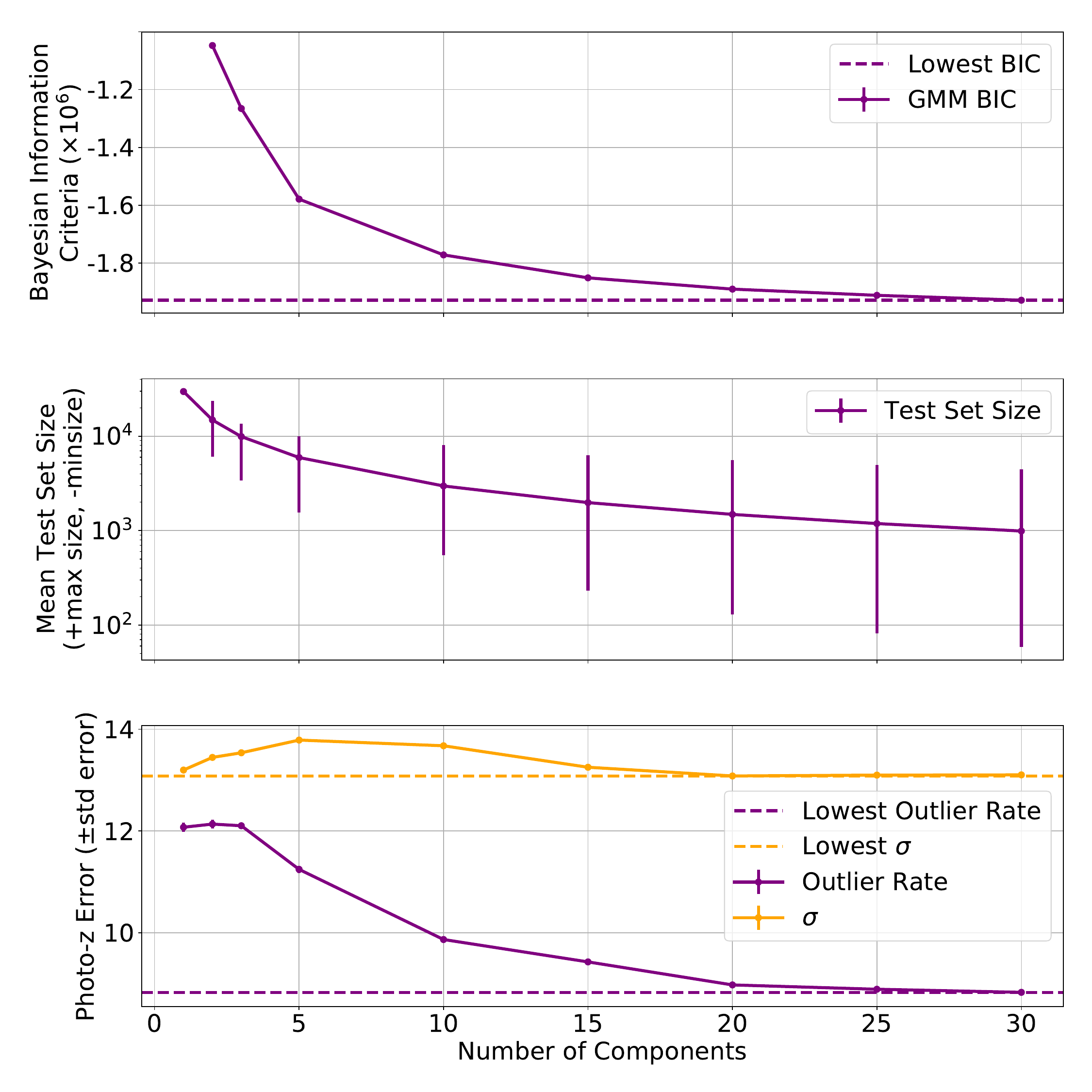}
    \caption{Optimisation of the Bayesian Information Criteria (top), Test Size (middle), and Outlier Rate (bottom) across a range of components.}
    \label{fig:gmm_components}
\end{figure}

Figure~\ref{fig:gmm_components} shows the \ac{BIC} (Equation~\ref{eqn:bic}; top panel) with error bars denoting the standard error of each component, the average test size of each component with the error bars denoting the minimum and maximum test set size for each component (middle), and the photometric error in the form of both the outlier rate (Equation~\ref{eqn:outlier}), and the accuracy (Equation~\ref{eqn:accuracy}), with error bars denoting the standard error (bottom).

Figure~\ref{fig:gmm_components} shows that while the $\mathrm{n_{comp}}$ is being optimised for lowest \ac{BIC}, this has the additional benefit of lowering the resulting redshift estimation error (Figure~\ref{fig:gmm_components}; middle and bottom panels) --- showing that the clusters being identified by the \ac{GMM} algorithm are meaningful in the following redshift estimation. A value of 30 is chosen for the $\mathrm{n_{comp}}$, despite the \ac{BIC} continuing to decline beyond this point. However, the number of sources in the smaller clusters defined by the \ac{GMM} becomes too small to adequately train a GPz model. 

\begin{figure*}
    \centering
    \includegraphics[trim=0 0 0 0, width=\textwidth]{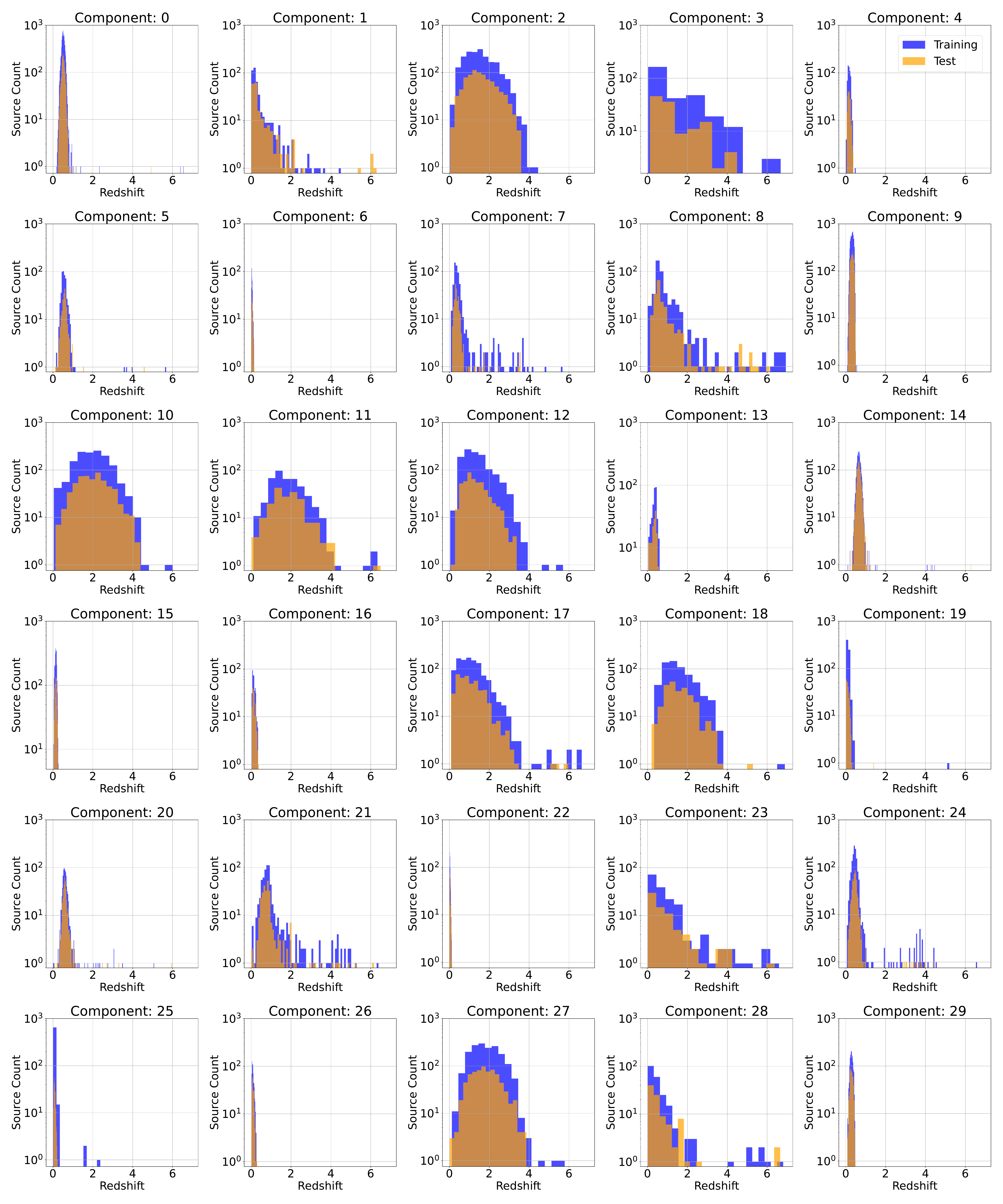}
    \caption{Redshift distribution of the 30 \ac{GMM} components. In each case the vertical axis shows the count. }
    \label{fig:component_redshift}
\end{figure*}

Once the data are segmented into 30 components (an example from one random seed is shown in Figure \ref{fig:component_redshift}), a GPz\footnote{\url{https://github.com/cschreib/gpzpp}} model is trained for each component. The GPz algorithm is based around sparse \acp{GP}, which attempt to model the feature space provided using Gaussian components.

\subsection{Training Method}
\label{sec:ml_training}

\ac{ML} algorithms are typically set up and trained following one of two procedures:

\begin{enumerate}
    \item Training / Validation / Test Splits
    \begin{itemize}
        \item The data is split into training, testing and validation sets (for this work, the data are split into 50\%/20\%/30\% subsets). The \ac{ML} algorithm is trained on the training set, with model hyper-parameters optimised for the validation set. Once optimised, the test set is used to estimate the model's generalisatibility. 
        \item This method is utilised by the ANNz and GMM algorithms
    \end{itemize}
    \item $k$-Fold Cross-Validation
    \begin{itemize}
        \item The dataset is split into two sets (for this work, the data are split into 70\%/30\% subsets), used as training and test sets. Differing from the first method, this method trains and optimises the \ac{ML} algorithms on the training set alone, before testing the optimised models on the test set. 
        \item The training set is split into $k_f$ subsets. $k_f$ models are trained on $k_f - 1$ subsets, and hyperparameters optimised and validated against the remaining subset. 
        \item In this work, we use a value of 5 for $k_f$, with the $k$-fold Cross-Validation algorithms used to optimise the hyperparameters of the \ac{kNN} and \ac{RF} algorithms. 
    \end{itemize}
\end{enumerate}

The externally developed software (ANNz and GPz) both operate using training/validation/test split datasets. This is preferable for large, mostly uniform distributions,  as it greatly reduces training time. However, for highly non-uniform distributions, the under-represented values are less likely to be involved in all stages of training, validation and testing. Hence, for the \ac{kNN} and \ac{RF} algorithms, where we control the training process, we choose the $k$-fold cross validation method of training and optimising hyper-parameters, to best allow the under-represented high-redshift sources to be present at all stages of training.

\subsubsection{Photometry Used in Training}

All algorithms use the same primary photometry --- $g$, $r$, $i$, $z$ optical magnitudes, and W1, W2, W3 and W4 infrared magnitudes. However, the different algorithms vary in how they treat the uncertainties associated with the photometry. For the simple \ac{ML} algorithms (\ac{kNN} and \ac{RF}), the uncertainties are ignored. ANNz computes their own uncertainties using a method based on the \ac{kNN} algorithm, outlined in \citet{oyaizuGalaxyPhotometricRedshift2008}, and GPz uses them directly in the fitting of the Gaussian Process. 

\subsubsection{Using ANNz and GPz for Classification}
\label{sec:annz_gpz_classification}

While the \texttt{Sci-Kit Learn} implementations of the \ac{kNN} and \ac{RF} algorithms have both regression and classification modes, there is no directly comparable classification mode for the ANNz and GPz algorithms. In order to compare them with the classification modes of the \ac{kNN} and \ac{RF} algorithms, we use the ANNz and GPz algorithms to predict  the median of the bin, in lieu of a category. The predictions are then re-binned to the same boundaries as the original bins, and the re-binned data compared.

\section{Results}

For clarity, we break our results up into three sub-sections --- Subsection~\ref{sec:regression_results} reports the results using the regression modes of each \ac{ML} method, Subsection~\ref{sec:classification_results} reports the results using the classification modes of each \ac{ML} method, and Subsection~\ref{sec:reg_vs_class_results} reports the comparison between the two modes.

\subsection{Regression Results}
\label{sec:regression_results}

The results using the regression modes of each \ac{ML} algorithm are summarised in Table~\ref{table:regressionResultsTable}. Table~\ref{table:regressionResultsTable} shows that the \ac{kNN} algorithm performs best in terms of both $\eta_{0.15}$ and $\eta_{2\sigma}$ outlier rates, while also performing similarly across other metrics -- although the \ac{GMM}+GPz algorithm provides the lowest $\sigma$. 

\begin{figure}[t]
    \centering
    \includegraphics[trim=0 0 0 0, width=\textwidth]{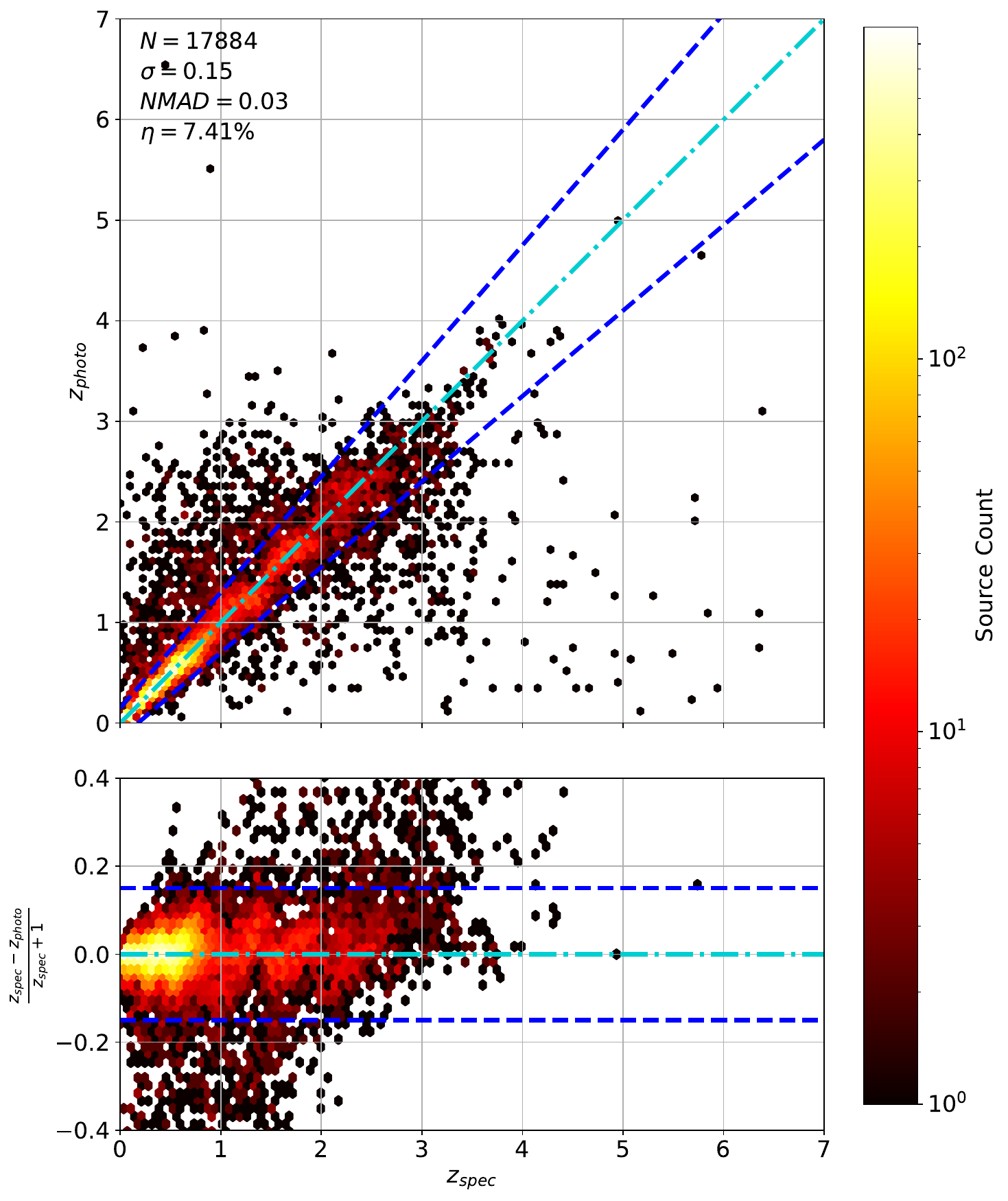}
    \caption{Comparison of spectroscopic and predicted values using kNN Regression. The x-axis shows the spectroscopic redshift, with the y-axis (Top) showing the redshift estimated by the \ac{ML} model.  The y-axis (Bottom) shows the normalised residual between spectroscopic and predicted values as a function of redshift. The turquoise dash-dotted line shows a perfect correlation, and the blue dashed lines show the boundaries accepted by the $\eta_{0.15}$ outlier rate. The colourbar shows the density of points per coloured point.  }
    \label{fig:knn_regress}
\end{figure}

\begin{figure}
    \centering
    \includegraphics[trim=0 0 0 0, width=\textwidth]{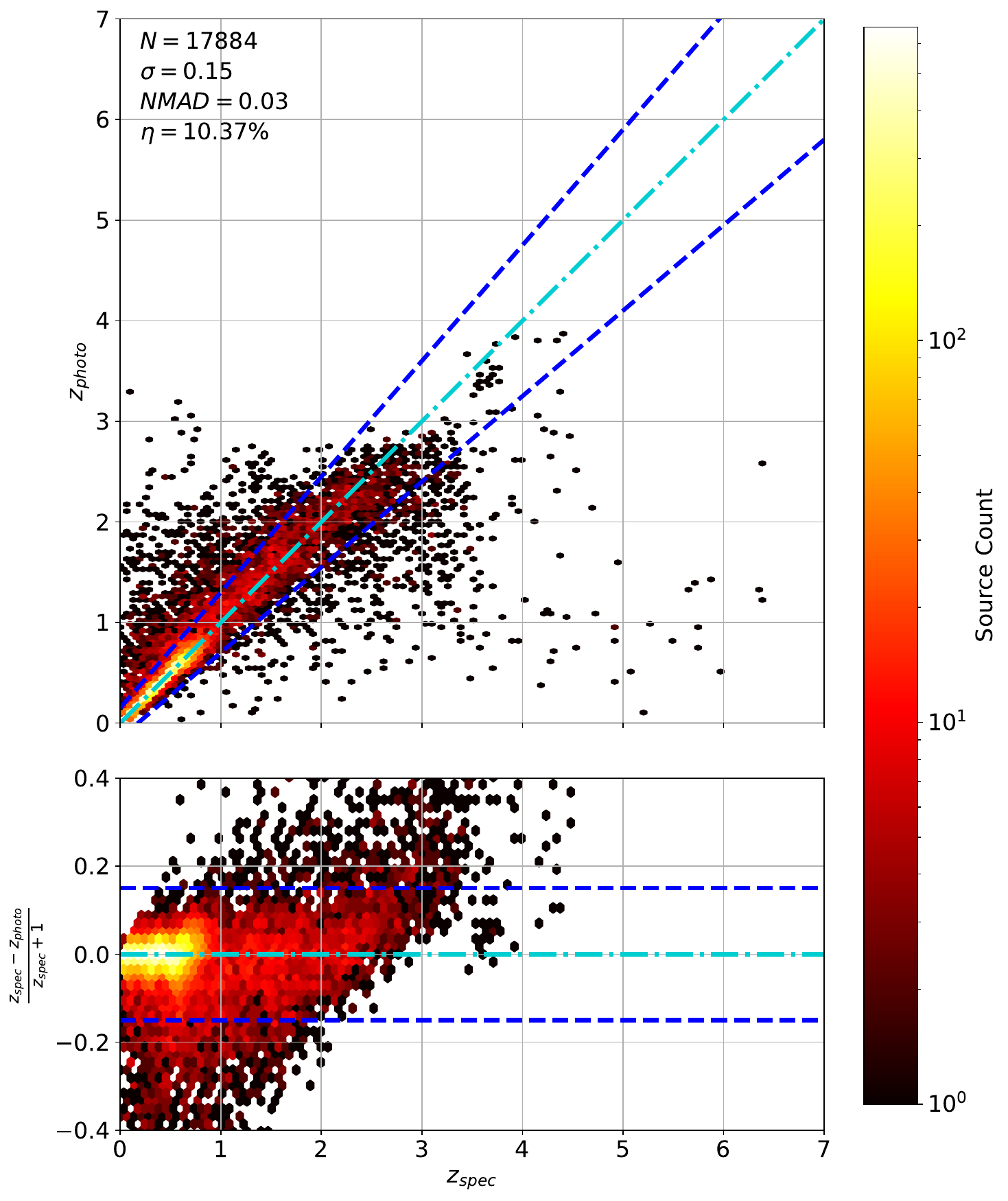}
    \caption{Same as Figure \ref{fig:knn_regress} but for RF Regression}
    \label{fig:rf_regress}
\end{figure}

\begin{figure}
    \centering
    \includegraphics[trim=0 0 0 0, width=\textwidth]{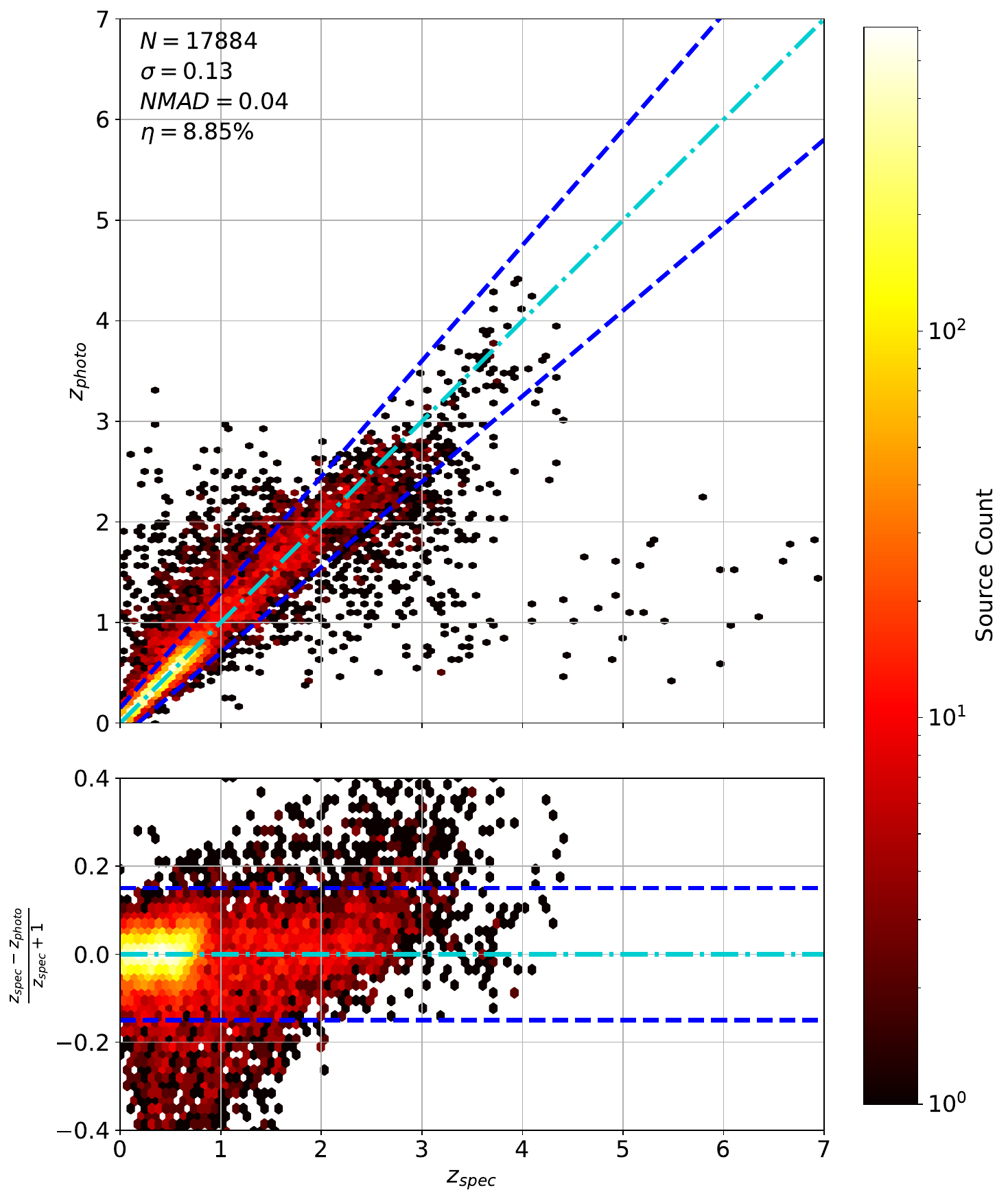}
    \caption{Same as Figure \ref{fig:knn_regress} but for ANNz Regression}
    \label{fig:annz_regress}
\end{figure}

\begin{figure}[t]
    \centering
    \includegraphics[trim=0 0 0 0, width=\textwidth]{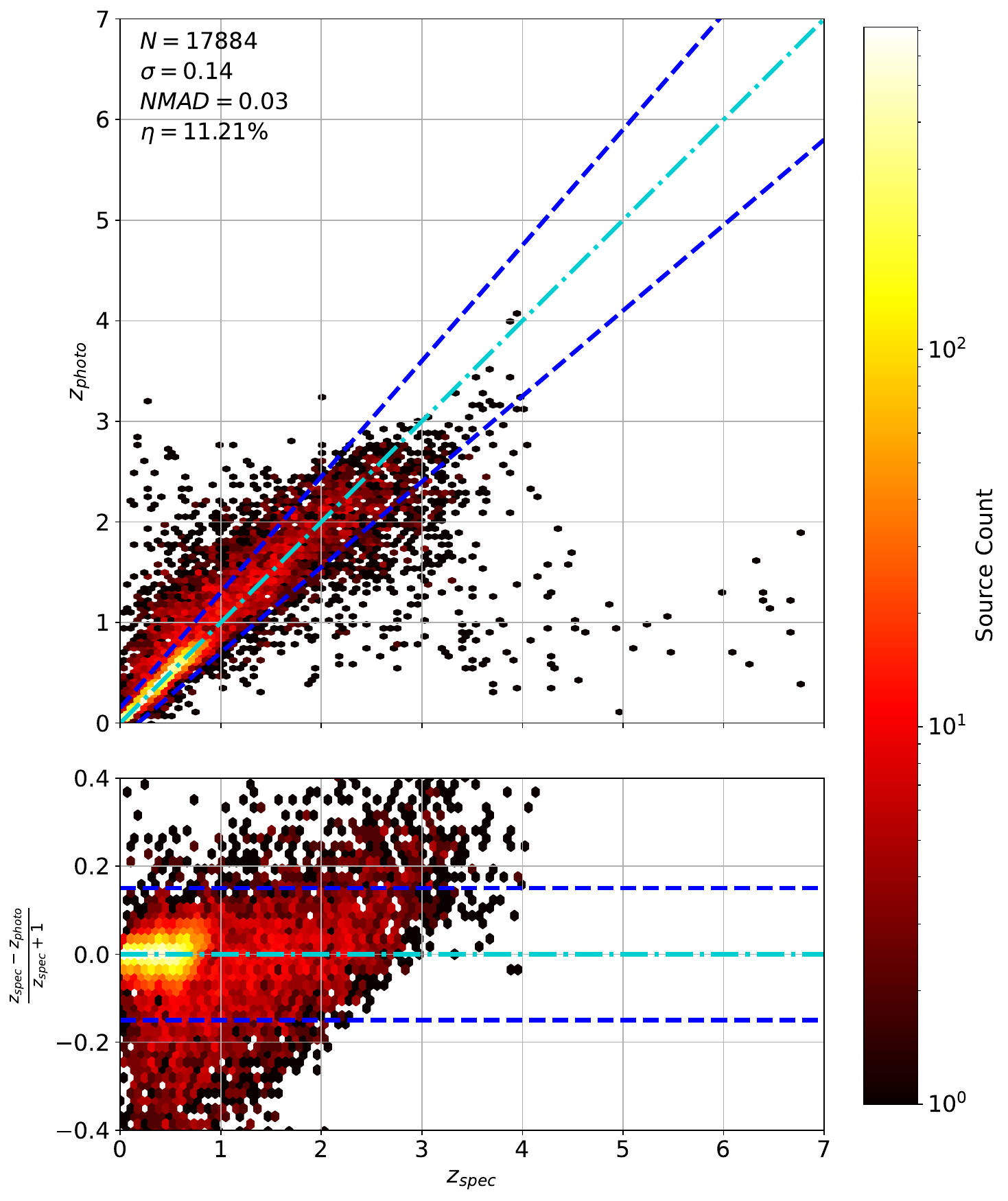}
    \caption{Same as Figure \ref{fig:knn_regress} but for GPz Regression}
    \label{fig:gpz_regress}
\end{figure}

Scatter plots  (Figures~\labelcref{fig:knn_regress,fig:rf_regress,fig:annz_regress,fig:gpz_regress}) show the results from each \ac{ML} algorithm where the x-axis of each panel shows the measured spectroscopic redshifts, the y-axis of the top panel shows the redshift predicted by the given \ac{ML} method and the bottom panel the normalised residuals. The dashed red line shows a perfect prediction, with the dashed blue lines highlighting the boundary set by the outlier rate. All figures use the same random seed, and the same test set. 

\begin{table*}
	\centering
    \caption{Regression results table comparing the different algorithms across the different error metrics (listed in the table footnotes). The best values for each error metric are highlighted in bold. }
    \label{table:regressionResultsTable}
    \begin{tabular}{cccccc}
    	\toprule
     Algorithm &  $\eta_{0.15}$\footnote{Catastrophic outlier rate, Equation~\ref{eqn:outlier}}  & $\eta_{2\sigma}$\footnote{2$\sigma$ outlier rate, Equation~\ref{eqn:outlier_2sigma}} & $\sigma$\footnote{Residual Standard Deviation, Equation~\ref{eqn:outlier_sigma}} & \ac{NMAD}\footnote{\acl{NMAD}, Equation~\ref{eqn:nmad}} & \ac{MSE}\footnote{\acl{MSE}, Equation~\ref{eqn:mse}}\\   
    \midrule
    \ac{kNN} & \textbf{7.26\% $\pm$ 0.02} & \textbf{3.86\% $\pm$ 0.01} & 0.1450 $\pm$ 0.0005 & 0.02930 $\pm$ 0.00003 & \textbf{0.1195 $\pm$ 0.0007} \\ 
    \ac{RF}  & 10.19\% $\pm$ 0.02 & 4.59\% $\pm$ 0.02 & 0.1472 $\pm$ 0.0004 & \textbf{0.02790 $\pm$ 0.00003} & 0.1235 $\pm$ 0.0007 \\ 
    ANNz     & 8.82\% $\pm$ 0.04  & 4.02\% $\pm$ 0.07 & 0.141 $\pm$ 0.003 & 0.0505 $\pm$ 0.0003 & 0.120 $\pm$ 0.005 \\ 
    GMM+GPz  & 9.92\% $\pm$ 0.13  & 4.81\% $\pm$ 0.03 & \textbf{0.1336 $\pm$ 0.0008} & 0.0382 $\pm$ 0.0004 & 0.126 $\pm$ 0.002\\ 
	\bottomrule
	\end{tabular}
\end{table*}

As shown in Figures~\labelcref{fig:knn_regress,fig:rf_regress,fig:annz_regress,fig:gpz_regress}, all algorithms suffer from the same issues --- overestimating the low-redshift sources ($z < 1$), while underestimating the high-redshift sources ($z > 3$). At the low-redshift end, the large majority of sources are estimated within the $\eta_{0.15}$ outlier rate by all algorithms, with all algorithms overestimating roughly the same number of sources. At high-redshift, the GPz algorithm performs worst, however, the small number of sources at high-redshift mean this does not significantly impact the error metrics.

\subsection{Classification Results}
\label{sec:classification_results}

The results using the classification modes of each \ac{ML} algorithm are summarised in Table~\ref{table:classificationResultsTable}. As with the Regression results in Section~\ref{sec:regression_results}, the \ac{kNN} algorithm produces the lowest $\eta_{0.15}$ rate, with the \ac{RF} algorithm being second best. All methods (aside from the \ac{RF} algorithm) have approximately the same $\sigma$. However, the \ac{RF} and GPz algorithms have a marginally lower \ac{NMAD}.

\begin{table*}
	\centering
    \caption{Classification results table comparing the different algorithms across the different error metrics (listed in the table footnotes). The best values for each error metric are highlighted in bold. }
    \label{table:classificationResultsTable}
    \begin{tabular}{cccccc}
    	\toprule
     Algorithm &  $\eta_{0.15}$\footnote{Catastrophic outlier rate, Equation~\ref{eqn:outlier}}  & $\eta_{2\sigma}$\footnote{2$\sigma$ outlier rate, Equation~\ref{eqn:outlier_2sigma}} & $\sigma$\footnote{Residual Standard Deviation, Equation~\ref{eqn:outlier_sigma}} & \ac{NMAD}\footnote{\acl{NMAD}, Equation~\ref{eqn:nmad}} & Accuracy\footnote{Accuracy, Equation~\ref{eqn:accuracy}}\\   
    \midrule
    \ac{kNN} & \textbf{6.21\% $\pm$ 0.02}  & \textbf{3.17\% $\pm$ 0.01} & 0.1499 $\pm$ 0.0005 & \textbf{0.02791 $\pm$ 0.00003} & \textbf{0.4165 $\pm$ 0.0003} \\ 
    \ac{RF}  & 7.69\% $\pm$ 0.02  & \textbf{3.17\% $\pm$ 0.01} & 0.1742 $\pm$ 0.0006 & 0.02819 $\pm$ 0.00003 & 0.3950 $\pm$ 0.0004 \\ 
    ANNz     & 8.72\% $\pm$ 0.05  & 4.41\% $\pm$ 0.03 & \textbf{0.1249 $\pm$ 0.0004} & 0.0507 $\pm$ 0.0002 & 0.355 $\pm$ 0.001 \\ 
    GMM+GPz  & 10.0\% $\pm$ 0.1 & 5.28\% $\pm$ 0.04 & 0.1304 $\pm$ 0.0009 & 0.0396 $\pm$ 0.0001 & 0.408 $\pm$ 0.002 \\ 
	\bottomrule
	\end{tabular}
\end{table*}

\begin{figure*}
    \centering
    \includegraphics[trim=0 0 0 0, width=\textwidth]{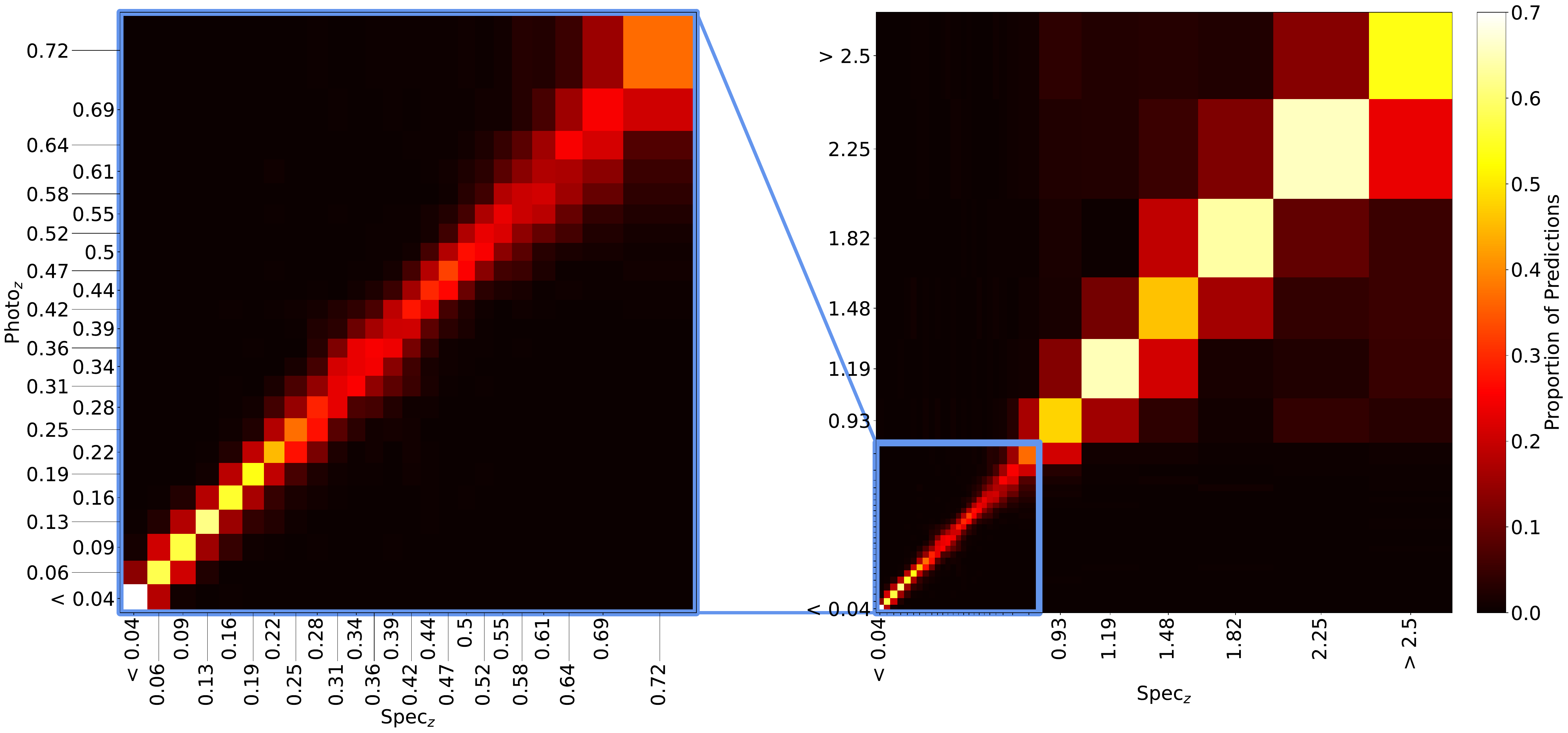}
    \caption{Confusion matrix showing the results using the \ac{kNN} classification algorithm. The size of the boxes are approximately scaled (with the exception of the final, highest redshift boxes) with the width of the classification bin. The x-axis shows the spectroscopic redshift, and the y-axis shows the predicted redshift. The left panel is an exploded subsection the overall right panel.}
    \label{fig:knn_class_scaled}
\end{figure*}

\begin{figure*}
    \centering
    \includegraphics[trim=0 0 0 0, width=\textwidth]{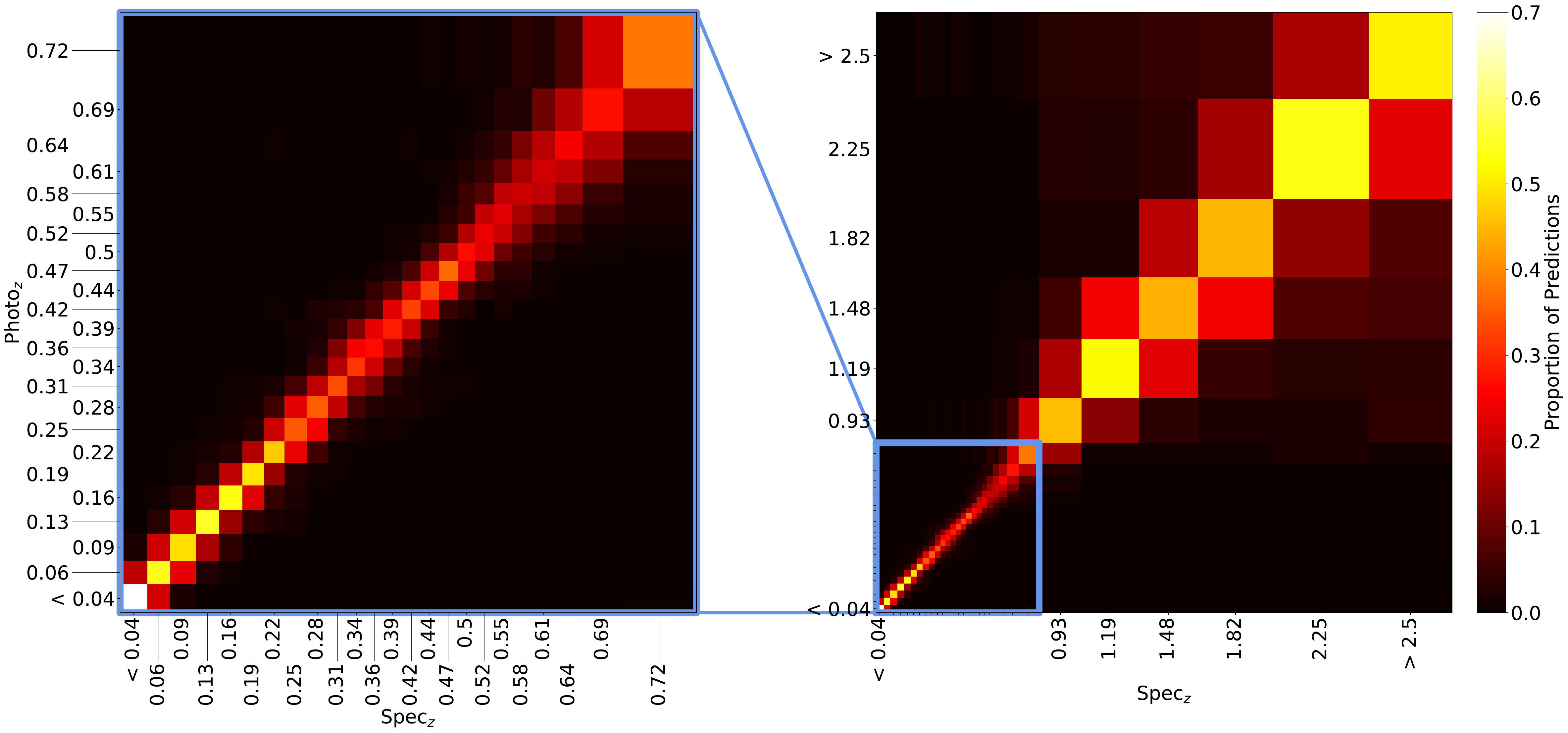}
    \caption{Same as Figure~\ref{fig:knn_class_scaled}, but for \ac{RF} Classification}
    \label{fig:rf_class_scaled}
\end{figure*}

\begin{figure*}
    \centering
    \includegraphics[trim=0 0 0 0, width=\textwidth]{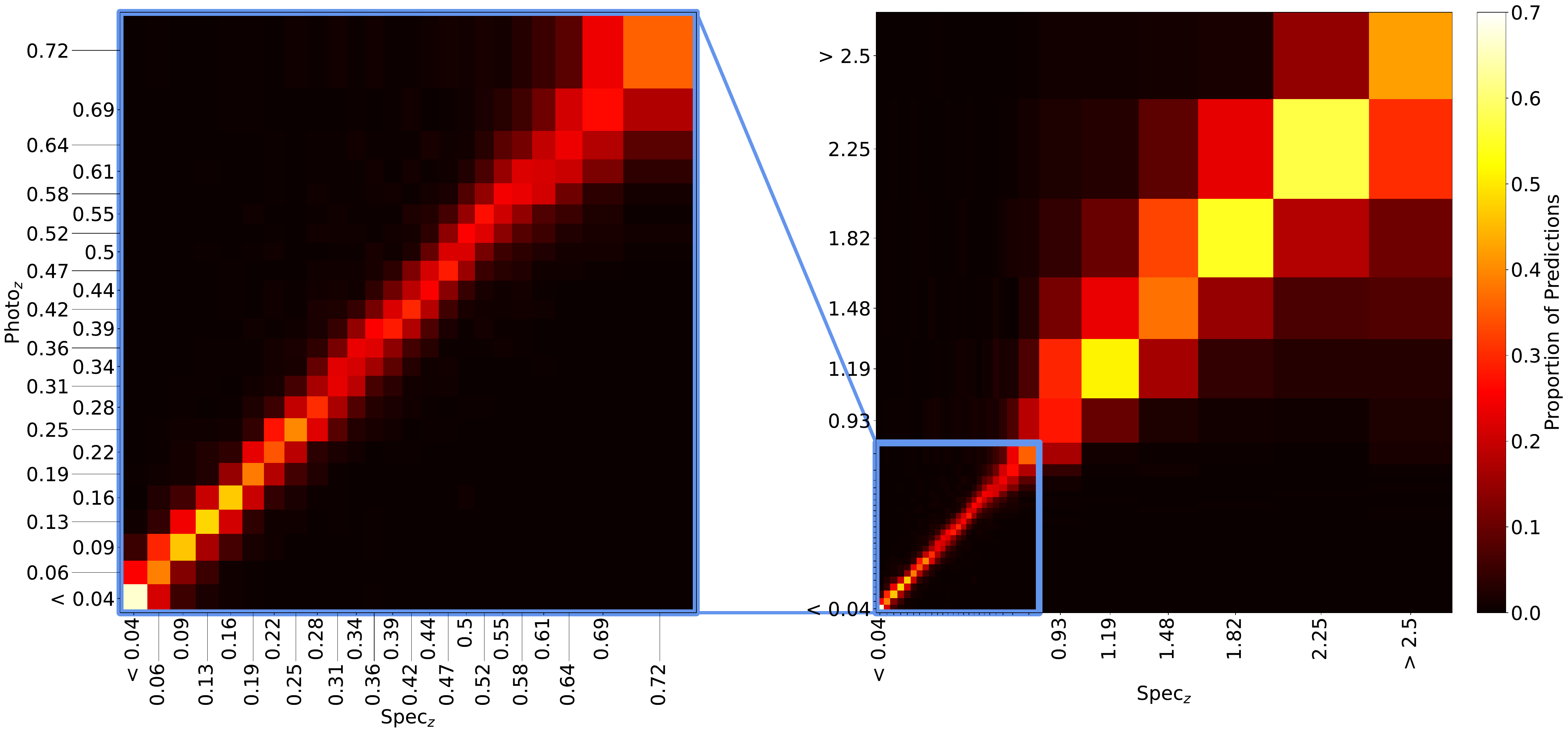}
    \caption{Same as Figure~\ref{fig:knn_class_scaled}, but for ANNz Classification}
    \label{fig:annz_class_scaled}
\end{figure*}

\begin{figure*}
    \centering
    \includegraphics[trim=0 0 0 0, width=\textwidth]{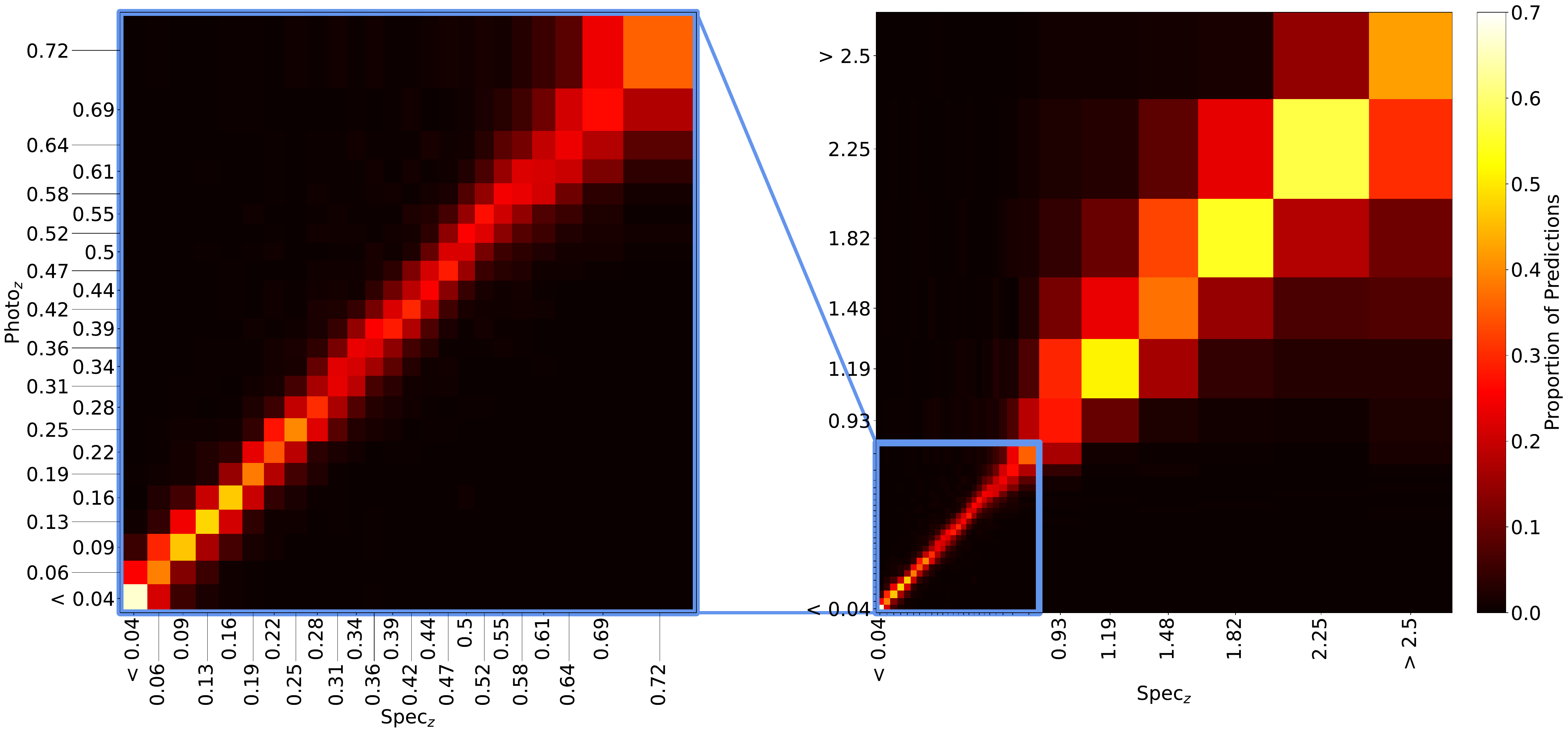}
    \caption{Same as Figure~\ref{fig:knn_class_scaled}, but for GPz Classification}
    \label{fig:gpz_class_scaled}
\end{figure*}

Plots showing the results from each \ac{ML} algorithm (Figures~\labelcref{fig:knn_class_scaled,fig:rf_class_scaled,fig:annz_class_scaled,fig:gpz_class_scaled}) show the scaled classification bins, with the x-axis showing the measured (binned) spectroscopic redshifts, and the y-axis showing the \ac{ML} classified bin for each source. While a perfect correlation along the diagonal would be ideal, the inherent error built into the $\eta_{0.15}$ error metric means that at low redshift, there might be many adjacent bins that are deemed ``acceptable'' redshift estimates, whereas at the highest redshift, there is only one possible bin a source can be classified into for it to be an acceptable estimate.

The \ac{kNN} algorithm correctly predicts the highest proportion of sources belonging to the highest redshift bin, though it should be noted that all algorithms struggle with assigning this under-represented class. While the width of the final bin means that sources that are not exactly classified are therefore incorrectly classified (unlike sources at the low-redshift end), in all cases, over 70\% of the highest redshift sources are placed in the highest two redshift bins. Alternatively, if these bins were to be combined, we would be able to say that the over 70\% of sources at $z > 2$ would be correctly classified. Further discussion is presented in Section~\ref{sec:combine_bins}.

\subsection{Regression vs Classification}
\label{sec:reg_vs_class_results}

\begin{figure}
    \centering
    \includegraphics[trim=0 0 0 0, width=\textwidth]{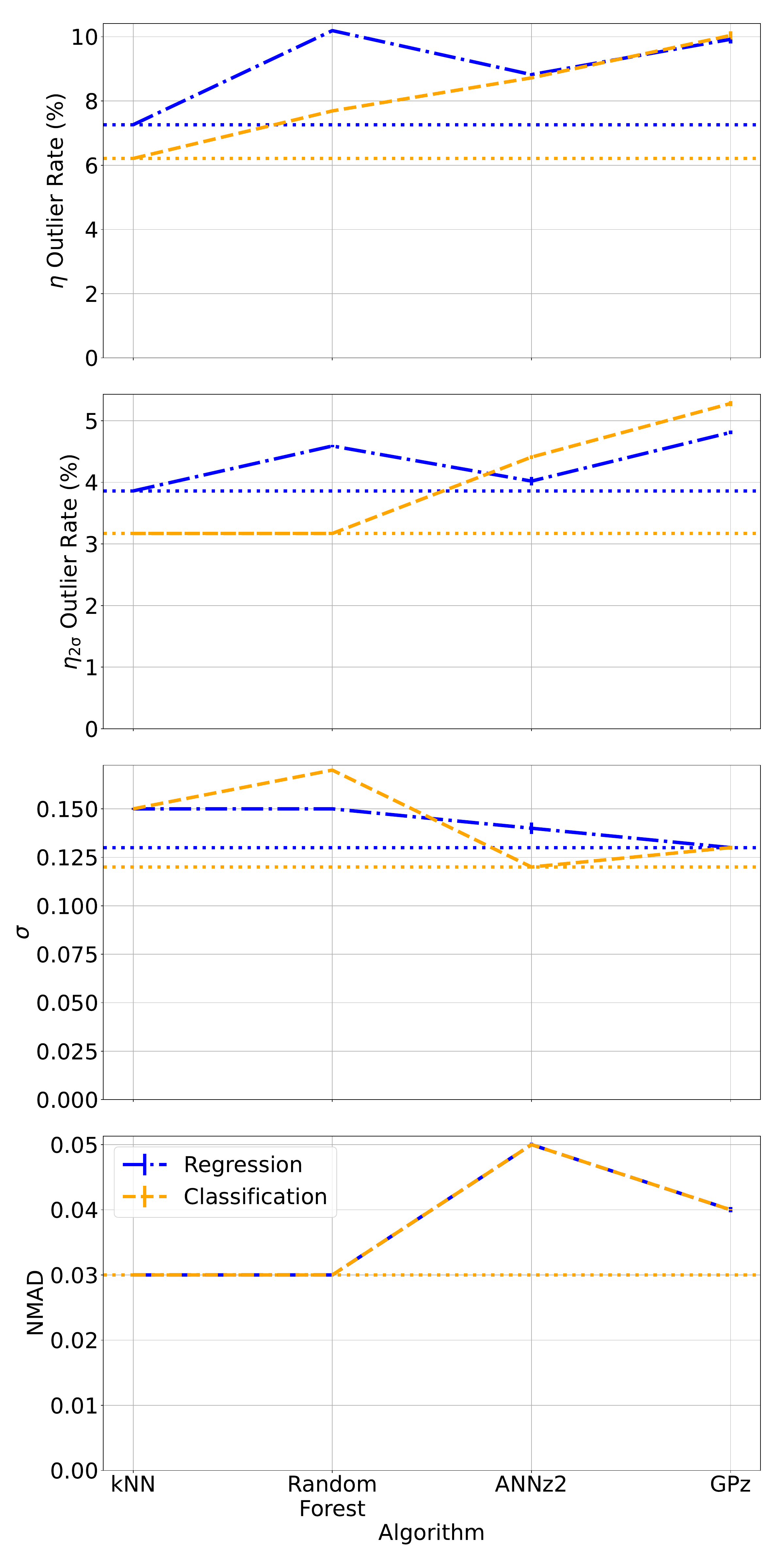}
    \caption{Comparison of the different algorithms in their regression and classification modes. In all cases, the lower the value the better, with the lowest value for each metric shown with a horizontal dotted line. }
    \label{fig:algorithm_comparison}
\end{figure}

When comparing the results in Tables~\ref{table:regressionResultsTable} and \ref{table:classificationResultsTable} (demonstrated in Figure~\ref{fig:algorithm_comparison}), we find that the binning of redshifts greatly improves the results using the \ac{RF} algorithm (in terms of $\eta_{0.15}$ outlier rate) while for other algorithms, it doesn't significantly alter the $\eta_{0.15}$ outlier rate.  The classification process does slightly reduce $\sigma$ for \ac{kNN} and ANNz algorithms, bringing them closer to the results from the GPz algorithm. 

\begin{figure*}
    \centering
    \includegraphics[trim=0 0 0 0, height=\textheight]{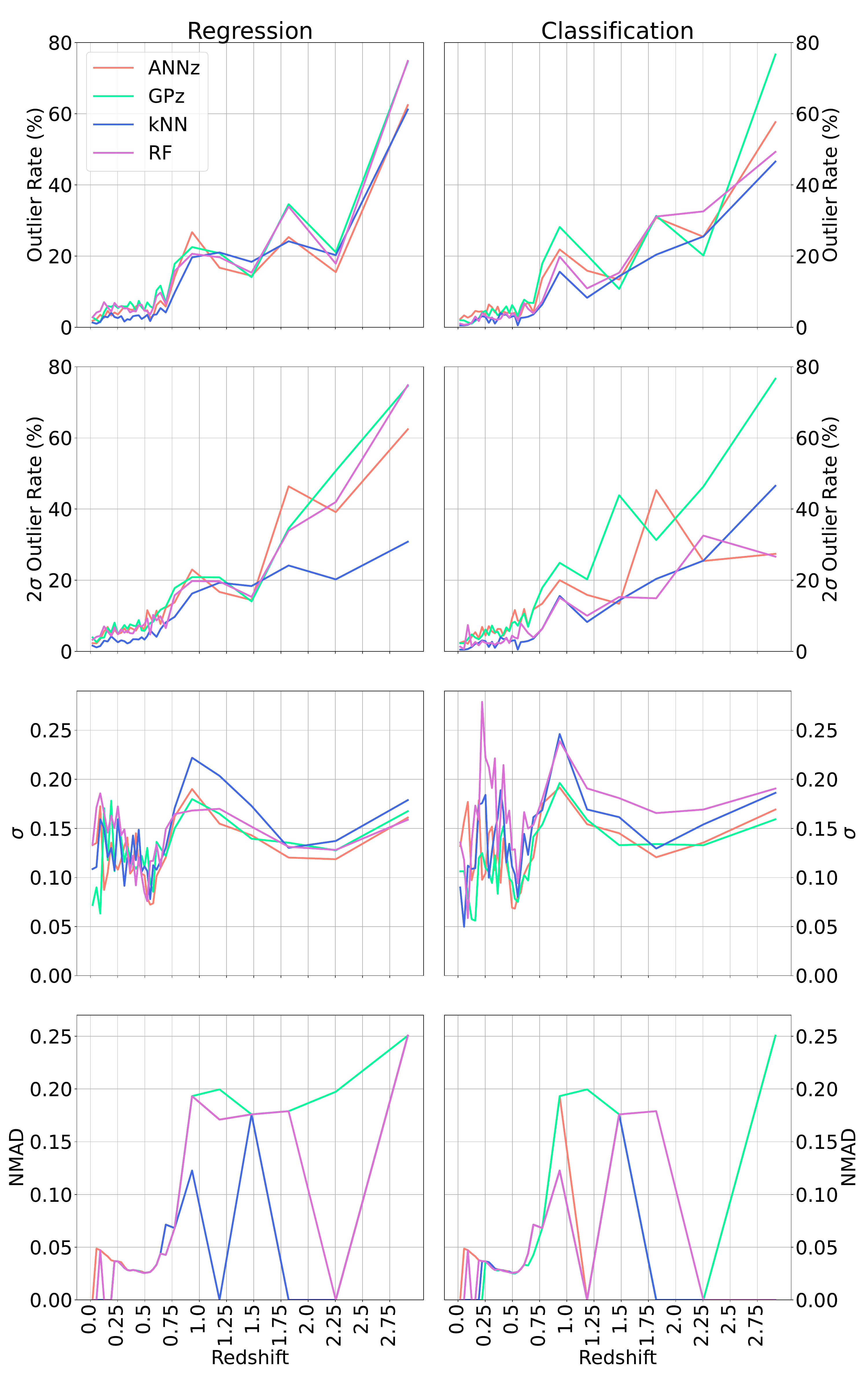}
    \caption{Comparing Regression with Classification over all methods, and all metrics. }
    \label{fig:reg_v_class}
\end{figure*}

When directly comparing the the algorithms in regression and classification mode across the different redshift bins (Figure~\ref{fig:reg_v_class}; showing the $\eta_{0.15}$ and $\eta_{2\sigma}$ outlier rates, the $\sigma$, and \ac{NMAD} as a function of redshift, comparing the Regression modes of each algorithm with the classification modes), we can see that in terms of $\eta_{0.15}$ outlier rate, the \ac{kNN}, \ac{RF} and ANNz algorithms significantly improve for the highest bin (mostly going from 60--80\%, to 40--60\% outlier rates). The average accuracy (both in terms of $\sigma$ and \ac{NMAD}) are comparable between regression and classification modes,

\section{Discussion}

We have found that all \ac{ML} algorithms  suffer from similar issues when estimating the redshift, regardless of the training data, or algorithm used: the redshifts of low-redshift sources are over-estimated (i.e they are predicted to have a higher redshift than their measured redshift), and those of high-redshift sources are under-estimated (i.e they are predicted to be at a lower redshift than their measured redshift suggests). 

In this work, we investigate the combination of heterogeneous datasets (with the impact shown in \ref{sec:appendix_data}), creating a training set with a higher median redshift in order to better sample the high-redshift space, and provide more acceptable redshift estimates to a higher redshift. We combine radio catalogues from the northern hemisphere with \ac{SDSS} optical photometry and spectroscopic redshifts, with radio catalogues from the southern hemisphere with \ac{DES} optical photometry and spectroscopic redshifts from the \ac{OzDES} survey, with the \ac{DES} photometry mapped to the \ac{SDSS} photometry using a third-order polynomial. We compare simple \ac{ML} algorithms in the \ac{kNN} (when using the more complex Mahalanobis distance metric, instead of the standard Euclidean distance metric) and \ac{RF} algorithms, with the much more complex ANNz and GPz (with GPz models trained on smaller subsets, modelled using a \ac{GMM}) --- a \ac{NN} based approach and \ac{GP} based approach respectively. 

We find that the \ac{kNN} algorithm provides the lowest $\eta_{0.15}$ outlier rates across both the Regression, and Classification modes, with outlier rates of $7.26\% \pm 0.02$ and $6.21\% \pm 0.017$ respectively, providing acceptable redshift estimates of $\sim93\%$ of radio sources with complete photometry, up to a redshift of $z\sim3$.

\subsection{Rigidity of $\eta_{0.15}$ outlier rate for Classification}
\label{sec:combine_bins}

The $\eta_{0.15}$ outlier rate is designed to be more accepting of errors as the source's redshift increases. By binning the data into 30 bins with equal numbers of sources, the classification tests break this acceptance as the predicted values of the higher redshift bins become significantly more spread than at low-redshift, to the point where sources are predicted as being outliers if the source is not classified into the exactly correct bin. There are multiple options to extend the flexibility of the $\eta_{0.15}$ outlier rate to this training regime, however, all have flaws. One method would be to adjust the outlier rate so that instead of determining catastrophic outliers based on a numeric value (i.e. 0.15 scaling with redshift), it allows a fixed number of predicted bins above and below the actual redshift bin of the source (i.e. a source can be predicted in the exactly correct bin, $\pm$ some number of bins, and still be considered an acceptable prediction). However, this would mean significant `fiddling' with the bin distribution to ensure that the original intention of the $\eta_{0.15}$ outlier rate is maintained (that, a source be incorrect by up to 0.15 --- scaling with redshift --- before it is considered a ``catastrophic failure''), and would defeat the initial purpose of presenting redshift estimation as a classification task --- creating a uniform distribution in order to better predict sources at higher redshift ranges that are under-represented in all training datasets. Another option would be to drop the $\eta_{0.15}$ outlier rate, and label any source that is predicted within an arbitrary number of bins (2-3 perhaps) of the correct bin as an acceptable estimate. However, this would severely penalise the low-redshift end of the distribution that is dense in sources, and would not be comparable across studies, as it would be impossible to ensure the redshift bins (both in distribution, and density) were similar across different datasets. The simplest alternative is to combine the highest two redshift bins, thereby allowing sources in those top two bins to be classified as either, and not be considered a catastrophic failure. 

\begin{figure*}
    \centering
    \includegraphics[trim=0 0 0 0, height=\textheight]{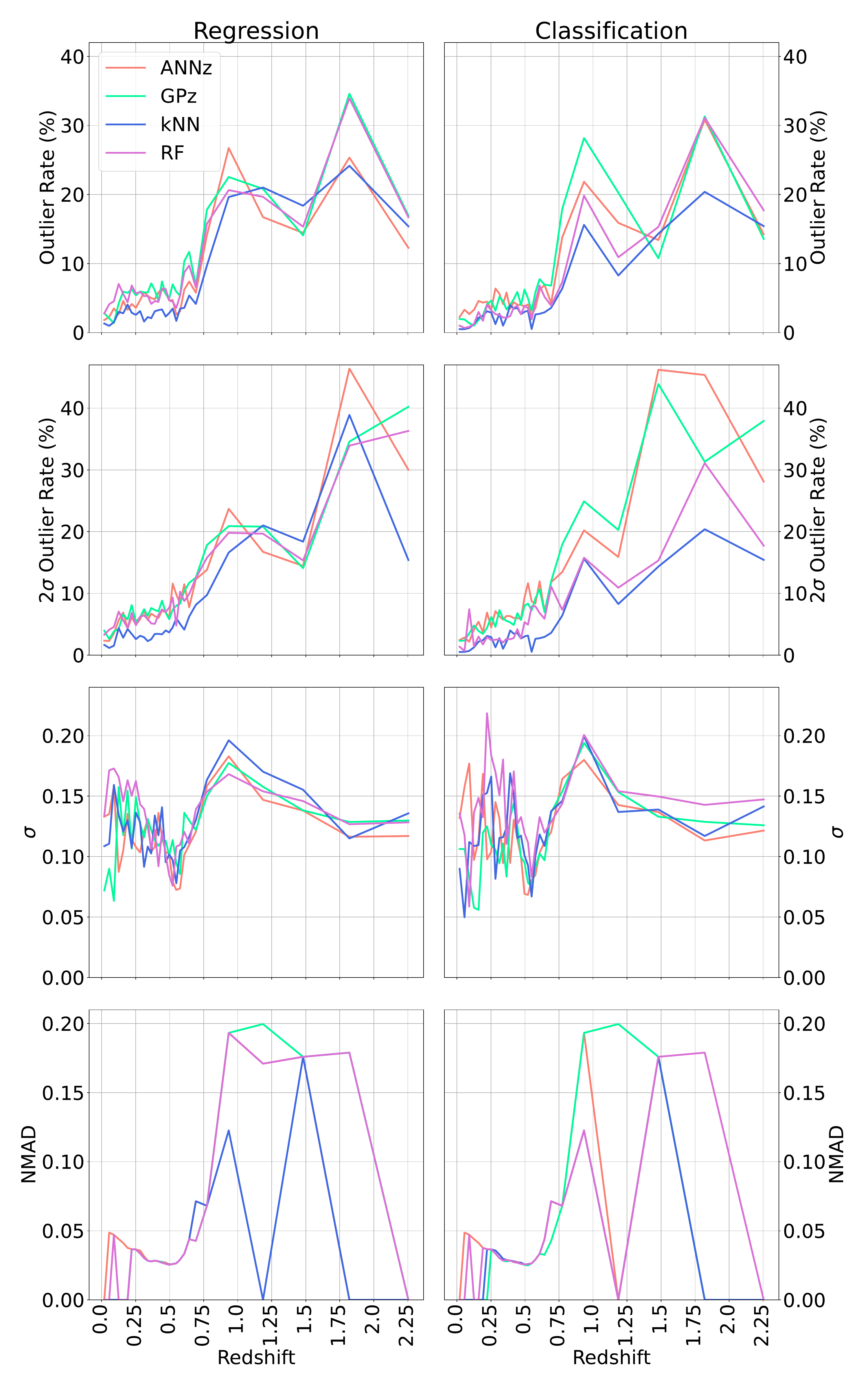}
    \caption{Comparing Regression with Classification over all methods, and all metrics. In this case, the highest redshift bin is combined with the second highest. Results following the combination of the highest two redshift bins discussed in Section~\ref{sec:combine_bins}.}
    \label{fig:reg_v_class_2}
\end{figure*}

In Table~\ref{table:classificationResultsTable_modified} and Figure~\ref{fig:reg_v_class_2}, we present alternatives to Table~\ref{table:classificationResultsTable} and Figure~\ref{fig:reg_v_class} based on the upper two bins being combined. 

\begin{table*}
	\centering
    \caption{Classification results table comparing the different algorithms across the different error metrics (listed in the table footnotes). The best values for each error metric are highlighted in bold. Results following the combination of the highest two redshift bins discussed in Section~\ref{sec:combine_bins}.}
    \label{table:classificationResultsTable_modified}
    \begin{tabular}{cccccc}
    	\toprule
     Algorithm &  $\eta_{0.15}$\footnote{Catastrophic outlier rate, Equation~\ref{eqn:outlier}}  & $\eta_{2\sigma}$\footnote{2$\sigma$ outlier rate, Equation~\ref{eqn:outlier_2sigma}} & $\sigma$\footnote{Residual Standard Deviation, Equation~\ref{eqn:outlier_sigma}} & \ac{NMAD}\footnote{\acl{NMAD}, Equation~\ref{eqn:nmad}} & Accuracy\footnote{Accuracy, Equation~\ref{eqn:accuracy}}\\   
    \midrule
    \ac{kNN} & \textbf{4.88\% $\pm$ 0.02}  & 2.92\% $\pm$ 0.01 & 0.1285 $\pm$ 0.0004 & \textbf{0.039259 $\pm$ 0.00004} & 0.4044 $\pm$ 0.0003 \\ 
    \ac{RF}  & 6.18\% $\pm$ 0.02  & \textbf{2.48\% $\pm$ 0.01} & 0.1445 $\pm$ 0.0004 & 0.03943 $\pm$ 0.00004 & 0.4081 $\pm$ 0.0004 \\ 
    ANNz     & 7.02\% $\pm$ 0.04  & 3.96\% $\pm$ 0.02 & \textbf{0.1178 $\pm$ 0.0004} & 0.0422 $\pm$ 0.0002 & 0.370 $\pm$ 0.001 \\ 
    GMM+GPz  & 7.9\% $\pm$ 0.1 & 4.51\% $\pm$ 0.03 & 0.1217 $\pm$ 0.0008 & 0.0392 $\pm$ 0.0001 & \textbf{0.424 $\pm$ 0.002} \\ 
	\bottomrule
	\end{tabular}
\end{table*}

The combination of redshift bins significantly decreases the $\eta_{0.15}$ outlier rate for all algorithms, with the \ac{kNN} algorithm still performing best (and dropping from 6.21\% to 4.88\%). 

\subsection{Comparison with Previous Work}

Comparison with previous works is difficult, as the selection criteria, such as source type and redshift distribution, can play a significant role in the final error metrics, with most studies aiming for the largest training samples. The motivation of finding the largest possible training set pushes studies into large scale surveys like the \ac{SDSS}, with millions of sources with spectroscopic redshifts available for use. For the testing of algorithms for the use on similar surveys like the \ac{DES} and \ac{DECaLS} surveys, or the \ac{LSST} being conducted at the Vera Rubin Observatory, this motivation is entirely appropriate. However, this workflow cannot be directly compared with algorithms trained and tested on datasets dominated by a specific subset of sources (for example, radio selected samples, which are typically dominated by difficult-to-estimate \acp{AGN}), at a significantly higher redshift. The closest comparison would with \citet{luken_2022}, which contains similar primary algorithms being tested (both this work and \citet{luken_2022} use the \ac{kNN} algorithm with a Mahalanobis distance metric, and the \ac{RF} algorithm, and compare both classification and regression modes), and similar photometry (both use 4 optical, and 4 infrared bands). Both studies are conducted on a radio-selected sample. However, while both are radio-selected, \citet{luken_2022} is restricted to the \ac{ATLAS} dataset and the narrower infrared bands of the \ac{SWIRE} survey, with a significantly smaller dataset and lower median redshift. The combination of the change in infrared photometry to the all-sky, wider-band AllWISE photometry, and the smaller, lower redshift training set used by \citet{luken_2022} leads to slightly lower $\eta_{0.15}$ outlier rates ($\sim6\%$ in \citet{luken_2022}, compared to $\sim7\%$ in this work when comparing the \ac{kNN} algorithm using Regression). However, due to the size and redshift distribution of the dataset compiled in this work, models trained are able to estimate the redshift of radio sources to a significantly higher redshift ($z < 3$, compared with $z < 1$).

\subsection{Algorithm Comparison}

The best performing algorithm (in terms of $\eta_{0.15}$ outlier rate) is the \ac{kNN} algorithm, despite the \ac{kNN} algorithm being significantly less complex than all other approaches tested, with all other methods algorithms combining the results of many models (\ac{RF} combining many Decision Trees, the ANNz algorithm combining 100 different, randomly initialised tree and \ac{NN} based models, and GPz including a pre-processing step using the \ac{GMM} algorithm). This may be due to the difference in the way the different algorithms are trained. While the \ac{RF}, ANNz and GPz algorithms are all methods training some kind of model to best represent the training set, the \ac{kNN} algorithm treats the data itself as the model, and is not trying to learn a representation. This subtle difference means that in cases where the test data is well-represented by the training data, and the number of features is small, the \ac{kNN} algorithm may out-perform more complex algorithms. The \ac{kNN} algorithm also has the added advantage that it doesn't need to try and account for the ``noise'' within astronomical data --- as long as the same types of noise present in the test data is also present in the training data, the \ac{kNN} algorithm doesn't need to handle it in any particular manner.

However, the \ac{kNN} algorithm has two major drawbacks. First, the \ac{kNN} algorithm is making the assumption that the test data follows all the same distributions as the training data. There is no way for the \ac{kNN} algorithm to extrapolate beyond its training set, whereas more complicated algorithms like GPz are --- to a small degree --- able to extend beyond the training sample. This means the \ac{kNN} algorithm is entirely unsuitable when the test set is not drawn from the same distributions as the training set. 

Secondly, like many \ac{ML} algorithms, there is no simple way to provide errors for estimates made by the \ac{kNN} algorithm in regression mode (the classification mode is able to produce a probability density function across the classification bins chosen). The ANNz algorithm is able to use the scatter in its \ac{ML} models predictions as a proxy for the error, and the GPz algorithm is based on the \ac{GP} algorithm, which inherently provides a probability density function --- a significant benefit for some science cases. 

The tension between best $\eta_{0.15}$ outlier rate, and ability to quantify errors is not trivial and is best left to the individual science case as to which algorithm is best suited to the chosen purpose. 

Finally, it is worth reiterating the differing error metrics being optimised between the different algorithms. The ANNz and GPz algorithms are both optimising error metrics favoured by their developers for their particular science needs. In this case, the different error metrics being optimised (like the $\sigma$) do not match the science needs of the \ac{EMU} project, with the $\eta_{0.15}$ outlier rate preferred. The effect of this means that we are comparing an error metric that is optimised in some algorithms (the \ac{kNN} and \ac{RF} algorithms), but not in others (the ANNz and GPz algorithms). This presents an inherent disadvantage to the ANNz and GPz algorithms, and and may contribute to their lower performance.

 \subsection{Estimating Confidence Intervals}
\label{sec:uncertain_estimates}

Both the ANNz and GPz algorithms explicitly estimate the uncertainty of any prediction made. GPz estimates uncertainties directly as a by-product of the Gaussian fitting in GPz. ANNz estimates its uncertainties as an additional step, where the 100 most similar galaxies from the training set to the test source are found using the $k$NN algorithm, biases for each estimated redshift calculated, and the $68^\mathrm{th}$ percentile taken as the uncertainty of the galaxy \citep{oyaizuGalaxyPhotometricRedshift2008,sadehANNz2PhotometricRedshift2016}.

The \ac{RF} algorithm is next simplest to identify uncertain estimates. Following \citet{wagerrfconfidence}\footnote{implemented as \url{https://contrib.scikit-learn.org/forest-confidence-interval/index.html}}, confidence intervals for \ac{RF} models can be estimated using the Jackknife method.

Finally, the $k$NN algorithm does not have a natural way of estimating the uncertainty of predictions. Similar to the method described in \citet{oyaizuGalaxyPhotometricRedshift2008} though, we can get an understanding of which estimates are likely to be uncertain by examining the similar galaxies. We can follow the below workflow to estimate the uncertainty of our predictions, noting that they are unlikely to be realistic uncertainties for the estimate, and more an estimate of how uncertain the ``model'' is of the prediction, given the data:

For every source in the test set:
\begin{enumerate}
    \item Identify the $k_n$ sources used in the estimation of the redshift.
    \item Use the same model to estimate the redshift of the above $k_n$ sources.
    \item Calculate the variance of the $k_n$ sources redshift estimates, and take the variance as the uncertainty for the prediction.
\end{enumerate}
We emphasise that this uncertainty is not an estimate of how well the redshift prediction of the test source fits the photometry --- it is purely an estimate of how varied the sources were that were used to make the initial estimate, with the implicit understanding being that the more varied the sources used to predict the redshift, the less likely the estimate is to be accurate. Additionally, there are no photometric uncertainties involved, so the uncertainty provided is further unlikely to be scientifically meaningful, beyond helping to identify potentially unreliable estimates.

\subsubsection{Removing Uncertain Predictions}
\label{sec:removing_uncertain}

Certainty thresholds for defining acceptable estimates are not, to the best of our knowledge, typically published. \citet{duncan_2022} suggests Equation~\ref{eqn:gpz_cutoff}:
\begin{equation}
\label{eqn:gpz_cutoff}
    \frac{\sigma_z}{(1 + z_{photo})} < 0.2
\end{equation}
where $\sigma_z$ is the uncertainty estimate from the GPz model, and $z_{photo}$ is the photometric redshift estimated from the same model. Unfortunately, given the different quantities the different uncertainties are designed to capture, Equation~\ref{eqn:gpz_cutoff} can only be used for the estimates measured using the GPz algorithm. 

For other algorithms we aim to find a (where possible) statistically sound method of removing the most uncertain estimates, while maintaining approximately the same number of `certain' sources in order to compare outlier rates with the certain GPz estimates. 

For the \ac{kNN} algorithm and uncertainties (defined in Section~\ref{sec:uncertain_estimates}), we can define Equation~\ref{eqn:knn_cutoff}:
\begin{equation}
\label{eqn:knn_cutoff}
    \frac{\sigma_z}{(1 + z_{photo})} < \frac{\sum_i (\sigma_{zi} - \bar{\sigma_z})^2}{n-1}
\end{equation}
where $\bar{\sigma_z}$ is the average uncertainty. 

No statistical method of determining a cutoff for the ANNz and \ac{RF} produced similar source counts as the GPz algorithm, and hence for this work we choose the following values (Equations~\ref{eqn:annz_cutoff} and \ref{eqn:rf_cutoff} respectively) in order to produce comparable outlier rates:
\begin{equation}
\label{eqn:annz_cutoff}
    \frac{\sigma_z}{(1 + z_{photo})} < 0.1
\end{equation}
\begin{equation}
\label{eqn:rf_cutoff}
    \sigma_z < 2.302
\end{equation}

Once these outliers are removed, the residual outlier rates for all methods drop significantly. We show the original outlier rates, the outlier rates of the `certain' predictions, and the outlier rates of the `uncertain' predictions shown in Table~\ref{table:certain_predictions} for all algorithms. Prediction plots similar to Figure~\ref{fig:knn_regress} for each subset and algorithm can be found in Figures~\ref{fig:appendix_certain_knn} to \ref{fig:appendix_certain_gpz} in \ref{appendix:appendix_certain}. 

\begin{table*}
	\centering
    \caption{Outlier rates of each error metric using their regression modes, showing the original outlier rate, the outlier rate of the subset of sources deemed `certain', and the outlier rate of the remaining sources.}
    \label{table:certain_predictions}
    \begin{tabular}{cccccc}
    	\toprule
     Algorithm &  Original $\eta_{0.15}$  & `Certain' Predictions $\eta_{0.15}$ & `Certain' Source Count & `Uncertain' Predictions $\eta_{0.15}$  & `Uncertain' Source Count \\   
    \midrule
    \ac{kNN} & 7.18\%  & 1.27\%  & 14,463  & 32.18\%  & 3,421 \\
    \ac{RF} & 9.96\%  & 5.23\%  & 14,610  & 31.09\%  & 3,274 \\
    ANNz & 8.85\%  & 3.10\%  & 14,697 & 35.36\%  & 3,187 \\
    GPz & 11.21\%  & 3.14\%  & 14,522  & 46.07\%  & 3,362 \\
	\bottomrule
	\end{tabular}
\end{table*}

As demonstrated, the removal of predictions with high uncertainty greatly improves the outlier rates of all algorithms, with the \ac{kNN} algorithm still performing best, the ANNz and GPz algorithms performing equally well, and the \ac{RF} algorithm performing worst. We do note, however, that the formal definition of uncertain sources by \citet{duncan_2022} is combined with very well defined uncertainties to make GPz estimates more robust and reliable, particularly when spectroscopic redshifts are not available in test fields of sufficient depth and quantity to help quantify reliability.

\subsection{Effects of Differing Radio Survey Depths}

Radio sources are typically more difficult to estimate the redshift of using \ac{ML} than optically-selected sources, as they tend to contain rarer sub-types of galaxies, and hence constructing a representative training sample is problematic. While all of the samples in our training set have been radio-selected, the depth of the radio survey used can play a part in what sub-types of galaxies are represented in the radio sample. As shown by \citet[][Figure 13; upper right panel]{vla-cosmos} at $\sim$110$\mu$Jy, radio samples stop being dominated by \ac{AGN}, and begin being dominated by \acp{SFG}, and hence would require additional \ac{SFG} samples in the training sample in order to best estimate these sources. While the majority ($\sim$90\%) of sources used in our training sample come from the \ac{RGZ} catalogues (drawn from the \ac{VLA} \ac{FIRST} survey) which have a sensitivity of $\sim$150\,$\mu$Jy, we include sources from the Stripe 82 region (\citet{Stripe_Hodge_2011}; RMS: 52\,$\mu$Jy, and \citet{Stripe_Prescott_2018}; RMS: 82\,$\mu$Jy), and the \ac{ATLAS} surveys (\citet{franzenATLASThirdRelease2015}; RMS: 14\,$\mu$Jy). These additional sources provide some coverage of the radio-faint parameter space, however, we acknowledge that the comparatively small number is inadequate to completely model the space. 

Future work will include more radio selected data from deep fields like \ac{COSMOS}, and the \ac{LOFAR} Deep Fields.

\section{Conclusion}

Machine Learning attempts for estimating the redshift of radio selected galaxies have significant benefits over traditional template fitting methods --- they don't require specifically developed templates, nor do they require the disentanglement of the black hole emission from the galaxy emission. However, the major downside is the requirement for a representative training sample --- a significant difficulty given the requirement for spectroscopic redshift measurements, and the typically significantly higher median redshift of radio surveys, when compared with optical surveys. 

By combining radio-selected data from the northern- and southern-hemisphere, we have created a larger sample of radio galaxies for training \ac{ML} algorithms. Once the \ac{DES} optical data was homogenised with the \ac{SDSS} optical photometry, current leading \ac{ML} algorithms were tested. We show that the \ac{kNN} algorithm --- in both regression and classification tests --- provides the lowest $\eta_{0.15}$ outlier rate, estimating $\sim92\%$ of radio-selected sources within an acceptable limit. The depth in redshift distribution of the assembled training set allows us to estimate the redshift of sources up to $z = 3$ before the results are dominated by random, under-estimated scatter. 

We show that we can use the classification modes of the tested \ac{ML} methods to identify $\sim76\%$ of sources at the highest two redshift bins ($z = 2.25$ and $z > 2.51$), providing a way of first identifying the highest redshift sources, before using the regression modes of the provided algorithms to estimate the redshift of the remaining sources more effectively. 

In this work, we show that the \ac{kNN} algorithm using the Mahalanobis distance metric performs best (i.e. minimises outlier rate) for the estimation of the redshift of radio galaxies.

\section{Data Availability}
All code used within this study is available at \url{https://github.com/kluken/PhotoZForHigh-ZRadioSurveysPASA}. Data are described and are available at \citet{csiro_data}.

\section{Acknowledgements}

We thank the anonymous referee for their time and thoughtful comments, as well as their prompt response in helping to improve the manuscript.

We also thank Sarah White for her helpful comments.

The Australia Telescope Compact Array is part of the Australia Telescope National Facility which is funded by the Australian Government for operation as a National Facility managed by CSIRO. We acknowledge the Gomeroi people as the traditional owners of the Observatory site.

This scientific work uses data obtained from Inyarrimanha Ilgari Bundara / the Murchison Radio-astronomy Observatory. We acknowledge the Wajarri Yamaji People as the Traditional Owners and native title holders of the Observatory site. CSIRO’s ASKAP radio telescope is part of the Australia Telescope National Facility (https://ror.org/05qajvd42). Operation of ASKAP is funded by the Australian Government with support from the National Collaborative Research Infrastructure Strategy. ASKAP uses the resources of the Pawsey Supercomputing Research Centre. Establishment of ASKAP, Inyarrimanha Ilgari Bundara, the CSIRO Murchison Radio-astronomy Observatory and the Pawsey Supercomputing Research Centre are initiatives of the Australian Government, with support from the Government of Western Australia and the Science and Industry Endowment Fund.

Based in part on data acquired at the Anglo-Australian Telescope. We acknowledge the traditional owners of the land on which the AAT stands, the Gamilaroi people, and pay our respects to elders past and present.

This project used public archival data from the Dark Energy Survey (DES). Funding for the DES Projects has been provided by the U.S. Department of Energy, the U.S. National Science Foundation, the Ministry of Science and Education of Spain, the Science and Technology FacilitiesCouncil of the United Kingdom, the Higher Education Funding Council for England, the National Center for Supercomputing Applications at the University of Illinois at Urbana-Champaign, the Kavli Institute of Cosmological Physics at the University of Chicago, the Center for Cosmology and Astro-Particle Physics at the Ohio State University, the Mitchell Institute for Fundamental Physics and Astronomy at Texas A\&M University, Financiadora de Estudos e Projetos, Funda{\c c}{\~a}o Carlos Chagas Filho de Amparo {\`a} Pesquisa do Estado do Rio de Janeiro, Conselho Nacional de Desenvolvimento Cient{\'i}fico e Tecnol{\'o}gico and the Minist{\'e}rio da Ci{\^e}ncia, Tecnologia e Inova{\c c}{\~a}o, the Deutsche Forschungsgemeinschaft, and the Collaborating Institutions in the Dark Energy Survey.

The Collaborating Institutions are Argonne National Laboratory, the University of California at Santa Cruz, the University of Cambridge, Centro de Investigaciones Energ{\'e}ticas, Medioambientales y Tecnol{\'o}gicas-Madrid, the University of Chicago, University College London, the DES-Brazil Consortium, the University of Edinburgh, the Eidgen{\"o}ssische Technische Hochschule (ETH) Z{\"u}rich,  Fermi National Accelerator Laboratory, the University of Illinois at Urbana-Champaign, the Institut de Ci{\`e}ncies de l'Espai (IEEC/CSIC), the Institut de F{\'i}sica d'Altes Energies, Lawrence Berkeley National Laboratory, the Ludwig-Maximilians Universit{\"a}t M{\"u}nchen and the associated Excellence Cluster Universe, the University of Michigan, the National Optical Astronomy Observatory, the University of Nottingham, The Ohio State University, the OzDES Membership Consortium, the University of Pennsylvania, the University of Portsmouth, SLAC National Accelerator Laboratory, Stanford University, the University of Sussex, and Texas A\&M University.

Based in part on observations at Cerro Tololo Inter-American Observatory, National Optical Astronomy Observatory, which is operated by the Association of Universities for Research in Astronomy (AURA) under a cooperative agreement with the National Science Foundation.

This work was performed on the OzSTAR national facility at Swinburne University of Technology. The OzSTAR program receives funding in part from the Astronomy National Collaborative Research Infrastructure Strategy (NCRIS) allocation provided by the Australian Government.

This publication has been made possible by the participation of more than 12,000 volunteers in the Radio Galaxy Zoo project. Their contributions are individually acknowledged at \url{http://rgzauthors.galaxyzoo.org}

Funding for the Sloan Digital Sky 
Survey IV has been provided by the 
Alfred P. Sloan Foundation, the U.S. 
Department of Energy Office of 
Science, and the Participating 
Institutions. 

SDSS-IV acknowledges support and 
resources from the Center for High 
Performance Computing  at the 
University of Utah. The SDSS 
website is www.sdss.org.

SDSS-IV is managed by the 
Astrophysical Research Consortium 
for the Participating Institutions 
of the SDSS Collaboration including 
the Brazilian Participation Group, 
the Carnegie Institution for Science, 
Carnegie Mellon University, Center for 
Astrophysics | Harvard \& 
Smithsonian, the Chilean Participation 
Group, the French Participation Group, 
Instituto de Astrof\'isica de 
Canarias, The Johns Hopkins 
University, Kavli Institute for the 
Physics and Mathematics of the 
Universe (IPMU) / University of 
Tokyo, the Korean Participation Group, 
Lawrence Berkeley National Laboratory, 
Leibniz Institut f\"ur Astrophysik 
Potsdam (AIP),  Max-Planck-Institut 
f\"ur Astronomie (MPIA Heidelberg), 
Max-Planck-Institut f\"ur 
Astrophysik (MPA Garching), 
Max-Planck-Institut f\"ur 
Extraterrestrische Physik (MPE), 
National Astronomical Observatories of 
China, New Mexico State University, 
New York University, University of 
Notre Dame, Observat\'ario 
Nacional / MCTI, The Ohio State 
University, Pennsylvania State 
University, Shanghai 
Astronomical Observatory, United 
Kingdom Participation Group, 
Universidad Nacional Aut\'onoma 
de M\'exico, University of Arizona, 
University of Colorado Boulder, 
University of Oxford, University of 
Portsmouth, University of Utah, 
University of Virginia, University 
of Washington, University of 
Wisconsin, Vanderbilt University, 
and Yale University.

\appendix
\section{Data Homogenisation}
\label{sec:appendix_data}
\acresetall

As demonstrated in Figure~\ref{fig:des_sdss_homogenisaton}, the difference in measured photometry can be significant between the \ac{DES} and \ac{SDSS} catalogues. In order to quantify how much of an impact this difference in photometry has, we present the following results using the \ac{kNN} algorithm. \ref{sec:appendix_train_sdss_test_des} and subsections show the effect of the data homogenisation discussed in Section~\ref{sec:data_homogenisation} when training on \ac{SDSS} photometry, and testing on \ac{DES} photometry. \ref{sec:appendix_train_des_test_sdss} and subsections show the effect of the data homogenisation when training on \ac{DES} photometry, and testing on \ac{SDSS} photometry. Finally, \ref{sec:appendix_comparison} directly compares these results.

\subsection{Training on SDSS Photometry, Testing on DES Photometry}
\label{sec:appendix_train_sdss_test_des}

This section is divided into two components --- \ref{sec:appendix_train_sdss_test_des_reg} and \ref{sec:appendix_appendix_train_sdss_test_des_class}. In these subsections, we demonstrate the results of using uncorrected and corrected photometry in regression and classification tests.

\subsubsection{Regression}
\label{sec:appendix_train_sdss_test_des_reg}

\begin{figure}
    \centering
    \includegraphics[trim=0 0 0 0, width=\textwidth]{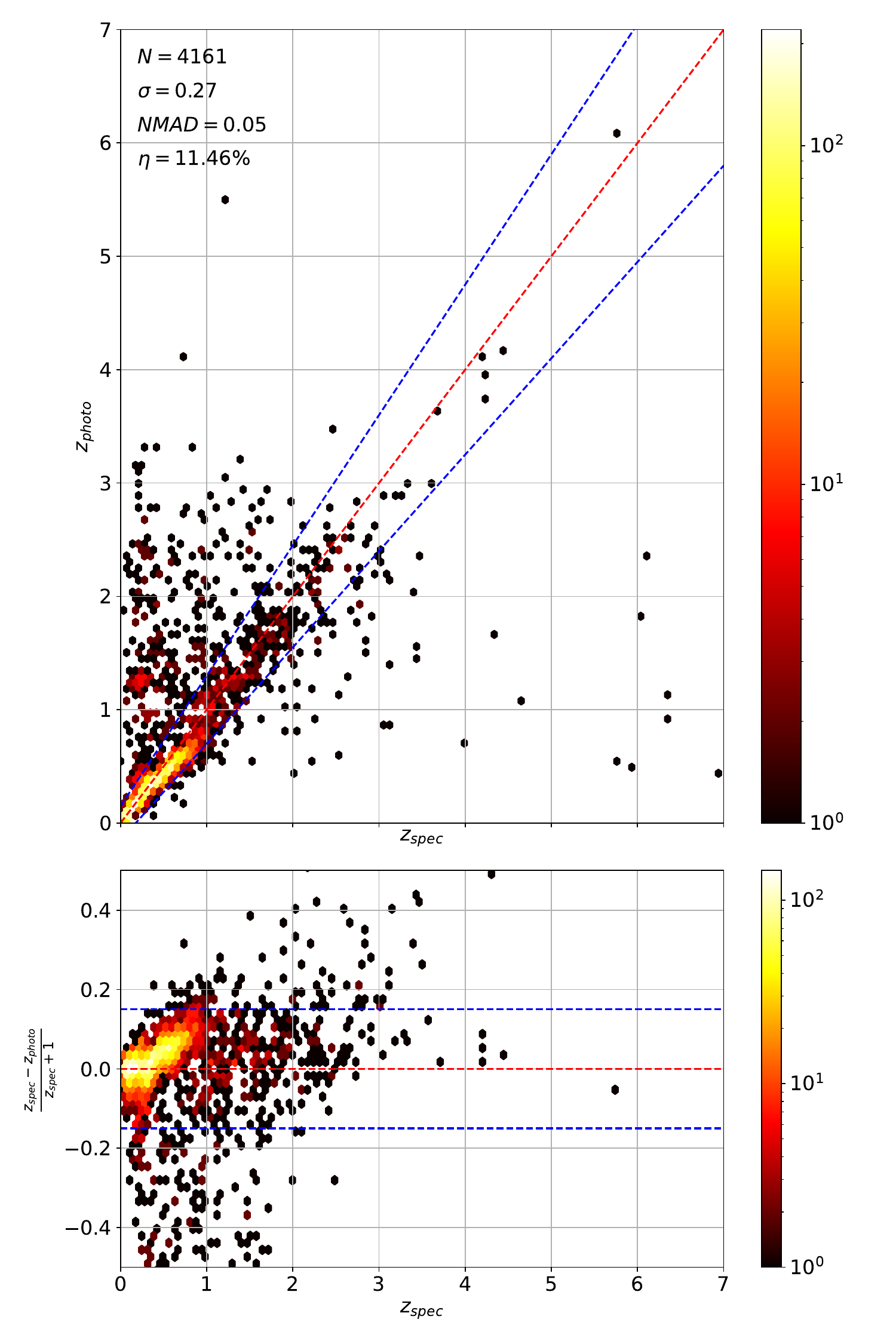}
    \caption{Training on SDSS, Testing on DES, with the same axes and notation as Figure 14}
    \label{fig:appendix_train_sdss_test_des_regress}
\end{figure}

\begin{figure}
    \centering
    \includegraphics[trim=0 0 0 0, width=\textwidth]{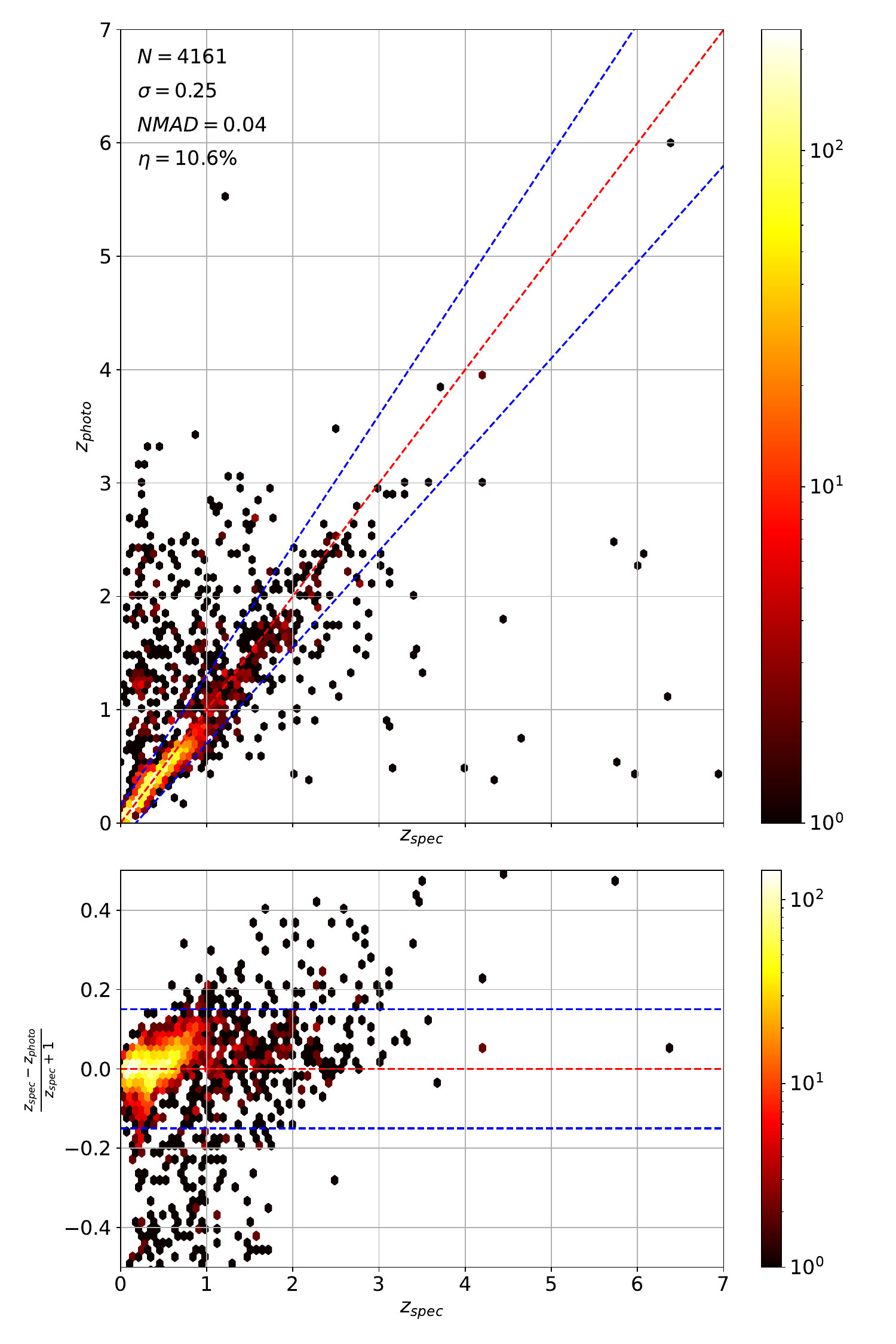}
    \caption{Training on SDSS, Testing on DES - Corrected, , with the same axes and notation as Figure 14}
    \label{fig:appendix_train_sdss_test_des_regress_corrected}
\end{figure}

Figure~\ref{fig:appendix_train_sdss_test_des_regress} and \ref{fig:appendix_train_sdss_test_des_regress_corrected} are of the same style as Figure~\ref{fig:knn_regress}. Figure~\ref{fig:appendix_train_sdss_test_des_regress} is the result of training on \ac{SDSS} photometry, and testing on \ac{DES} photometry. Figure~\ref{fig:appendix_train_sdss_test_des_regress_corrected} is the result of training on \ac{SDSS} photometry, and testing on corrected \ac{DES} photometry.

\subsubsection{Classification}
\label{sec:appendix_appendix_train_sdss_test_des_class}

\begin{figure*}
    \centering
     \begin{subfigure}[b]{0.49\textwidth}
        \centering
        \includegraphics[trim=0 0 0 0, width=\textwidth]{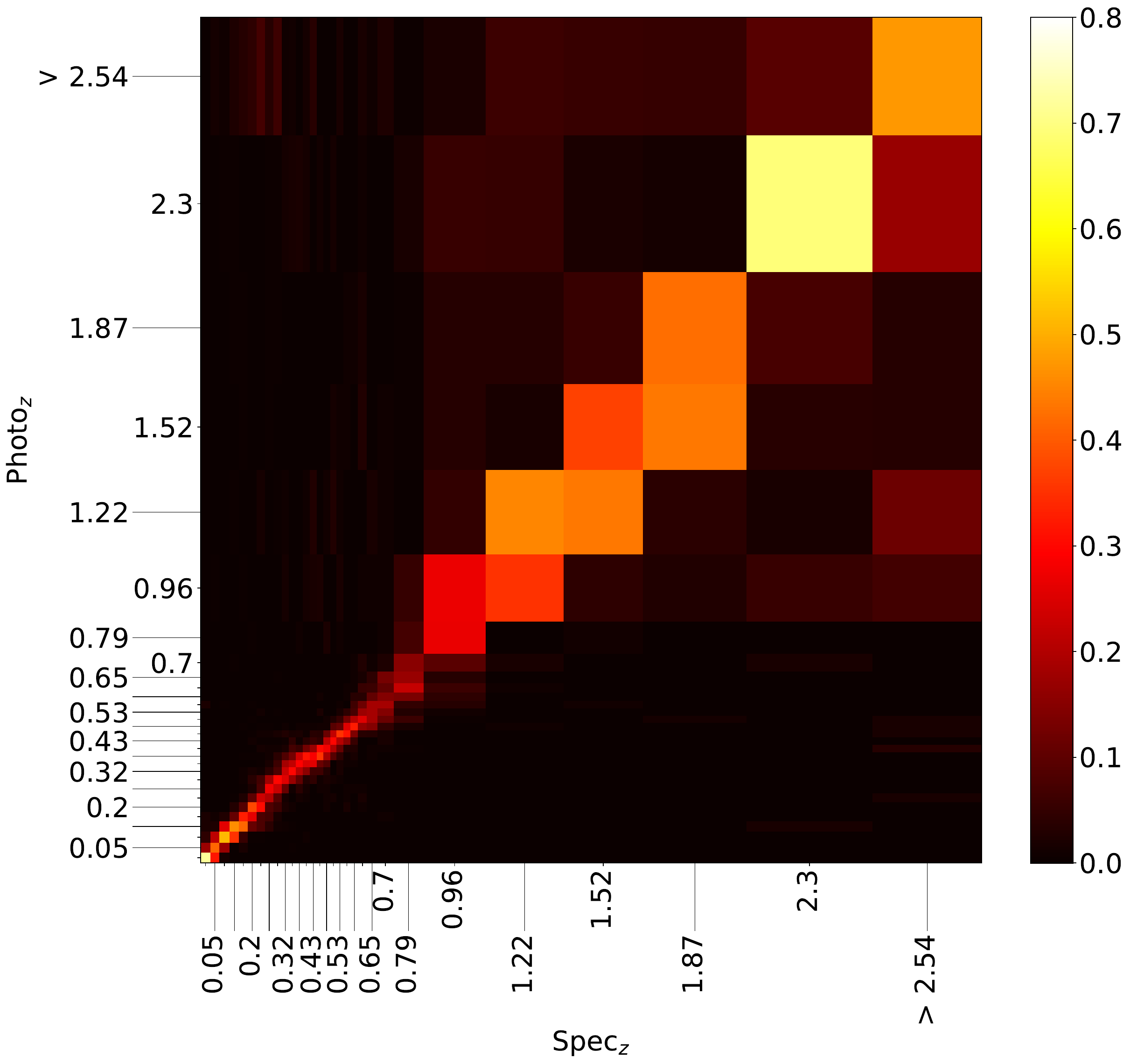}
        \caption{Training on SDSS, Testing on DES}
        \label{fig:appendix_train_sdss_test_des_class}
     \end{subfigure}
     \hfill
     \begin{subfigure}[b]{0.49\textwidth}
         \centering
         \includegraphics[trim=0 0 0 0, width=\textwidth]{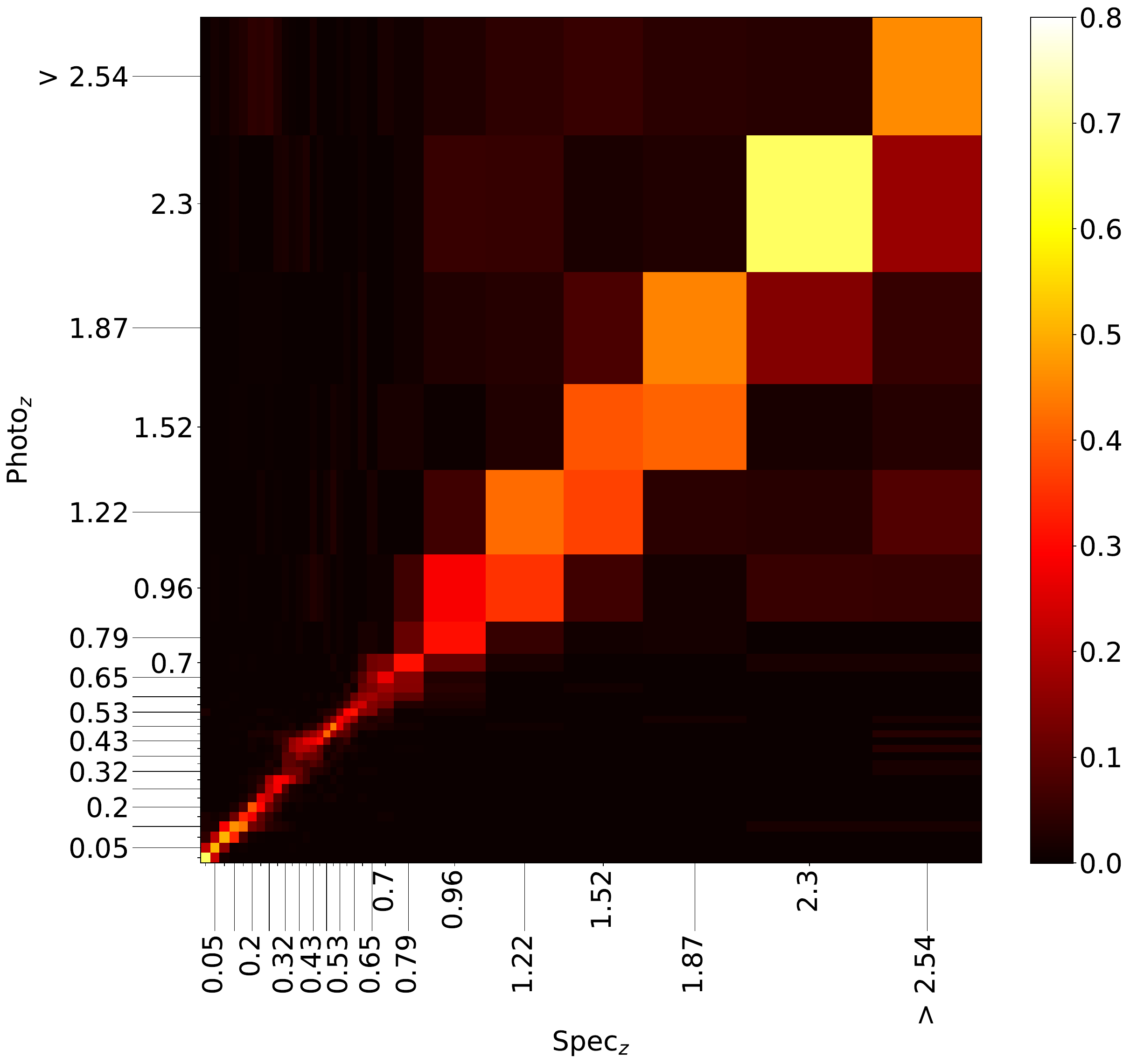}
        \caption{Training on SDSS, Testing on DES - Corrected}
        \label{fig:appendix_train_sdss_test_des_class_corrected}
    \end{subfigure}
\end{figure*}

Figure~\ref{fig:appendix_train_sdss_test_des_class} and \ref{fig:appendix_train_sdss_test_des_class_corrected} are of the same style as Figure~\ref{fig:knn_class_scaled}. Figure~\ref{fig:appendix_train_sdss_test_des_class} is the result of training on \ac{SDSS} photometry, and testing on \ac{DES} photometry. Figure~\ref{fig:appendix_train_sdss_test_des_class_corrected} is the result of training on \ac{SDSS} photometry, and testing on corrected \ac{DES} photometry.

\subsection{Training on DES Photometry, Testing on SDSS Photometry}
\label{sec:appendix_train_des_test_sdss}

This section is divided into two components --- \ref{sec:appendix_train_des_test_sdss_reg} and \ref{sec:appendix_train_des_test_sdss_class}. In these subsections, we demonstrate the results of using uncorrected and corrected photometry in regression and classification tests.

\subsubsection{Regression}
\label{sec:appendix_train_des_test_sdss_reg}

\begin{figure}
    \centering
    \includegraphics[trim=0 0 0 0, width=\textwidth]{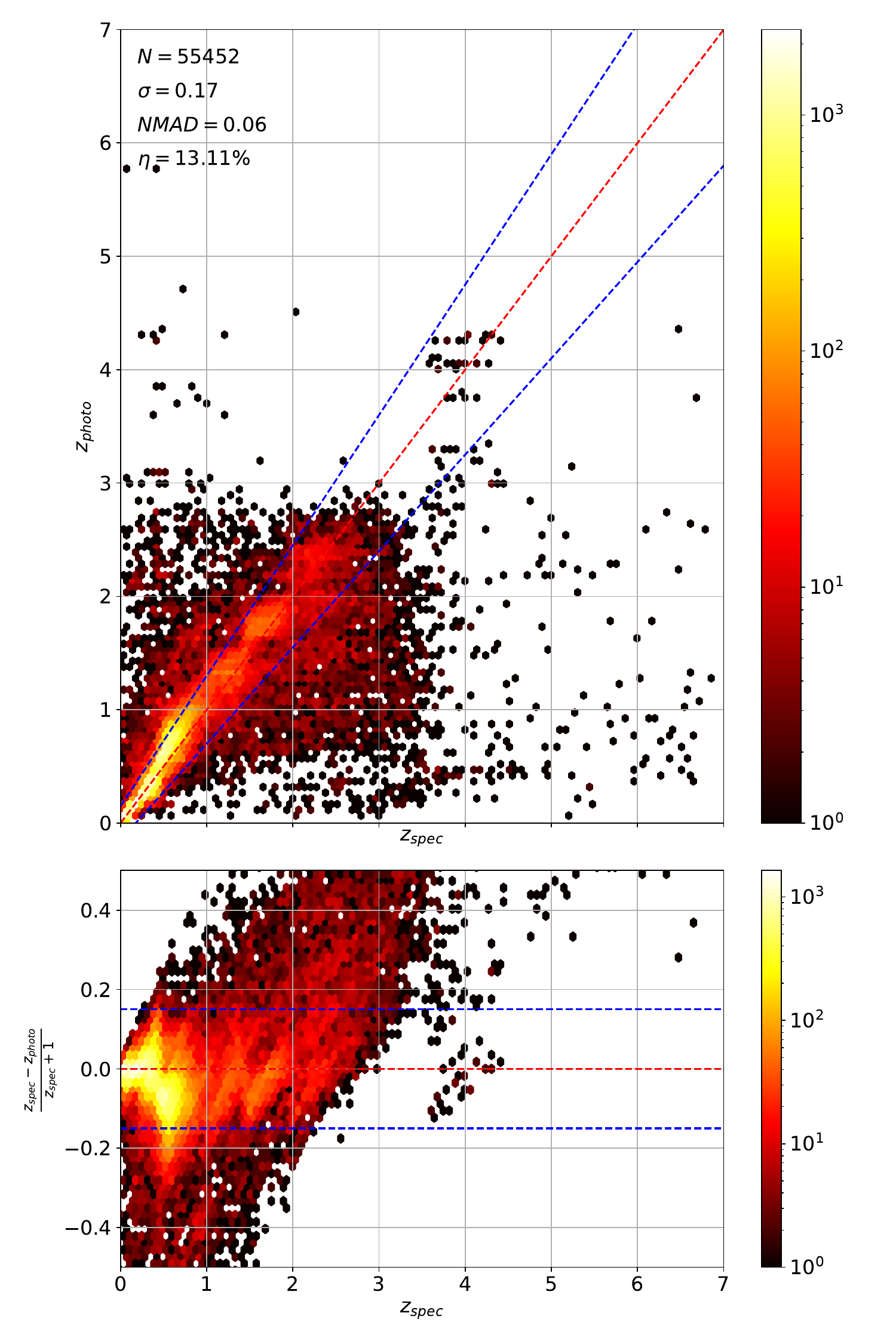}
    \caption{Training on DES, Testing on SDSS}
    \label{fig:appendix_train_des_test_sdss_regress}
\end{figure}

\begin{figure}
    \centering
    \includegraphics[trim=0 0 0 0, width=\textwidth]{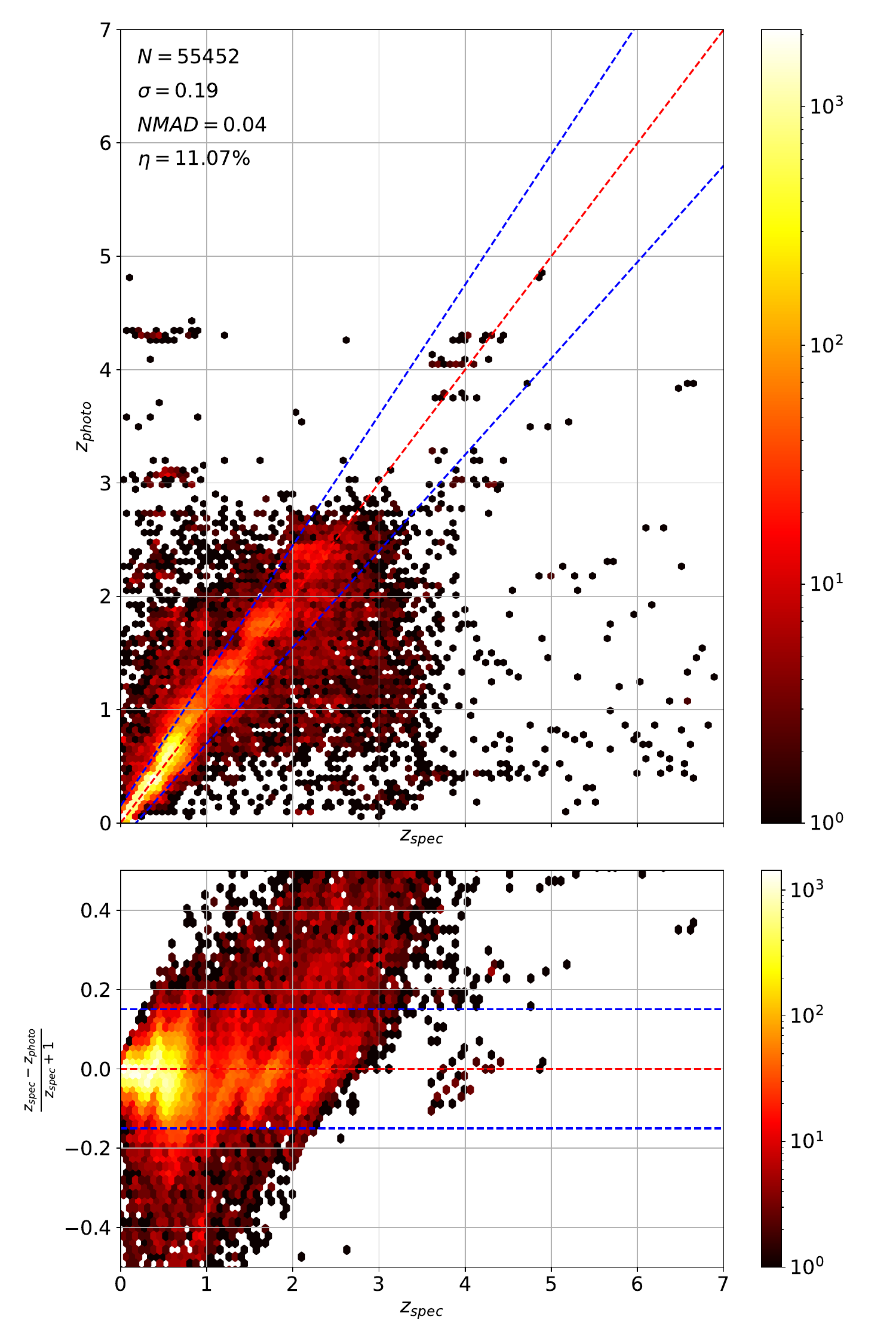}
    \caption{Training on DES, Testing on SDSS - Corrected}
    \label{fig:appendix_train_des_test_sdss_regress_corrected}
\end{figure}

Figure~\ref{fig:appendix_train_des_test_sdss_regress} and \ref{fig:appendix_train_des_test_sdss_regress_corrected} are of the same style as Figure~\ref{fig:knn_regress}. Figure~\ref{fig:appendix_train_des_test_sdss_regress} is the result of training on \ac{SDSS} photometry, and testing on \ac{DES} photometry. Figure~\ref{fig:appendix_train_des_test_sdss_regress_corrected} is the result of training on \ac{SDSS} photometry, and testing on corrected \ac{DES} photometry.

\subsubsection{Classification}
\label{sec:appendix_train_des_test_sdss_class}

\begin{figure*}
    \centering
         \begin{subfigure}[b]{0.49\textwidth}
         \centering
         \includegraphics[trim=0 0 0 0, width=\textwidth]{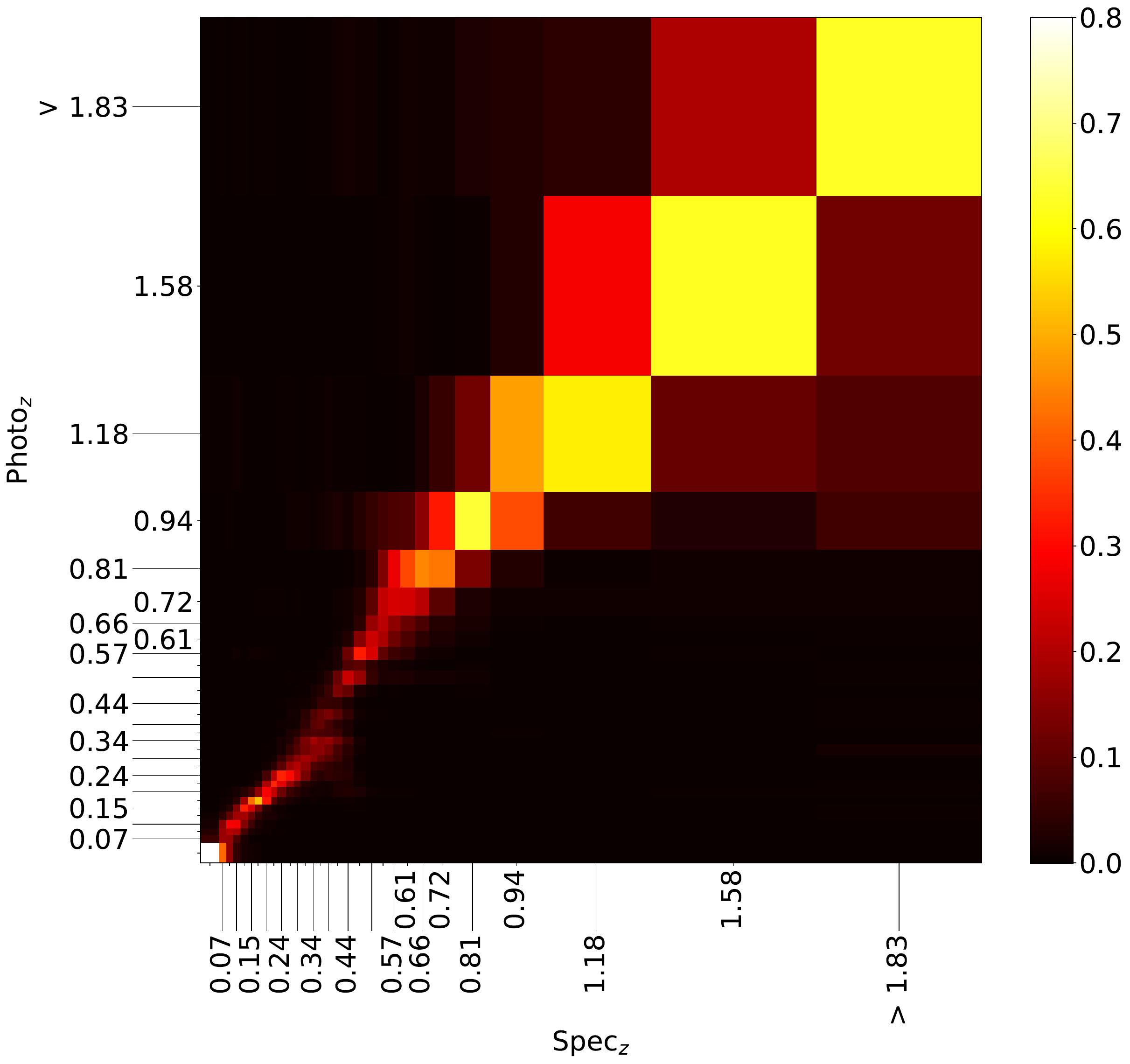}
         \caption{Training on DES, Testing on SDSS}
         \label{fig:appendix_train_des_test_sdss_class}
     \end{subfigure}
     \hfill
     \begin{subfigure}[b]{0.49\textwidth}
         \centering
         \includegraphics[trim=0 0 0 0, width=\textwidth]{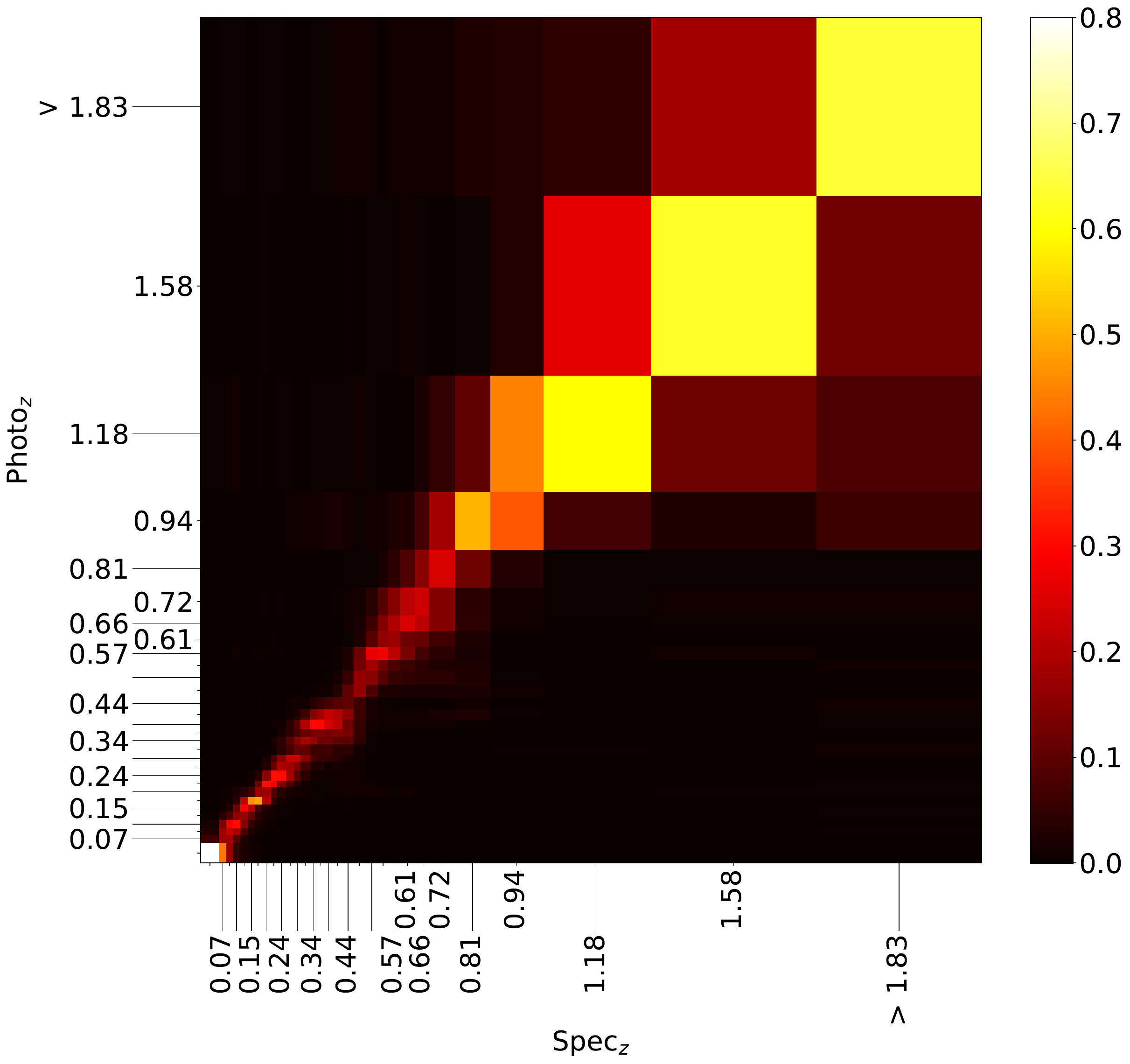}
         \caption{Training on DES, Testing on SDSS - Corrected}
        \label{fig:appendix_train_des_test_sdss_class_corrected}
     \end{subfigure}
\end{figure*}

Figure~\ref{fig:appendix_train_des_test_sdss_class} and \ref{fig:appendix_train_des_test_sdss_class_corrected} are of the same style as Figure~\ref{fig:knn_class_scaled}. Figure~\ref{fig:appendix_train_des_test_sdss_class} is the result of training on \ac{SDSS} photometry, and testing on \ac{DES} photometry. Figure~\ref{fig:appendix_train_des_test_sdss_class_corrected} is the result of training on \ac{SDSS} photometry, and testing on corrected \ac{DES} photometry.

\subsection{Comparison between Uncorrected, and Corrected data}
\label{sec:appendix_comparison}

\begin{figure*}
    \centering
    \includegraphics[trim=0 0 0 0, width=\textwidth]{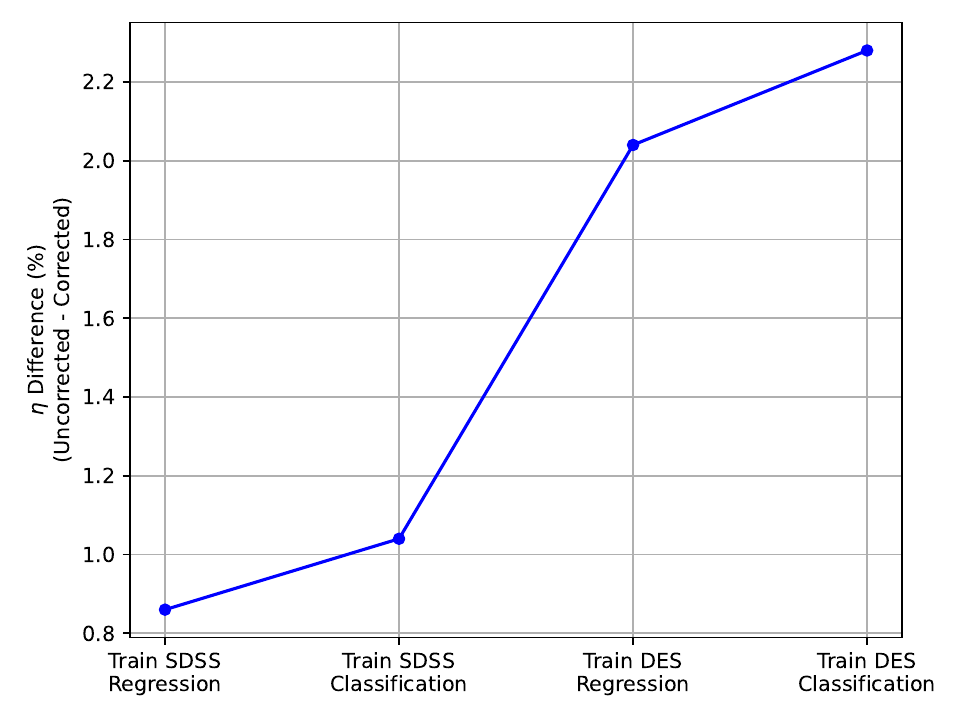 }
    \caption{Comparison of the $\eta$ outlier rates when trained on corrected and uncorrect data.}
    \label{fig:appendix_compare}
\end{figure*}

In all tests, homogenising the \ac{DES} photometry to the \ac{SDSS} improved the outlier rates. Table~\ref{table:appendix_compare} and Figure~\ref{fig:appendix_compare} demonstrate that the outlier rate improves by $\sim1-2\%$ for all tests. 

\begin{table}
	\centering
    \caption{Comparison between predictions using the \ac{kNN} algorithm, trained on one subset of data (either the northern \ac{SDSS} photometry, or the southern \ac{DES} photometry) and tested on the other. }
    \label{table:appendix_compare}
    \begin{tabular}{ccccc}
    	\toprule
     Training &  Test    & Method         & $\eta$       & $\eta$ \\   
     Set      &  Set     &                & Uncorrected  & Corrected \\
    \midrule
    \ac{SDSS} & \ac{DES} & Regression     & 11.46\%      & 10.6\%  \\ 
    \ac{SDSS} & \ac{DES} & Classification & 8.56\%       & 7.52\%  \\ 
    \ac{DES} & \ac{SDSS} & Regression     & 13.11\%      & 11.07\%  \\ 
    \ac{DES} & \ac{SDSS} & Classification & 12.97\%      & 10.69\%  \\ 
	\bottomrule
	\end{tabular}
\end{table}

\section{Comparing All, with Certain and Uncertain Predictions}
\label{appendix:appendix_certain}

\begin{figure*}
    \centering
    \includegraphics[trim=0 0 0 0, width=\textwidth]{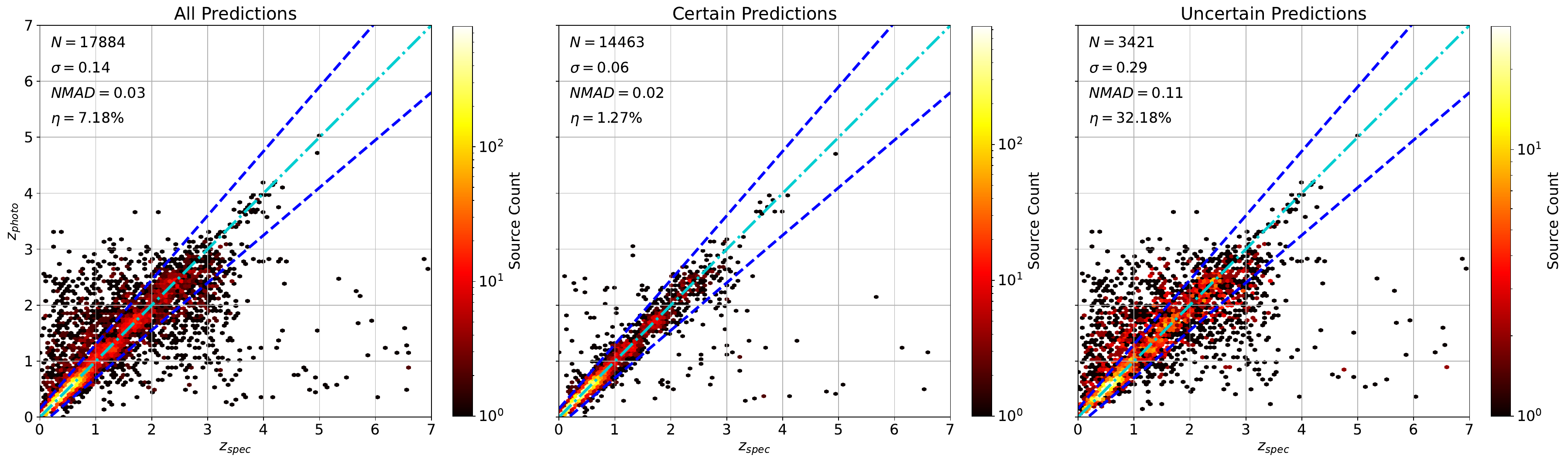 }
    \caption{Similar to the top panel of Figure~\ref{fig:knn_regress}, comparing all of the predictions (left), with the predictions deemed `certain' (middle) and the predictions deemed `uncertain' (right) using the \ac{kNN} algorithm. }
    \label{fig:appendix_certain_knn}
\end{figure*}

\begin{figure*}
    \centering
    \includegraphics[trim=0 0 0 0, width=\textwidth]{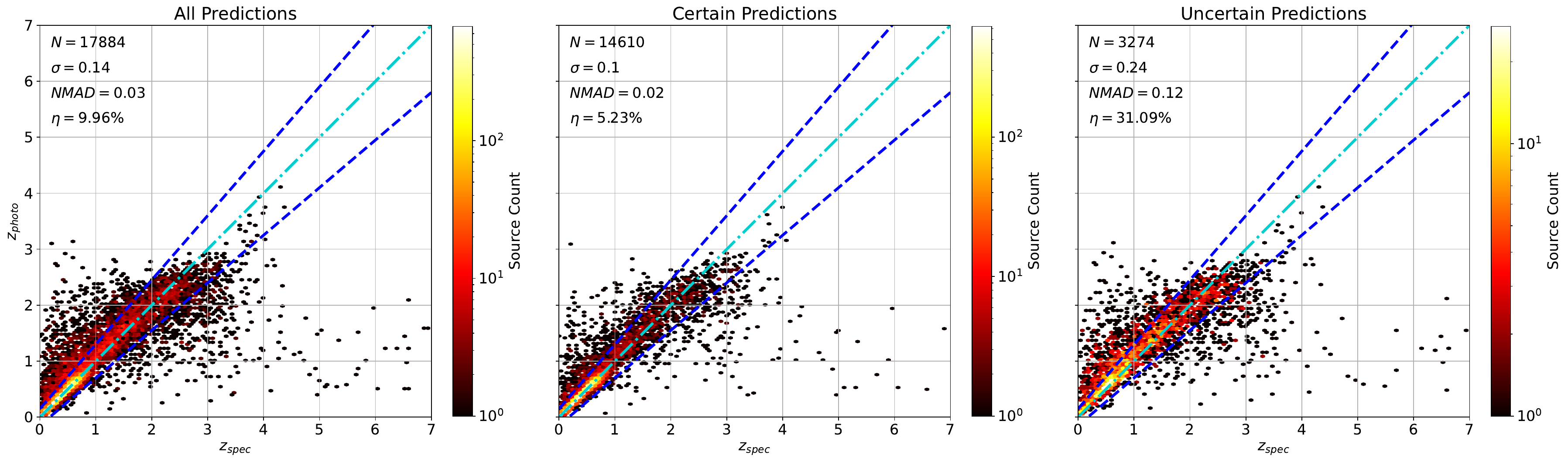 }
    \caption{As with Figure~\ref{fig:appendix_certain_knn}, using the \ac{RF} algorithm. }
    \label{fig:appendix_certain_rf}
\end{figure*}

\begin{figure*}
    \centering
    \includegraphics[trim=0 0 0 0, width=\textwidth]{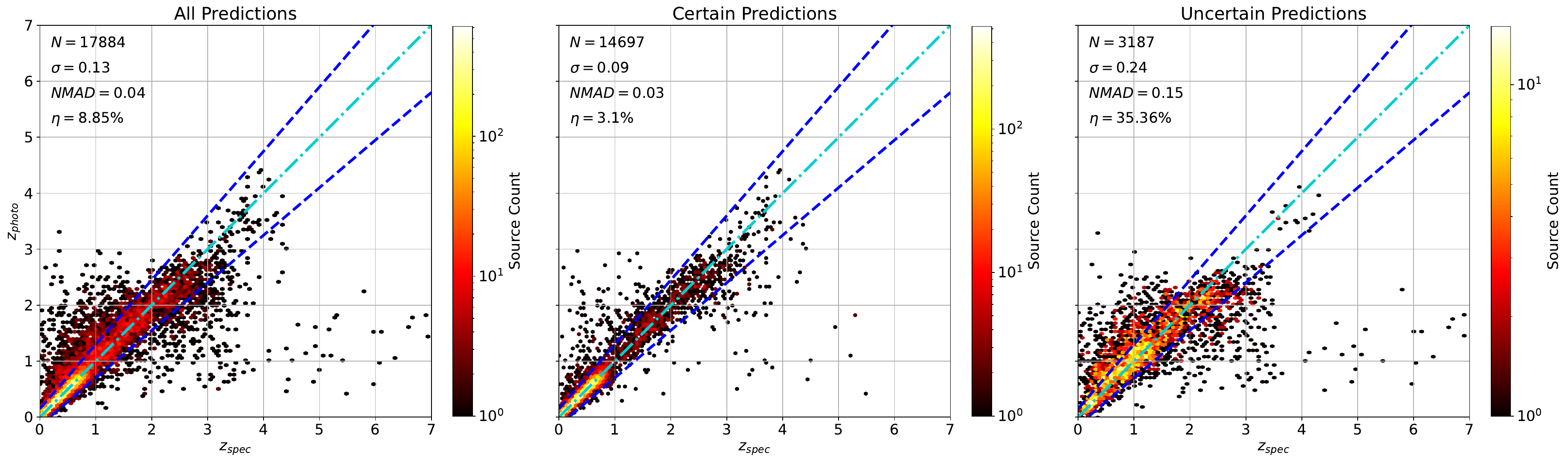 }
    \caption{As with Figure~\ref{fig:appendix_certain_knn}, using the ANNz algorithm. }
    \label{fig:appendix_certain_annz}
\end{figure*}

\begin{figure*}
    \centering
    \includegraphics[trim=0 0 0 0, width=\textwidth]{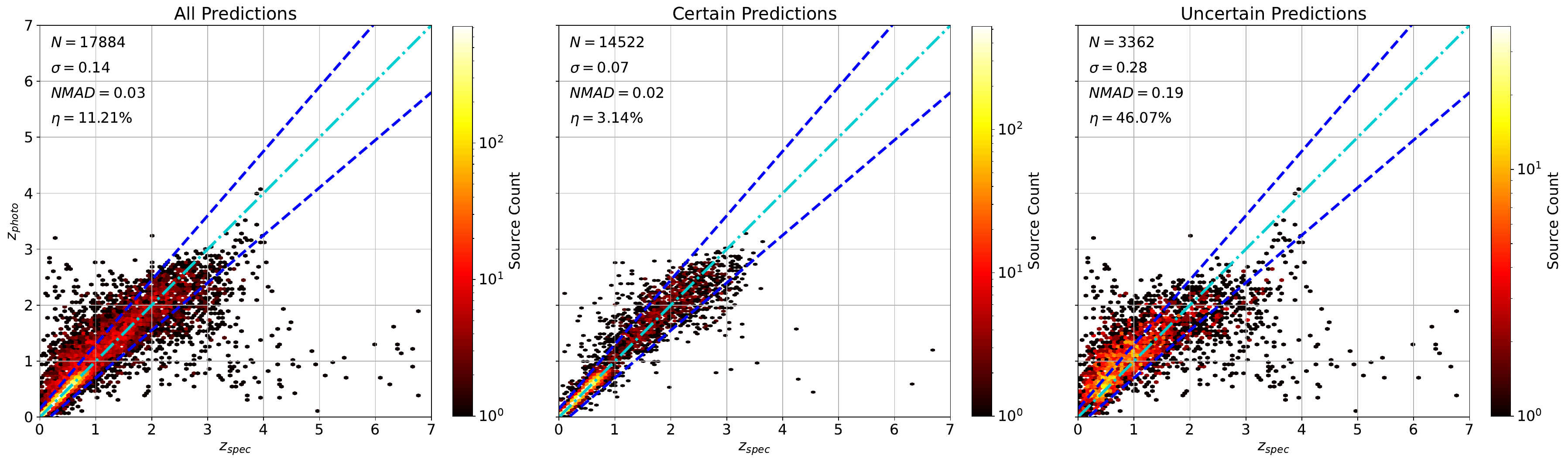 }
    \caption{As with Figure~\ref{fig:appendix_certain_knn}, using the GPz algorithm. }
    \label{fig:appendix_certain_gpz}
\end{figure*}

Figures~\ref{fig:appendix_certain_knn}, \ref{fig:appendix_certain_rf}, \ref{fig:appendix_certain_annz}, and \ref{fig:appendix_certain_gpz} show plots similar to the top panel of Figure~\ref{fig:knn_regress}, allowing for comparisons between all predictions, just those predictions deemed `certain' by the criteria in Section~\ref{sec:removing_uncertain}, and those that don't meet the criteria, and are therefore deemed `uncertain'. Across all algorithms, much of the scatter between the predicted and measured redshift is removed from the `certain' sample, with all algorithms benefiting across all error metrics. The kNN algorithm retains the lowest outlier rate. However,  the majority of its `certain' sources lie between $0 < z < 2.5$, with few sources beyond $z > 2.5$. The ANNz and GPz algorithm perform next best in terms of outlier rate, though the GPz algorithm performs better in both $\sigma$ and \ac{NMAD}. The GPz algorithm removes most $z > 3$ sources, with the ANNz algorithm extending up to $z < 4$. The \ac{RF} algorithm performs worst, with predictions capped at $ z < 3$.

\bibliography{bibliography}

\end{document}